\documentclass[acmsmall,screen]{acmart}

\usepackage{tikz}
\usepackage{amsmath}
\usepackage{tabularx}
\usepackage{caption}
\usepackage{multirow}
\usepackage{multicol}
\usepackage{makecell}
\usepackage{adjustbox}
\usepackage{subfigure}
\usepackage{booktabs}

\usepackage{amsmath,amssymb,amsfonts}
\usepackage{filecontents}
\usepackage{tabularx}
\usepackage{listings}
\usepackage{multirow}
\usepackage{multicol}
\usepackage{xcolor}
\usepackage{cleveref}
\usepackage{acronym}
\usepackage{xspace,url}
\usepackage{bbding}
\usepackage{pifont}
\usepackage{wasysym}
\usepackage{tcolorbox}
\usepackage{colortbl}

\usepackage[ruled,linesnumbered]{algorithm2e}
\usepackage[figuresleft]{rotating}

\usepackage{pdflscape}

\newcommand{\etal}{\textit{et al}.~}
\newcommand{\ie}{\textit{i}.\textit{e}.~}
\newcommand{\eg}{\textit{e}.\textit{g}.~}
\newcommand{\trainbeforepartition}{training-before-partition\xspace}
\newcommand{\partitionbeforetrain}{partition-before-training\xspace}

\acrodef{fl}[FL]{Federated Learning}
\acrodef{gdpr}[GDPR]{the General Data Protection Regulation}
\acrodef{asr}[ASR]{attack success rate}
\acrodef{tsdp}[TSDP]{\textit{TEE-shielded DNN partition}}
\acrodef{mia}[MIA]{\textbf{Membership Inference Attack}}
\acrodef{ms}[MS]{\textbf{Model Stealing}}
\acrodef{dp}[DP]{Differential Privacy}
\acrodef{mpc}[MPC]{Multi-Party Computation}
\acrodef{he}[HE]{Homomorphic Encryption}

\newcommand{\diff}[1]{{#1}} %
\newcommand{\minor}[1]{{#1}} %

\newcommand{\sys}{\textsc{TEESlice}\xspace} %
\newcommand{\tool}{\textsc{TEESlice}\xspace} %
\newcommand{\nettailor}{\textsc{NetTailor}\xspace}
\newcommand{\lora}{\textsc{LoRA}\xspace}

\newcommand{\F}{Fig.}
\newcommand{\E}{Eq.}
\newcommand{\T}{Table}
\renewcommand{\S}{Sec.}
\newcommand{\A}{Alg.}
\newcommand{\App}{Appx.}

\newcommand{\parh}[1]{\noindent\textbf{#1}}

\newif\iflongappendix
\newif\ifentireresult
\entireresulttrue %

\newcommand{\tsdp}{TSDP\xspace}%
\newcommand{\shieldwhole}{shielding-whole-model\xspace}%
\newcommand{\noshield}{white-box\xspace}%

\SetCommentSty{mycommfont}

\newcommand\diagfil[4]{%
  \multicolumn{1}{p{#1}}{\hskip-\tabcolsep
  $\vcenter{\begin{tikzpicture}[baseline=0,anchor=south west,inner sep=0pt,outer sep=0pt]
  \path[use as bounding box] (0,0) rectangle (#1+2\tabcolsep,\baselineskip);
  \node[minimum width={#1+2\tabcolsep},minimum height=\baselineskip+\extrarowheight+\belowrulesep+\aboverulesep,fill=#2] (box)at(0,-\aboverulesep) {};
  \fill [#3] (box.south west)--(box.north east)|- cycle;
  \node[anchor=center] at (box.center) {#4};
  \end{tikzpicture}}$\hskip-\tabcolsep}}

\AtBeginDocument{%
  \providecommand\BibTeX{{%
    \normalfont B\kern-0.5em{\scshape i\kern-0.25em b}\kern-0.8em\TeX}}}

\setcopyright{acmlicensed}
\copyrightyear{2024}
\acmYear{2024}
\acmDOI{XXXXXXX.XXXXXXX}

\acmJournal{TOSEM}

\begin{document}

\title{TEESlice: Protecting Sensitive Neural Network Models in Trusted Execution Environments When Attackers have Pre-Trained Models}

\author{Ding Li}
\email{ding_li@pku.edu.cn}

\author{Ziqi Zhang}
\email{ziqi\_zhang@pku.edu.cn}
\authornote{Corresponding Author.}

\author{Mengyu Yao}
\email{mengyuyao@stu.pku.edu.cn}

\author{Yifeng Cai}
\email{caiyifeng@pku.edu.cn}

\author{Yao Guo}
\email{yaoguo@pku.edu.cn}
\author{Xiangqun Chen}
\email{cherry@pku.edu.cn}
\affiliation{%
  \institution{Key Laboratory of High-Confidence Software Technologies (MOE), School of Computer Science, Peking University}
  \country{China}
}

\renewcommand{\shortauthors}{Ding Li et al.}
\renewcommand{\shorttitle}{TEESlice: Protecting Sensitive Neural Network Models in Trusted Execution Environments When Attackers have Pre-Trained Models}
\begin{abstract}
Trusted Execution Environments (TEE) are used to safeguard on-device models. However, directly employing TEEs to secure the entire DNN model is challenging due to the limited computational speed. Utilizing GPU can accelerate DNN's computation speed but commercial widely-available GPUs usually lack security protection. To this end, scholars introduce \ac{tsdp}, a method that protects privacy-sensitive weights within TEEs and offloads insensitive weights to GPUs. Nevertheless, current methods do not consider the presence of a knowledgeable adversary who can access abundant publicly available pre-trained models and datasets. This paper investigates the security of existing methods against such a knowledgeable adversary and reveals their inability to fulfill their security promises. Consequently, we introduce a novel partition before training strategy, which effectively separates privacy-sensitive weights from other components of the model. Our evaluation demonstrates that our approach can offer full model protection with a computational cost reduced by a factor of 10. 
In addition to traditional CNN models, we also demonstrate the scalability to large language models. Our approach can compress the private functionalities of the large language model to lightweight slices and achieve the same level of protection as the shielding-whole-model baseline.

\end{abstract}



\maketitle
\section{Introduction}

\diff{Deep Neural Networks (DNNs) and recent Large Language Models (LLMs) have emerged as a significant category of intelligent software for user devices. These applications are capable of executing a diverse range of complex AI tasks such as voice assistants~\cite{yad2023va}, image recognition~\cite{drolia2017ir}, and natural language processing~\cite{desai2020lightweight}. Nevertheless, the deployment of intelligent software on user devices presents a novel attack surface in comparison to cloud-based services: \textit{The detailed information of intelligent software (e.g., model weight values) is exposed to potential malicious users of the device.} By having access to this white-box information, adversaries can easily achieve high attack accuracy with significantly lower costs for common attacks like \ac{ms} and \ac{mia}~\cite{hu2022membership,orekondy2019knockoff,jagielski2020high,papernot2016towards,carlini2019the,leino2020stolen}. These attacks pose a serious threat to the security of the intelligent software, its intellectual property, and the sensitive data privacy (\eg training data privacy) of the software owners. Hence, a primary goal in fortifying on-device intelligent software is to {thwart} adversaries from obtaining the white-box information, thereby {transforming} straightforward and effective white-box \ac{ms} and \ac{mia} attacks into black-box (considerably more challenging) scenarios~\cite{hu2022membership,mo2020darknetz,hou2021model,sun2020shadownet}.}

\diff{
In this paper, we focus on protecting DNN models on clients' devices, which are equipped with Trusted Execution Environments (TEEs) and low-grade commercial GPUs. TEEs are commonly used to safeguard on-device intelligent software~\cite{hanzlik2021mlcapsule,lee2019occlumency,kim2020vessels,li2021lasagna}. Like traditional software is protected, TEEs ensure that sensitive data (e.g., private keys) are kept separate from the system environment, making it inaccessible to formidable adversaries like malicious operating systems and administrators. Compared to other methods of protection at the algorithmic level, such as \ac{mpc}~\cite{juvekar2018gazelle}, \ac{he}~\cite{gilad2016cryptonets}, Regularization~\cite{nasr2018machine}, and \ac{dp}~\cite{dwork2014DP}, TEE-based security imposes lower computational overhead on mobile and IoT devices while preserving the accuracy of the secured models~\cite{hu2022membership,tramer2019slalom}. Nevertheless, applying TEEs directly to safeguard entire DNN models poses challenges because low-grade commercial GPUs (\eg GeForce RTX 4090 and RTX A6000) do not provide the functionality of TEE. Although some recent high-end GPUs (\eg Nvidia Hopper GPU Architecture~\cite{NvidiaH100}) provide the functionality of confidential computing, their prices are too high for ordinary model users. An Nvidia H100 GPU is over 15$\times$ more expensive than a GeForce RTX 4090\footnote{At Jun 2024, the price of an H100 GPU is about \$30,000, while the price of a GeForce RTX 4090 is less than \$2,000}. Attempting to shield the complete DNN model within a TEE (\shieldwhole) could result in a 50x reduction in the model's speed.
}

\diff{While safeguarding an entire deep learning model using TEEs may not be practical for on-device scenarios, recent research suggests safeguarding the privacy-sensitive and critical components of the model to ensure both high utility and security simultaneously. Specifically, a concept known as \ac{tsdp} has been proposed. This approach involves splitting a large DNN model into two components: a \textit{privacy-sensitive} component, which is small and contains vital information, and a \textit{privacy-insensitive} component, which is larger and holds less critical data. The privacy-sensitive part operates within TEEs, while the privacy-insensitive part runs on GPUs~\cite{mo2020darknetz,hou2021model,shen2022soter,sun2020shadownet}. The rationale behind \ac{tsdp} is akin to securing conventional software with TEEs, where the privacy sensitive portion (e.g., private keys) is compact and can be protected by TEEs, while the larger portion of the software (e.g., the remaining codebase) operates outside of TEEs~\cite{lazard2018teeshift}.}

\diff{Current \ac{tsdp} approaches generally assume that the portion off-loaded to the GPU does not reveal sensitive information of DNN models. These methods employ retraining techniques to show that even if an attacker uses this portion for \ac{ms} or \ac{mia}, the reconstructed DNN model only achieves a similar accuracy to a black-box baseline~\cite{mo2020darknetz,hou2021model,shen2022soter,sun2020shadownet}, which is significantly lower than the white-box accuracy. Previous \ac{tsdp} studies rely on empirical experiments to prove that the disclosed model components do not leak significantly more information than a black-box interface~\cite{hou2021model,sun2020shadownet,mo2020darknetz,shen2022soter}.}

\diff{This paper examines the security promises provided by current \ac{tsdp} solutions in the presence of a more sophisticated and cunning adversary in the age of large language models. Specifically, we explore a realistic threat scenario where \textit{the adversary can leverage readily available public information from the Internet, such as pre-trained models and public datasets}~\cite{chen2021teacher,chen2022copy,wang2018with}. With the prevalence of large language models, it has become common for software developers to utilize publicly accessible models to accelerate the development of proprietary software. Previous studies have demonstrated that these public models can be exploited to compromise private software~\cite{Sitawarin2023DefendingAT}. To undermine \tsdp, attackers can use public information to scrutinize outsourced model components and obtain more information on privacy beyond simply analyzing black-box output, thus undermining the security guarantees of \ac{tsdp}. However, none of the existing methods has thoroughly assessed their security guarantees in the presence of public information. Therefore, we contend that it is crucial to systematically evaluate the security assurances of \ac{tsdp} solutions under this threat landscape.}

\diff{To investigate the security of \ac{tsdp} methods, our initial step involves conducting a comprehensive review of the existing literature on \ac{tsdp}. We analyze publications released from 2018 to 2023 in reputable conferences such as IEEE S\&P, MobiSys, ATC, ASPLOS, RTSS, MICRO, AAAI, ICLR, ICML, PETs, MICRO, and TDSC. Each paper's technical approaches are scrutinized, and we classify them into five distinct categories based on their primary contributions. These categories include fortifying deep layers (\ding{172}), fortifying shallow layers (\ding{173}), fortifying high-magnitude weights (\ding{174}), fortifying intermediate layers (\ding{175}), and fortifying non-linear layers (\ding{176}). Subsequently, we choose one exemplary paper for each category and proceed to implement its technical methodology.}

\diff{After categorizing the existing \ac{tsdp} approaches, we perform a thorough security assessment using a more powerful adversary that has access to public-pre-trained models. Both Membership Inference (\ac{ms}) and Model Inversion Attacks (\ac{mia}) are carried out against the representative \ac{tsdp} solutions we reviewed, and the attack accuracy is compared against two baselines: the black-box baseline (\shieldwhole) offers the highest security assurance but the lowest utility, while the white-box baseline (where the entire Deep Neural Network model is offloaded outside of TEE) provides the highest utility but lacks security protection. The experiment results reveal that current \ac{tsdp} methods inadvertently expose significant private information to attackers through offloaded model weights, allowing attacks of almost white-box quality against TEE-protected models. The accuracy of \ac{ms} attacks on existing \ac{tsdp} solutions is {$3.85\times$ -- $4.56\times$} higher than that of the black-box (\shieldwhole) baseline. On the contrary, the unprotected \noshield baseline demonstrates a {$4.57\times$} higher accuracy compared to the \shieldwhole configuration. The results for \ac{mia} attacks show a similar trend, with existing \ac{tsdp} methods exhibiting $1.16\times$ -- {$1.36\times$} higher \ac{mia} accuracy than the \shieldwhole baseline, while the accuracy for the \noshield setup is {$1.37\times$} higher.}

\diff{Furthermore, we found that significant challenges were faced in improving the security of established \ac{tsdp} methods without fundamentally altering their approaches. For example, we evaluated the effectiveness of \ac{ms}/\ac{mia} attacks using various setups of current \ac{tsdp} techniques. Identifying an optimal configuration that balances a DNN model's performance with security requirements proved to be particularly challenging. Specifically, achieving a high level of tolerance to attacks requires distinct settings to configure the protected component when protecting different models and datasets. Thus, a thorough empirical process is essential to determine the ideal configuration customized to specific models and datasets within all existing \ac{tsdp} strategies. However, conducting such empirical analyses is excessively costly due to the large number of possible combinations of models and datasets.}

\diff{During our literature survey and empirical evaluation, we found that the
fundamental weakness of existing \tsdp approaches is that they follow a
\textit{\trainbeforepartition} strategy. This involves first training a private
model with a public pre-trained model and private data, and then separating the
model into two parts: a shielded part that runs in TEEs, and an offloaded part
that runs out of TEEs. Since training occurs before model partitioning,
privacy-related weights may likely pervade the entire model. Therefore, it is hard for existing \ac{tsdp} solutions to accurately \textit{isolate} privacy-related
weights, creating potential attack surfaces. }

\diff{In order to enhance the security of \ac{tsdp} solutions against the new threat model, we introduce a novel \ac{tsdp} framework named \sys. This framework effectively separates privacy-sensitive weights from outsourced weights during the inference phase. Unlike the \trainbeforepartition approach used in prior research, \sys employs a \partitionbeforetrain strategy. This method involves initially dividing a DNN model into a backbone and several private segments, utilizing publicly pre-trained models as the backbone, and then training the segments with private data. Consequently, \sys effectively isolates privacy-related weights from offloaded weights and ensures the protection of all privacy-sensitive weights in TEEs.}

\diff{The primary difficulty in implementing the \partitionbeforetrain approach lies in guaranteeing that individual segments are of a manageable size for execution in TEEs without compromising on accuracy. To address this challenge, we suggest employing a dynamic pruning method. Initially, the private segments are trained with larger sizes to ensure they possess adequate model capacity for achieving high accuracy. Subsequently, the algorithm automatically adjusts the segment sizes to stay below a specified threshold of accuracy loss. Through this process, \sys is able to identify the optimal configuration, or "sweet spot," that minimizes the number of segments (computation) within the TEE while preserving the accuracy level of the non-partitioned model.}

\diff{Our evaluation indicates that \sys\ surpasses existing \ac{tsdp} methods in terms of both security assurance and utility cost. It is challenging for attackers to extract sensitive information through the analysis of model structures, demonstrating that \sys\ achieves a security level equivalent to the \shieldwhole\ baseline with a computational cost that is $10\times$ lower compared to alternative \tsdp\ solutions, in both experimental and real-world scenarios. Additionally, \sys\ attains a high level of security with minimal trade-offs. Statistical analysis reveals no discernible differences in accuracy between the protected \sys\ model and the original unpartitioned model. Furthermore, the outsourced public backbone does not enhance the efficacy of attacks. Our evaluation also shows that \sys\ can effectively protect large language models with LoRA.}
The contribution of this paper can be summarized as follows:

\begin{itemize}

    \item We systematically evaluate the security guarantee of previous \tsdp solutions using two representative attacks, \ac{ms} and \ac{mia}, and reveal the security issues of these solutions.
    \item We illustrate the difficulty in improving the security of previous \tsdp approaches without substantially changing their methodologies. 

    \item We propose \sys, a novel \tsdp\ solution for DNN inference that isolates privacy from off-loaded model parts to provide a strong security guarantee using TEEs and cryptographic primitives. Our detailed evaluation shows that \sys offers a high security guarantee with moderate overhead and no accuracy loss.
\end{itemize}

\minor{
This paper is an extended version of a conference paper~\cite{zhang2024no}. The conference paper categorized existing \tsdp solutions, evaluated their security on three representative models, and proposed \tool on the CNN models. This paper includes additional content compared with the conference paper. 
First, this paper conducts a more comprehensive review of existing \tsdp solutions, including the scenarios, threat model, design insight, evaluated attacks, outsourced data security, and limitations. 
Second, this paper includes more experiments on the security evaluation of existing \tsdp work and demonstrates the scalability of the observation.
Third, this paper proposes an extended approach of \tool that can be applied to large language models. Our evaluation demonstrates the effectiveness of the approach to protect large language models.
}

\noindent \textbf{Availability.} The artifacts are available
at~\cite{TEESliceArtifact} \diff{and ~\cite{TEESliceLLM}.} 

\noindent \textbf{Overview.} In \S~\ref{sec:threat_model}, we will introduce the background and the threat model. In \S~\ref{sec:evaluate_existing_solutions}, we survey existing TSDP solutions and evaluate their defense effectiveness. Based on the vulnerability in \S~\ref{sec:evaluate_existing_solutions}, in \S~\ref{sec:dilemma}, we further reveal that it is difficult to mitigate the vulnerability straightforwardly. In \S~\ref{sec:approach}, we summarize the fundamental reason for the weaknesses of existing \tsdp and propose our solution, TEESlice. In \S~\ref{sec:teeslice_experiment}, we comprehensively evaluate TEESlice with other \tsdp solutions. At last, we present threats to validity (\S~\ref{sec:threats_to_validity}), related work (\S~\ref{sec:other_related_work}), and discussion (\S~\ref{sec:discussion}).

\section{Background and Threat Model}
\label{sec:threat_model}

\subsection{Background}

\parh{Trusted Execution Environment (TEE).}
A Trusted Execution Environment (TEE) is an isolated hardware enclave that
stores and processes sensitive data. \diff{The trusted enclave can be implemented as a
part of the process address space (Intel SGX~\cite{mckeen2013innovative}), a
virtual machine (AMD SEV~\cite{kaplan2016amd}, Intel TDX~\cite{IntelTDX}, ARM
CCA~\cite{ArmCCA}, and HyperEnclave~\cite{jia2022hyperenclave}), or a separate
system apart from the normal system (TrustZone~\cite{alves2004trustzone}). TEE
provides a high level of security guarantee against privileged attackers. The
defendable adversaries include OS-level process and malicious party who has
physical access to the platform (\eg maintenance staff). Evaluating the security guarantee of TEEs in front of various attacks (\eg side-channel attacks~\cite{li2022systematic,wang2017leaky,fei2021security,nilsson2020survey}) and enhancing TEE security~\cite{shih2017t,liu2015oblivm,oleksenko2018varys} are important topics in the security community. Recently, Nvidia's Hopper architecture provides a confidential computing mode to protect data privacy~\cite{NvidiaH100}. However, such a feature is only available in recent high-end GPUs and is not supported by low-grade commercial GPUs (\eg Nvidia RTX 4090 and Nvidia Ampere architecture). 
}

In this paper, we
follow prior work and deem TEE as \textit{a secure area on a potential adversary
host device (including
GPUs)}~\cite{mo2020darknetz,hou2021model,shen2022model,sun2020shadownet}. It
means \textit{the data, code, and the whole computation process} inside TEEs are
secure. Although there are side-channel attacks that may leak sensitive data
from TEE, they are out of our consideration. 

\diff{
\parh{Security of DNN-based Intelligent Software.}
With the fast development and wide deployment of intelligent software,
researchers found that its security and privacy become an important concern. The
intelligent software can be vulnerable to various attacks, leading to severe
consequences such as privacy leakage~\cite{shokri2017membership}, intellectual
property theft~\cite{orekondy2019knockoff}, and
misbehavior~\cite{jagielski2020high}. In this paper, we focus on two
representative attacks: \textit{Model Stealing} (MS) and \textit{Membership
Inference Attack} (MIA). MS aims at the functionality of the target intelligent
software, while MIA aims at the private training data. MS builds a surrogate
model with a similar functionality as the target software by either directly
stealing the model weights~\cite{rakin2022deepsteal,zhu2021hermes,sun2021mind}
or training a local
model~\cite{orekondy2019knockoff,jagielski2020high,shen2022model}. MIA predicts
whether a sample is in the training dataset of the target
software~\cite{melis2019exploiting, papernot2016towards}.}

\diff{
\parh{Availability of Public Models and Data.}
There are various public DNN models on the Internet that can be
downloaded for free~\cite{pytorchhub,
tensorflowhub,caffemodelzoo,tensorflow_tl}. Companies upload
trained models online to facilitate the development of the AI
community~\cite{pytorchmodelzoo,tensorflowModelGarden} and help individual
developers to train models with low cost~\cite{zhuang2021a}. Such models are
usually trained on public datasets~\cite{deng2009imagenet,lin2014coco} and
contain no private information of the model owner. Recent research on ML
security shows that these public models can be leveraged to improve both ML
attacks and defenses~\cite{wang2018with,chen2021teacher,mo2021ppfl}.}

\begin{figure}[!t]
\centering
\includegraphics[width=0.8\linewidth]{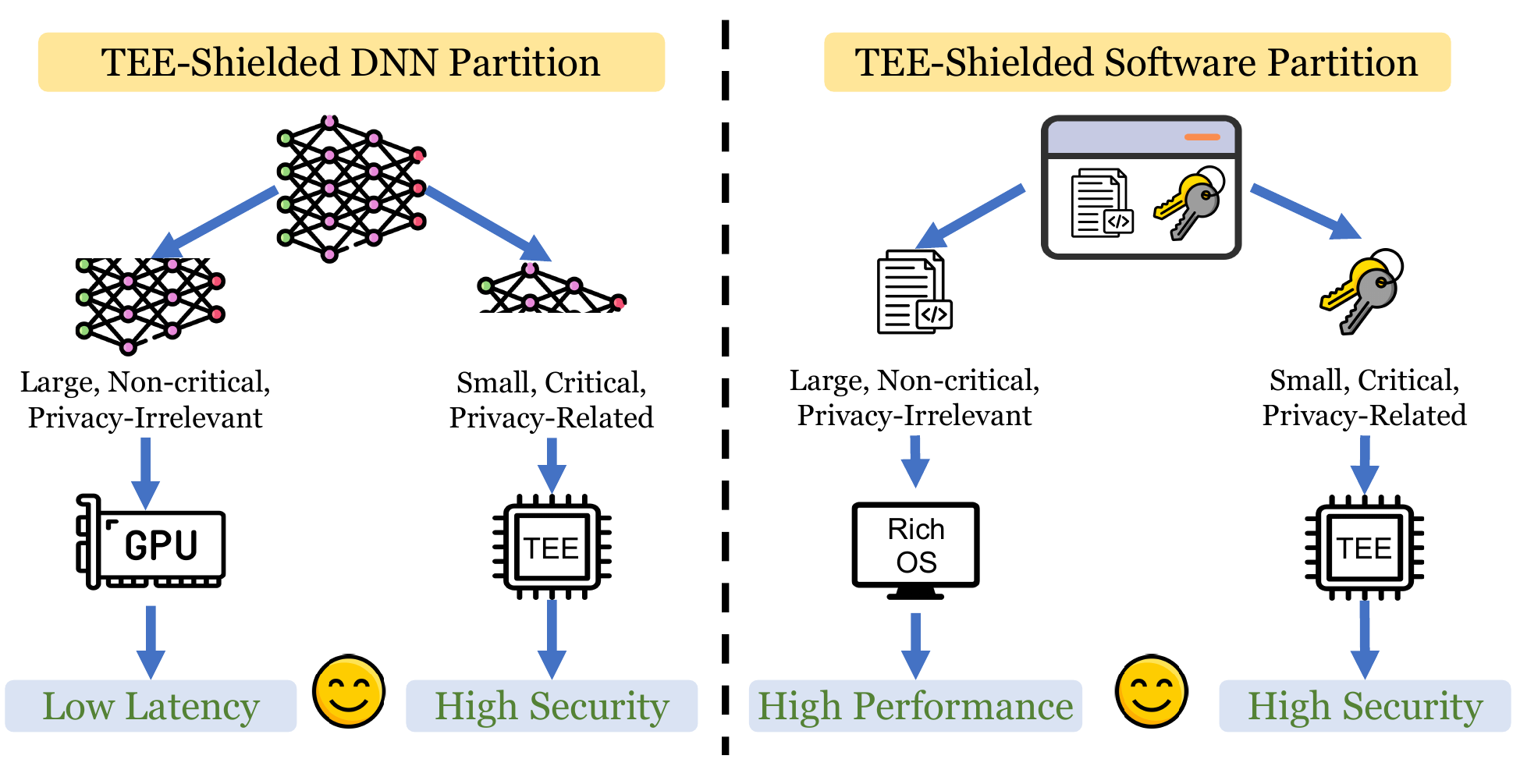}
\caption{An illustration of TSDP solutions.}
\label{fig:TSDP}
\vspace{-15pt}
\end{figure}

\parh{TSDP Solutions.}
\diff{
TSDP aims to make full use of both the security of TEEs and the computation
power of GPUs. As shown in \F~\ref{fig:TSDP}, the idea of TSDP is similar to
TEE-shielded software partition, which only protects security-sensitive part of the application to reduce the trusted computation base~\cite{lind2017glamdring,tsai2020civet,atamli2015securing,atamli2017framework}. For traditional software, a common practice to
use TEE is to select the privacy-sensitive keys or libraries and shield them
inside TEEs, while leaving the rest parts to the untrusted rich OS. In this way, the
software can achieve both high security and high performance. Similarly, \tsdp
solutions select the important DNN layers/weights and shield them inside TEEs,
while leaving the rest parts to the untrusted GPU. 
}

\subsection{Threat Model} 
\label{sec:threat_model}

\parh{Defender's Goal.}
TSDP solutions (and the defenders) aim to provide a black-box label-only protection against MS/MIA
by shielding partial DNN models inside TEEs. The motivation is to reduce
inference latency of the straightforward black-box protection that shields the
whole model inside TEEs (increase latency by up to
50$\times$~\cite{tramer2019slalom}).
The security goal of \tsdp solutions is to \textit{downgrade white-box MS/MIA
against on-device models to black-box label-only
attacks}~\cite{mo2020darknetz,hou2021model,sun2020shadownet,shen2022soter}. Such
degeneration is important and practical for deployed DNN models in production
environments. For MS, \tsdp solutions enforce accurate, cheap (usually taking
negligible number of queries) white-box
attacks~\cite{zhu2021hermes,rakin2022deepsteal} to expensive (usually taking
tens of thousands of queries) and inaccurate black-box
attacks~\cite{orekondy2019knockoff}. For MIA, \tsdp solutions provide a
deployment framework to guarantee differential privacy requirements with little
accuracy sacrifice~\cite{hu2022membership}.
We consider the security guarantee of a
black-box baseline, where TEE shields the whole DNN model and only returns
prediction labels, as the \textit{upper bound} security protection offered by
\ac{tsdp}
approaches~\cite{mo2020darknetz,hou2021model,shen2022soter,sun2020shadownet}. We
however, do \textit{not} aim to mitigate information leakage from TEE completely
outputs (i.e., prediction labels).

\parh{Attacker's Goal.}
The goal of the attacker is to steal the secrets of the model that is deployed on the user's device. Specifically, the attacker wants first to steal the model weights (white-box information) to copy the model functionality and then infer the training data privacy based on the stolen model weights.

\parh{Defender's Capability.}
We assume the defender (the model owner) can utilize public models and datasets on the Internet to help the defense~\cite{wang2018with,chen2021teacher,pal2019a}. The defender can also utilize TEEs on the user's device to protect the sensitive part of the model. The defender can remotely attest that the private model part is successfully deployed in the TEE.

\parh{Attacker's Capability.}
Similar to the defender, we assume the attacker can utilize public information to improve the accuracy of the model or attacks, which is a realistic setting for modern on-device learning tasks~\cite{pytorchhub,
tensorflowhub,tensorflow_tl,pytorchmodelzoo,zhuang2021a,deng2009imagenet,wang2018with,chen2021teacher}. The attacker can infer the architecture of the whole protected model, or an equivalent one, based on the
public information, such as the inference results or the unprotected model part,
with existing
techniques~\cite{chen2021teacher,chen2022copy,hou2021model,mo2020darknetz,hashemi2021darknight}.
Besides, we assume that the attacker can query the victim model for limited times
(less than 1\% of the training data), a practical assumption shared by related
work~\cite{rakin2022deepsteal,hua2018reverse,yan2020cache}. For simplicity, we
denote the victim model as $M_{\rm vic}$, the public model as $M_{\rm pub}$, and
the surrogate model produced by model stealing as $M_{\rm sur}$.
We assume the attacker can utilize malicious software or even control the OS to perform the attack. This setting is realistic because the model is deployed to the user's device. We do not consider physical attacks (\eg voltage~\cite{maistri2014electromagnetic,tang2017clkscrew,murdock2020plundervolt,lipp2021platypus} and electromagnetic~\cite{mangard2016cache}) against TEE because although prior work has demonstrated that such attacks are very strong~\cite{munoz2023survey}, the attack cost (time cost and human effort) is high, and the cost may not be worth the gain.

\section{Literature Review}

\begin{table}[!htbp]
    \caption{A summary of TSDP solutions. We
covered the top-tier conferences over the last five years across diverse fields,
including computer security, mobile computing, machine learning, and computer
systems. For
each paper, we summarized seven aspects: the scenario, threat model, applied stage, design
principles, design scenarios, evaluated attacks, the security of outsourced data, and potential limitations. }
    \centering
    \label{tbl:literature_long}
    \begin{adjustbox}{width=\linewidth}

    \begin{tabular}{@{}lcclcllllll@{}}
    \toprule
                        & \multirow{2}{*}{Solution}                                            & \multirow{2}{*}{Venue}                                  & \multicolumn{1}{c}{\multirow{2}{*}{Scenario}}                                        & \multirow{2}{*}{Stage}                                       & \multicolumn{1}{c}{\multirow{2}{*}{Threat Model}}                                                                       & \multicolumn{1}{c}{\multirow{2}{*}{Design Principle}}                                                                      & \multicolumn{1}{c}{\multirow{2}{*}{Evaluated Attacks}}                                                  & \multicolumn{2}{c}{Outsourced Data Security}                                                                                                                                                   & \multicolumn{1}{c}{\multirow{2}{*}{Limitation}}                                                                \\ \cmidrule(lr){9-10}
                        &                                                                      &                                                         & \multicolumn{1}{c}{}                                                                 &                                                              & \multicolumn{1}{c}{}                                                                                                    & \multicolumn{1}{c}{}                                                                                                       & \multicolumn{1}{c}{}                                                                                    & \multicolumn{1}{c}{Confidentiality}                                                           & \multicolumn{1}{c}{Integrity}                                                          & \multicolumn{1}{c}{}                                                                                           \\ \midrule
    \multirow{20}{*}{\rotatebox{90}{Single-Point Partition}}  & \begin{tabular}[c]{@{}c@{}}Ternary \\ Model\\ Partition~\cite{gu2018yerbabuena}\end{tabular} & \begin{tabular}[c]{@{}c@{}}Arxiv \\ 2018\end{tabular}   & \begin{tabular}[c]{@{}l@{}}DNN outsourcing\\ on cloud servers\end{tabular}           & Inference                                                    & \begin{tabular}[c]{@{}l@{}}Server is untrusted\\ Protect input and label\end{tabular}                                   & \begin{tabular}[c]{@{}l@{}}Only shallow intermediate \\ features can be used to \\ reconstruct input\end{tabular}          & Input Reconstruction                                                                                    & -                                                                                             & -                                                                                      & \multirow{14}{*}{\begin{tabular}[c]{@{}l@{}}Protection \\ is based on \\ heuristic \\ observation\end{tabular}} \\ \cmidrule(lr){2-10}
                        & Origami~\cite{narra2019origami}                                                              & \begin{tabular}[c]{@{}c@{}}Arxiv \\ 2019\end{tabular}   & \begin{tabular}[c]{@{}l@{}}DNN outsourcing\\ on cloud servers\end{tabular}           & Inference                                                    & \begin{tabular}[c]{@{}l@{}}Server is untrusted\\ Protect input privacy\end{tabular}                                     & \begin{tabular}[c]{@{}l@{}}Feature maps after first \\ several layers can not be \\ used to reconstruct input\end{tabular} & \begin{tabular}[c]{@{}l@{}}GAN-based\\ input reconstruction\end{tabular}                                & -                                                                                             & -                                                                                      &                                                                                                                \\ \cmidrule(lr){2-10}
                        & DarkneTZ~\cite{mo2020darknetz}                                                             & \begin{tabular}[c]{@{}c@{}}MobySys \\ 2020\end{tabular} & \begin{tabular}[c]{@{}l@{}}Deploy DNN to \\ edge devices\end{tabular}                & Inference                                                    & \begin{tabular}[c]{@{}l@{}}Device owner is untrusted. \\ Protect model privacy\end{tabular}                             & \begin{tabular}[c]{@{}l@{}}Only deep layers can expose\\ membership information\end{tabular}                               & \begin{tabular}[c]{@{}l@{}}White-box\\ membership inference\end{tabular}                                & -                                                                                             & -                                                                                      &                                                                                                                \\ \cmidrule(lr){2-10}
                        & Shredder~\cite{mireshghallah2020shredder}                                                             & \begin{tabular}[c]{@{}c@{}}ASPLOS \\ 2020\end{tabular}  & \begin{tabular}[c]{@{}l@{}}DNN outsourcing\\ on cloud servers\end{tabular}           & Inference                                                    & \begin{tabular}[c]{@{}l@{}}Server is untrusted.\\ Protect user input.\end{tabular}                                      & \begin{tabular}[c]{@{}l@{}}Learn additive noise to \\ protect privacy in the \\ internal features.\end{tabular}            & Mutual information                                                                                      & Additive noise                                                                                & -                                                                                      &                                                                                                                \\ \cmidrule(lr){2-10}
                        & Serdab~\cite{elgamal2020serdab}                                                               & \begin{tabular}[c]{@{}c@{}}CCGRID \\ 2020\end{tabular}  & \begin{tabular}[c]{@{}l@{}}DNN outsourcing\\ cloud servers\end{tabular}              & Inference                                                    & \begin{tabular}[c]{@{}l@{}}Server is untrusted.\\ Protect user input.\end{tabular}                                      & \begin{tabular}[c]{@{}l@{}}Feature maps after first \\ several layers are dissimilar \\ to the input\end{tabular}          & -                                                                                                       & -                                                                                             & -                                                                                      &                                                                                                                \\ \cmidrule(l){2-11} 
                        & eNNclave~\cite{schlogl2020ennclave}                                                             & \begin{tabular}[c]{@{}c@{}}AISec \\ 2020\end{tabular}   & \begin{tabular}[c]{@{}l@{}}DNN outsourcing\\ on cloud servers\end{tabular}           & Inference                                                    & \begin{tabular}[c]{@{}l@{}}Server is untrusted.\\ Protect model privacy\end{tabular}                                    & \begin{tabular}[c]{@{}l@{}}Keep all confidential \\ model parameters in \\ the last (dense) layers\end{tabular}            & -                                                                                                       & -                                                                                             & -                                                                                      & \begin{tabular}[c]{@{}l@{}}Accuracy \\ loss\end{tabular}                                                       \\ \cmidrule(l){2-11} 
                        & PPFL~\cite{mo2021ppfl}                                                                 & \begin{tabular}[c]{@{}c@{}}MobySys \\ 2021\end{tabular} & Federated learning                                                                   & \begin{tabular}[c]{@{}c@{}}Training\\ Inference\end{tabular} & \begin{tabular}[c]{@{}l@{}}Both server and clients \\ are untrusted.\\ Protect model privacy.\end{tabular}              & \begin{tabular}[c]{@{}l@{}}Putting parameters in \\ the TEE of edge devices \\ and servers\end{tabular}                    & \begin{tabular}[c]{@{}l@{}}Data reconstruction\\ Property inference\\ Membership inference\end{tabular} & -                                                                                             & -                                                                                      & \begin{tabular}[c]{@{}l@{}}Difficult to \\ scale to large \\ DNN models\end{tabular}                           \\ \midrule
    \multirow{35}{*}{\rotatebox{90}{Multi-Point Partition}} & Slalom~\cite{tramer2019slalom}                                                               & \begin{tabular}[c]{@{}c@{}}ICLR \\ 2018\end{tabular}    & \begin{tabular}[c]{@{}l@{}}DNN outsourcing\\ on cloud servers\end{tabular}           & Inference                                                    & \begin{tabular}[c]{@{}l@{}}Server is untrusted\\ Protect input privacy\end{tabular}                                     & \begin{tabular}[c]{@{}l@{}}Outsource computation \\ from a TEE to a \\ co-located GPU\end{tabular}                         & -                                                                                                       & One-Time-Pad                                                                                  & \begin{tabular}[c]{@{}l@{}}Freivalds\\ Algorithm\end{tabular}                          & \begin{tabular}[c]{@{}l@{}}Pre-compute\\ OTP\end{tabular}                                                      \\ \cmidrule(l){2-11} 
                        & AegisDNN~\cite{xiang2021aegisdnn}                                                             & \begin{tabular}[c]{@{}c@{}}RTSS \\ 2021\end{tabular}    & \begin{tabular}[c]{@{}l@{}}Deploy DNN in \\ safety-critical \\ systems\end{tabular}  & Inference                                                    & \begin{tabular}[c]{@{}l@{}}Server is untrusted.\\ The correctness of model\\ output can be corrupted\end{tabular}       & \begin{tabular}[c]{@{}l@{}}Find out critical layers \\ for the requested level \\ of dependability\end{tabular}            & Fault injection attack                                                                                  & -                                                                                             & -                                                                                      & \begin{tabular}[c]{@{}l@{}}Do not protect\\ privacy\end{tabular}                                               \\ \cmidrule(l){2-11} 
                        & DarKnight~\cite{hashemi2021darknight}                                                            & \begin{tabular}[c]{@{}c@{}}MICRO \\ 2021\end{tabular}   & \begin{tabular}[c]{@{}l@{}}DNN outsourcing\\ on cloud servers\end{tabular}           & \begin{tabular}[c]{@{}c@{}}Training\\ Inference\end{tabular} & \begin{tabular}[c]{@{}l@{}}Server is untrusted.\\ Protect input privacy and \\ computation integrity\end{tabular}       & \begin{tabular}[c]{@{}l@{}}Use matrix masking \\ to obfuscate input data\end{tabular}                                      & \begin{tabular}[c]{@{}l@{}}Collusion attack\\ from multiple GPUs\end{tabular}                           & \begin{tabular}[c]{@{}l@{}}Linearly combine \\ the inputs and add\\ random noise\end{tabular} & \begin{tabular}[c]{@{}l@{}}Redundant\\ computation\end{tabular}                        & \begin{tabular}[c]{@{}l@{}}Protection fails \\ when all GPUs \\ collude\end{tabular}                           \\ \cmidrule(l){2-11} 
                        & Goten~\cite{lucien2021goten}                                                                & \begin{tabular}[c]{@{}c@{}}AAAI \\ 2021\end{tabular}    & \begin{tabular}[c]{@{}l@{}}DNN outsourcing\\ on cloud servers\end{tabular}           & \begin{tabular}[c]{@{}c@{}}Training\\ Inference\end{tabular} & \begin{tabular}[c]{@{}l@{}}Server is untrusted.\\ Protect input privacy.\end{tabular}                                   & \begin{tabular}[c]{@{}l@{}}Use additive secret \\ sharing to protect privacy\end{tabular}                                  & -                                                                                                       & \begin{tabular}[c]{@{}l@{}}Additive secret\\ share\end{tabular}                               & -                                                                                      & \begin{tabular}[c]{@{}l@{}}Require two \\ TEEs\end{tabular}                                                    \\ \cmidrule(l){2-11} 
                        & Magnitude~\cite{hou2021model}                                                            & \begin{tabular}[c]{@{}c@{}}TDSC \\ 2022\end{tabular}    & \begin{tabular}[c]{@{}l@{}}Deploy DNN to\\ edge devices\end{tabular}                 & Inference                                                    & \begin{tabular}[c]{@{}l@{}}Device owner is untrusted\\ Protect model privacy\end{tabular}                               & \begin{tabular}[c]{@{}l@{}}ML model degrades \\ severely when adding \\ noise to a few weights\end{tabular}                & \begin{tabular}[c]{@{}l@{}}Retraining-based\\ model stealing\end{tabular}                               & One-Time-Pad                                                                                  & -                                                                                      & \begin{tabular}[c]{@{}l@{}}Pre-compute\\ OTP\end{tabular}                                                      \\ \cmidrule(l){2-11} 
                        & SOTER~\cite{shen2022soter}                                                                & \begin{tabular}[c]{@{}c@{}}ATC \\ 2022\end{tabular}     & \begin{tabular}[c]{@{}l@{}}Deploy DNN to \\ third-party \\ edge devices\end{tabular} & Inference                                                    & \begin{tabular}[c]{@{}l@{}}Edge device can \\ be attacked.\\ Protect model privacy \\ and output integrity\end{tabular} & \begin{tabular}[c]{@{}l@{}}Many DNN layers have\\ associativity property\end{tabular}                                      & Model stealing                                                                                          & \begin{tabular}[c]{@{}l@{}}Morph the\\ intermediate results\end{tabular}                      & \begin{tabular}[c]{@{}l@{}}Fingerprint\\ challenge\end{tabular}                        & \begin{tabular}[c]{@{}l@{}}Morph operation \\ maybe cracked\end{tabular}                                       \\ \cmidrule(l){2-11} 
                        & GINN~\cite{asvadishirehjini2022ginn}                                                                 & \begin{tabular}[c]{@{}c@{}}CODASPY\\ 2022\end{tabular}  & \begin{tabular}[c]{@{}l@{}}DNN outsourcing\\ on cloud servers\end{tabular}           & \begin{tabular}[c]{@{}c@{}}Training\\ Inference\end{tabular} & \begin{tabular}[c]{@{}l@{}}Server is untrusted\\ Protect output integrity\end{tabular}                                  & \begin{tabular}[c]{@{}l@{}}Incorporate randomized \\ verification into the \\ training process\end{tabular}                & Backdoor attack                                                                                         & -                                                                                             & \begin{tabular}[c]{@{}l@{}}Gradient clip and\\ randomized \\ verification\end{tabular} & \begin{tabular}[c]{@{}l@{}}Do not protect\\ privacy\end{tabular}                                               \\ \cmidrule(l){2-11} 
                        & 3LegRace~\cite{niu20223legrace}                                                             & \begin{tabular}[c]{@{}c@{}}PETs \\ 2022\end{tabular}    & \begin{tabular}[c]{@{}l@{}}DNN outsourcing\\ on cloud servers\end{tabular}           & \begin{tabular}[c]{@{}c@{}}Training\\ Inference\end{tabular} & \begin{tabular}[c]{@{}l@{}}Server is untrusted\\ User input is private\end{tabular}                                     & \begin{tabular}[c]{@{}l@{}}Decompose data and \\ model into trusted \\ and untrusted parts\end{tabular}                    & \begin{tabular}[c]{@{}l@{}}Model inversion\\ Gradient inversion\end{tabular}                            & Add DP noise                                                                                  & -                                                                                      & \begin{tabular}[c]{@{}l@{}}Internal feature\\ is not encrypted\end{tabular}                                    \\ \cmidrule(l){2-11} 
                        & ShadowNet~\cite{sun2020shadownet}                                                            & \begin{tabular}[c]{@{}c@{}}S\&P \\ 2023\end{tabular}    & \begin{tabular}[c]{@{}l@{}}Deploy DNN to\\ edge devices\end{tabular}                 & Inference                                                    & \begin{tabular}[c]{@{}l@{}}Device owner is untrusted.\\ Protect model privacy\end{tabular}                              & \begin{tabular}[c]{@{}l@{}}Transform the weights \\ of linear layers before \\ outsourcing\end{tabular}                    & \begin{tabular}[c]{@{}l@{}}Black-box \\ model stealing\end{tabular}                                     & One-Time-Pad                                                                                  & \begin{tabular}[c]{@{}l@{}}Freivalds\\ Algorithm\end{tabular}                          & \begin{tabular}[c]{@{}l@{}}Pre-compute\\ OTP\end{tabular}                                                      \\ \cmidrule(l){2-11} 
                        & NNSplitter~\cite{zhou2023nnsplitter}                                                           & \begin{tabular}[c]{@{}c@{}}ICML \\ 2023\end{tabular}    & \begin{tabular}[c]{@{}l@{}}Deploy DNN to \\ edge devices\end{tabular}                & Inference                                                    & \begin{tabular}[c]{@{}l@{}}Device owner is untrusted.\\ Protect model privacy\end{tabular}                              & \begin{tabular}[c]{@{}l@{}}Learn to split a model \\ into two parts: obfuscated \\ model and model secrets\end{tabular}     & Model stealing                                                                                          & -                                                                                             & -                                                                                      & \begin{tabular}[c]{@{}l@{}}Expose large \\ amount of privacy \\ on GPU\end{tabular}                            \\ \cmidrule(l){2-11} 
                        & MirrorNet~\cite{liu2023mirrornet}                                                            & \begin{tabular}[c]{@{}c@{}}ICCAD \\ 2023\end{tabular}   & \begin{tabular}[c]{@{}l@{}}Deploy DNN to \\ edge devices\end{tabular}                & Inference                                                    & \begin{tabular}[c]{@{}l@{}}Device owner is untrusted.\\ Protect model privacy\end{tabular}                              & \begin{tabular}[c]{@{}l@{}}A monitor in TEE rectifies \\ the output of a low-accuracy \\ backbone model on GPU\end{tabular} & Model stealing                                                                                          & -                                                                                             & -                                                                                      & \begin{tabular}[c]{@{}l@{}}Backbone leaks\\ partial privacy\end{tabular}                                       \\ \bottomrule %
    \end{tabular}

    \end{adjustbox}

    \end{table}

In this section, we systematically summarize existing literature on TSDP. We
covered the top-tier conferences over the last five years across diverse fields,
including computer security, mobile computing, machine learning, and computer
systems. We also included the papers that are cited by top-tier papers. In total, we identified and reviewed
17 papers. Table~\ref{tbl:literature_short} summarizes the reviewed papers. For
each paper, we summarized seven aspects: the scenario, threat model, applied stage, design
principles, design scenarios, evaluated attacks, the security of outsourced data, and potential limitations. 
Scenario means the application scenario for which the partition method is designed. Stage means this partitioning method applies to the DNN inference stage, training stage, or both. The threat model refers to which part of the system is deemed malicious and what DNN attribute TSDP wants to protect. The design principle means the insight of the defense and the goal of partitioning. The evaluated attacks refer to what kind of attacks are evaluated in the original paper. The outsourced data security includes two aspects: the data confidentiality and the computation integrity. Confidentiality means the content of transferred data is properly protected, and computation integrity means the computation results in the untrusted environment are validated by the TEE. At last, the limitation refers to the aspects that this solution sacrifices for partition security.

\subsection{Categorization}

\begin{figure}[t]
\centering
\includegraphics[width=0.4\linewidth]{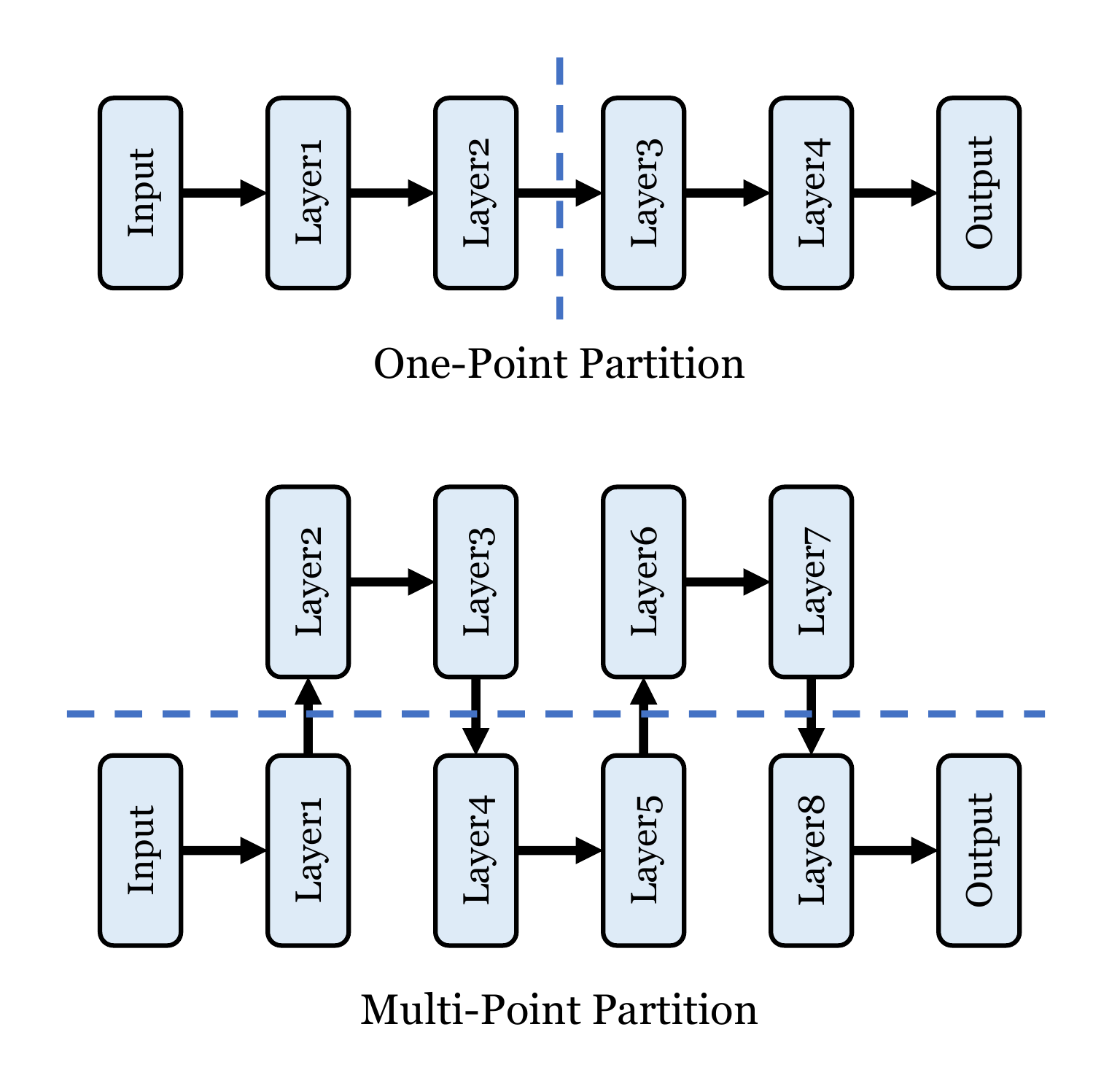}
\caption{Two types of DNN partition.}
\label{fig:TSDP_illustration}
\end{figure} 

We categorize existing solutions into two categories: \textbf{Single-Point
Partition} and \textbf{Multi-Point Partition}. Single-Point Partition solutions
only partition the DNN model at one point and produce two parts. One part is
shielded by the TEE and the other part is deployed on the GPU. The internal
feature at the split point is transmissed between the TEE and the GPU. 
Multi-Point Partition solutions partition the DNN model at multiple points and
produce multiple segments. Some segments are shielded by the TEE and other
segments are deployed on the GPU. Most multi-point partition solutions shield
non-linear layers (e.g. ReLU layers) in the TEE and deploy linear layers (e.g.
FC layers and convolutional layers) on the GPU. This is because non-linear
layers only occupies marginal computation (less than 5\%~\cite{tramer2019slalom}) and are
difficult to be encrypted. 
We can observe that in early years of this direction
(before 2021), most solutions are Single-Point Partition solutions. After 2021,
people start to focus on Multi-Point Partition solutions.

For single-point partition, the advantage is the straightforward design and
implementation. The disadvantage is the lack of guarantee on the input privacy.
Although it is heuristically evaluated that the transmissed feature does not
leak input privacy or membership information, there is no formal proof. It is
difficult to provide a formal proof because the feature is usually a
high-dimensional vector and it is challenging to define the privacy leakage associated with 
such a vector.

For multi-point partition, the advantage is that it provides flexible partition
design which can be integrated with other obfuscation or cryptographic
solutions. In this way, multi-point partition can provide a stronger privacy
guarantee than single-point partition. Thus we can observe that most recent
solutions are multi-point partition solutions. The disadvantage is that the
design and implementation are more complex.

\subsection{Survey Summary}
In this subsection, we will summarize the observations from the survey from seven aspects: the scenario, threat model, applied stage, design
principles, design scenarios, evaluated attacks, the security of outsourced data, and potential limitations. 

\parh{Scenario.}
Most papers focus on the cloud scenario for the target scenario, where the DNN model is deployed on the cloud and the client sends the input to the cloud for inference. The model owners can be the same as the cloud servers or third-party owners. In this setting, the main goal is to protect the user input data. This cloud-based setting is a major scenario for ML as a service because many internet companies use their cloud platforms to provide ML services.
Some papers focus on the edge scenario, where the DNN model is deployed on the user's device, and users directly use the edge device to run the DNN model. This is because in some scenarios, the input data is highly sensitive, and the data owners don't allow the data to leave their own server. The users' devices can either be mobile (such as smartphones) or a desktop computer located in the user's organization. 
These two scenarios are the most common scenarios for providing ML services.

\parh{Stage.}
For the applied stage of DNN model, most papers focus on the inference stage,
where the DNN model is already trained and the DNN model is used to process the
input. The reason that much training data that is used to train the DNN models is publicly available (e.g., Imagenet~\cite{deng2009imagenet} dataset) is that the training phase was not considered private. However, after several years, the training phase is regarded as private due to two reasons. First, as the size of the model increases, the amount of the public dataset may not be enough to train larger models. Developers need to utilize more prolific private data to train the larger model. Second, DNN model is applied to many personalized tasks where the training data is private. For the training phase, the required amount of computation is larger than the inference because training needs to compute additional gradients and save intermediate feature maps. TSDP is more important for the training stage because the disadvantage of TEE is amplified due to the large amount of computation. The techniques to protect the training phase can also be used to protect the inference phase because the required operations during the inference are a subset of the training's operations.

\parh{Threat Model.}
The threat model can be categorized into two types: 1) the cloud server in the cloud scenario is malicious, and 2) the users' device is malicious. For the first case, the goal of TSDP is to protect the uploaded user input and sometimes the output label of the input~\cite{gu2018yerbabuena}. The users do not want their private data to be leaked to the cloud server when using the remote model. For the second case, the goal of TSDP is to protect the model security on the user's devices. Because the user controls his device, TSDP does not need to protect the input data. On the contrary, the user has the motivation to steal the downloaded model and steal the intellectual property in the model. Besides protecting data privacy, some papers aim to protect the integrity of DNN output in some safety-critical scenarios on the edge~\cite{xiang2021aegisdnn,shen2022soter}. AegisDNN aims to defend against the adversarial fault injection attack in intelligence autonomy such as automatic driving and smart robotics~\cite{xiang2021aegisdnn}. SOTER aims to output integrity in mission-critical applications such as autopilot navigation and home monitoring~\cite{shen2022soter}.

\parh{Design Principle.}
Many TSDP solutions are motivated by heuristic observations that some parts of the DNN model are more important for certain attacks. Some papers find that the output features of shallow layers can be used to reconstruct the input sample, so they use TEE to secure the shallow layers~\cite{elgamal2020serdab,gu2018yerbabuena,mo2021ppfl,narra2019origami}. DarkneTZ~\cite{mo2020darknetz} and PPFL~\cite{mo2021ppfl} use experimental results to demonstrate that only the information of the last layers (gradients and outputs) can be used to perform membership inference attack, so these two papers use TrustZone to protect several last layers on the edge device. 
Some papers use learning-based approaches to learn the protection schemes. Shredder~\cite{mireshghallah2020shredder} learns additive noises to protect the output features of the shallow layers. NNSplitter~\cite{zhou2023nnsplitter} learns to split a model into the trusted part and the untrusted part. 
To protect model functionality, Magnitude~\cite{hou2021model} takes the widely used observation that the large-magnitude model weights are more important for model functionality. 
Many obfuscation schemes to protect the security of outsourced layers and internal features of TSDP are also motivated by empirical observations. One practical obfuscation technique is matrix permutation, which permutes the rows of weight matrix and internal features~\cite{sun2020shadownet,shen2022soter}. ShadowNet~\cite{sun2020shadownet} uses weight matrix permutation and random noise to protect model functionality. SOTER~\cite{shen2022soter} uses the associativity of the convolution and linear operations to obfuscate the outsourced model weights. 
Some papers are motivated by cryptographic techniques to protect model information with theoretical guarantees. Slalom~\cite{tramer2019slalom} adds one-time-pads to the internal features to encrypt the layer outputs. Goten~\cite{lucien2021goten} uses additive secret sharing to protect the computation phase of linear layers. 3LegRace~\cite{niu20223legrace} adds differential private noise to the layer output to prevent the attacker from recovering the input.

\parh{Evaluated Attacks.}
For the cloud scenario, input reconstruction and model inversion are the most frequently evaluated attacks because these two attacks can directly recover the input data~\cite{narra2019origami,mo2021ppfl,gu2018yerbabuena}. Membership inference attack is also widely used in the edge scenario to evaluate how much the model leaks the information of the training data~\cite{mo2020darknetz,mo2021ppfl,niu20223legrace}. Some paper also uses the gradient information to perform gradient-based membership inference~\cite{mo2020darknetz} and data reconstruction~\cite{niu20223legrace}. For the leakage of model functionality, model stealing attack is the most popular~\cite{sun2020shadownet,shen2022soter,hou2021model,zhou2023nnsplitter,liu2023mirrornet}. The attacker uses the offloaded layer weights to train the surrogate model and evaluate the performance difference between the surrogate model and the original victim model.

\parh{Outsourced Data Security.}
The security of outsourced data mainly consists of two parts: confidentiality and integrity. Confidentiality means the content of transferred data is properly protected, and computation integrity means the computation results in the untrusted environment are validated by the TEE. We will discuss the security for one-point partition and multi-point partition, respectively.

\parh{Security of One-Point Partition Techniques.}
The confidentiality for single-point partition mainly refers to the transmissed
feature between TEE and GPU. The integrity, on the other hand, refers to the correctness of the DNN
part on the GPU. However, due to the complexity of the DNN model after single-point
partition, it is difficult to guarantee confidentiality and integrity. The
splitted part on the GPU usually has similar computation complexity (the types of
layer operations are the same) as the original DNN model. This splitted part
usually contains both linear layers and non-linear layers. As existing encryption and verification solutions face difficulty in supporting non-linear layers, it is challenging to guarantee confidentiality and integrity for single-point partition.

As observed in Table~\ref{tbl:literature_short}, only one single-point partition
solution (Shredder~\cite{mireshghallah2020shredder}) uses additive noise to
partially protect the confidentiality. However, the goal of this solution is to
reduce the amount of sensitive information in the transmissed feature, rather
than to provide a confidentiality guarantee. For computation integrity on the
GPU part, there is no existing solution that can provide a formal guarantee.

\parh{Security of Multi-Point Partition Techniques.}
Confidentiality and integrity are important for multi-point partition and are
considered by most solutions. It is because, without these formal guarantees, the adversary can recover the private information in the TEE and beat the
protection scheme. A multi-point partition usually puts the linear layers on the
GPU and keep non-linear layers in the TEE. This design makes it possible to
apply existing encryption and verification solutions. In the next part of this
section, we will discuss potential attack and defense strategies that are
adopted by existing solutions.

As TSDP solutions assume that the GPU is untrusted or compromised by the
adversary, the correctness of GPU computation may be jeopardized. The adversary
can modify the GPU output (TEE input) and observe GPU input (TEE output) to
infer the confidential information. For example, the adversary can use zero-th
order gradient optimization~\cite{zhang2022how} by slightly modifying the TEE input and
observing how the TEE output changes to gradually infer the protected information.

\parh{Integrity of Multi-Point Partition Techniques.}
To prevent this attack, existing solutions usually use redundant computation to
verify the integrity of GPU computation. Such defense sends the same input to
GPU for multiple times at random intervals and check if the GPU output is the same. If the
GPU outputs are not consistent, the TEE will abort the computation. However, adversaries can still escape this defense by observing the GPU input
and making the same modification to the same GPU input, thereby escaping the
detection. To prevent this attack, existing solutions usually add noisy mask to
the GPU input. Even for two identical GPU inputs, the masked inputs are
different, preventing the adversary from identifing which GPU inputs are the same.

The advantage of redundant computation is that it is simple to implement and
flexible in adjusting the detection frequency. The defender can control how often
the TEE checks the integrity and try not to affect the system performance too
much.

\parh{Confidentiality of Multi-Point Partition Techniques.}
Even if the computation in the TEE is shielded, if the defender does not protect
the confidentiality of the transmissed feature, the adversary can still infer or
reconstruct the confidential information. Take a simple example where TEE shields
a convolutional layer whose input and output are not protected. The adversary on the
GPU can observe and collect many input-output pairs and reconstruct the weights
in the TEE by solving a linear equation. For a more complex example where a DNN
model segment in the TEE consists of mutiple layers (including both linear and
non-linear layers), the adversary can use input-output pairs as an additional
supervised signal to train shadow models. The trained model has a similar
functionality as the shielded model segment. Thus, protecting the
confidentiality of the transmissed feature is important for TSDP solutions.
Without confidentiality guarantee, the TSDP protection becomes vulnerable and easily compromised.

To protect the confidentiality of the transmissed feature, existing TSDP
solutions mainly use two strategies: matrix obfuscation and encryption based on
one-time-pad. The insights of both strategies are similar. TEE uses homomorphic
obfuscation or encryption techniques to convert plaintext GPU input into a
ciphertext input. The GPU computes the ciphertext output and sends it back to the TEE.
The TEE can then decrypt the ciphertext output and recover the plaintext output. Besides,
both strategies require that the outsourced layer on the GPU is linear. Thus, multi-point partition solutions usually put the linear layers on the GPU
and keep non-linear layers in TEE.

For matrix obfuscation, the defender permutates the matrix of model weights by
the channel dimension and sends the permuted matrix to the GPU. The GPU can uses the
obfuscated model weights to compute the obfuscated output and sends it back to
the TEE. The TEE can recover the feature dimension by de-permutating the obfuscated
output and recover the original output. In this way, the GPU does not know the
original order of the matrix and thus can not reuse the functionality of model
weights. The advantage of matrix obfuscation is that it is light-weight and do
not introduce much communication and computation overhead. The disadvantage is
the comparatively weak confidentiality guarantee. As the original weight values remain unchanged, the provided security guarantee is less robust than the encryption
schemes.

For encryption based on one-time-pad, the defender first quantizes the GPU input to
integer field and uses one-time-pad to encrypt it. The GPU computes the
results on the encrypted input and sends the encrypted output back to the TEE. The TEE
can decrypt the output by substracting a factor and recover the original output.
This factor is the output on which the linear layer takes the padding as input.
The advantage of one-time-pad encryption is the high security guarantee and
online efficiency. If one-time-pad is never reused and the padding size is equal
to the size of plaintext, the encryption is unbreakable~\cite{OTP}. At the online
phase, the additional overhead only involves adding and subtracting the padding,
thus it is efficient. However, the disadvantage is that the encryption introduces
large offline computation overhead or communication overhead. To recover the
plaintext output from the ciphertext output, the TEE needs to precompute the
substraction factor, which has a same level of computation complexity as
performing the linear layer. This precomputation can be performed in two ways:
1) TEE performs the computation during the offline phase, or 2) the model owner (or
data owner) performs the computation on a remote server and sends the result to
the TEE. The first method introduces large computation overhead on the TEE at the
offline phase, while the second method introduces significant communication overhead. Therefore,
the strong security guarantee of one-time-pad encryption comes with a higher
offline overhead.

\parh{Limitations.}
The limitation of the solutions that are motivated by empirical evaluation and heuristic design is that they lack a theoretic guarantee. Although existing empirical evaluation demonstrates that the partition and the protection are secure, there may be a stronger attacker in the future who can recover more privacy from these partition solutions. For the defenses that rely on the one-time-pad to protect confidentiality, the limitation is that such a one-time-pad should be computed in an offline phase (or remotely) and can not be reused. This limitation puts more burden on the storage system or communication system. For the obfuscation techniques that use matrix permutation~\cite{sun2020shadownet} and morph~\cite{shen2022soter}, the limitation is that they do not protect the original weight values and such obfuscation may be cracked. For the per-weight-based partition, such as Magnitude~\cite{hou2021model} and NNSplitter~\cite{zhou2023nnsplitter}, the limitation is that the outsourced weight values in plaintext still leak a large amount of privacy.

\section{Evaluating Existing \tsdp Defenses}
\label{sec:evaluate_existing_solutions}

Based on the literature survey, we select the TSDP techniques that can be applied to our threat model introduced in \S~\ref{sec:threat_model}. We categorize these techniques into five different types of defenses. 
Then, we implement a representative technique from each category and evaluate
their security via \ac{ms} and \ac{mia}. We assess if they were sufficiently
secure against the launched attacks, and we harvest empirical observations to
present lessons from the evaluation.

\begin{table}[!htbp]
    \caption{A fine-grained taxonomy of existing \tsdp\ solutions. We mark
    \colorbox{blue!30}{representative works} empirically assessed in this study.
    Other works are ignored in our empirical evaluation and are just part of the
    literature review. 
    }
    \centering
    \label{tbl:literature_short}
    \begin{adjustbox}{max width=0.5\linewidth}
    \begin{tabular}{ccc}
    \hline
    Literature                                                 & Conference/Journal & Category                                                                                 \\ \hline
    \colorbox{blue!30}{DarkneTZ}~\cite{mo2020darknetz}            & MobiSys 2020 & \multirow{4}{*}{\begin{tabular}[c]{@{}c@{}}Shielding\\ Deep Layers\end{tabular}}         \\
    Shredder \cite{mireshghallah2020shredder} & ASPLOS 2020  &                                                                                          \\
    PPFL \cite{mo2021ppfl}                    & MobiSys 2021 &                                                                                          \\
    MirrorNet \cite{liu2023mirrornet} & ICCAD 2023  &                                                                                          \\\hline
    Yerbabuena \cite{gu2018yerbabuena}        & Arxiv 2018   & \multirow{3}{*}{\begin{tabular}[c]{@{}c@{}}Shielding\\ Shallow Layers\end{tabular}}      \\
    Origami \cite{narra2019origami}           & Arxiv 2019   &                                                                                          \\
    \colorbox{blue!30}{Serdab} \cite{elgamal2020serdab}           & CCGRID 2020  &                                                                                          \\\hline
    \colorbox{blue!30}{Magnitude}~\cite{hou2021model}             & TDSC 2022    & \multirow{2}{*}{\begin{tabular}[c]{@{}c@{}}Shielding\\ Large-Mag. Weights\end{tabular}}                   \\
    NNSplitter~\cite{zhou2023nnsplitter}             & ICML 2023    &                    \\\hline
    AegisDNN \cite{xiang2021aegisdnn}         & RTSS 2021    & \multirow{2}{*}{\begin{tabular}[c]{@{}c@{}}Shielding\\ Intermediate Layers\end{tabular}} \\
    \colorbox{blue!30}{SOTER} \cite{shen2022soter}                & ATC 2022     &                                                                                          \\\hline
    \colorbox{blue!30}{ShadowNet} \cite{sun2020shadownet}         & S\&P 2023   & \begin{tabular}[c]{@{}c@{}}Shielding Non-Linear \\ Layers \& Obfuscation\end{tabular}                   \\\hline
    Slalom~\cite{tramer2019slalom}            & ICLR 2018    & -                                                                                        \\
    eNNclave \cite{schlogl2020ennclave}       & AISec 2020   & -                                                                                        \\ 
    DarKnight \cite{hashemi2021darknight}     & MICRO 2021   & -                                                                                        \\
    Goten \cite{lucien2021goten}              & AAAI 2021    & -                                                                                        \\
    GINN \cite{asvadishirehjini2022ginn}      & CODASPY 2022 & -                                                                                        \\
    3LegRace \cite{niu20223legrace}       & PETs 2022   & -                                                                                        \\ 
    \hline

    \end{tabular}

    \end{adjustbox}

    \end{table}

\subsection{Fine-Grained Defense Taxonomy}
\label{sec:literature_category}

\begin{figure*}[!t]
    \centering
    \includegraphics[width=\linewidth]{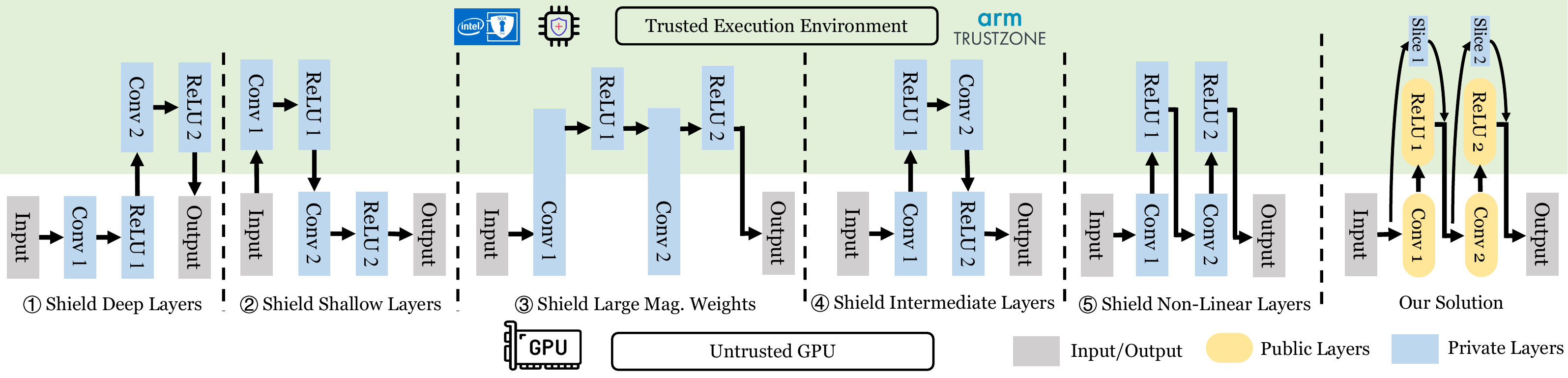}
    \caption{An illustration of different \tsdp solutions on a four-layer DNN.
    \colorbox{blue!15}{Blue} squares are privacy-related layers, and
    \colorbox{yellow!80}{yellow} rounded squares are privacy-irrelevant (public)
    layers. \ding{172} shields two deep layers (Conv2 and ReLU2) and \ding{173}
    shields two shallow layers (Conv1 and ReLU1). \ding{174} shields the
    large-magnitude weight of each layer. \ding{175} shields two random
    intermediate layers (ReLU1 and Conv2). \ding{176} shields non-linear layers
    (ReLU1 and ReLU2) and obfuscates other layers (Conv1 and Conv2). Our solution
    (introduced in \S~\ref{sec:approach}) shields privacy-related
    slices and non-linear layers of the public backbone model.}
    \label{fig:model_partition_categories}
    \end{figure*}

After inspecting the detailed design of all the prior TSDP solutions, we found that 
six papers do not meet the requirements under our threat model. Among them
three are unable to defend models against \ac{ms}
(DarKnight~\cite{hashemi2021darknight}, Slalom~\cite{tramer2019slalom},
GINN~\cite{asvadishirehjini2022ginn}, and 3LegRace~\cite{niu20223legrace}), one
has a stronger assumption that requires two TEEs to verify each other
(Goten~\cite{lucien2021goten}), and the last one decreases DNN accuracy
significantly (eNNclave~\cite{schlogl2020ennclave}; details in
\S~\ref{sec:experiment:trade-off}). We then divide the remaining ten papers into
five categories depending on the \tsdp schemes.
\F~\ref{fig:model_partition_categories} schematically illustrates each category
using a sample 4-layer DNN model (including two convolution layers and two ReLU
layers). We also include our proposed approach (details in
\S~\ref{sec:approach}) for comparison. The five categories are as follows:

\noindent \ding{172} \underline{Shielding Deep Layers} partitions the DNN
according to the layer depth and places the layers close to the output layer in
the TEE. In \F~\ref{fig:model_partition_categories}, two deepest layers (Conv2
and ReLU2) are shielded.

\noindent \ding{173} \underline{Shielding Shallow Layers} partitions the DNN
according to the layer depth and places the layers close to the input layer in
the TEE. In \F~\ref{fig:model_partition_categories}, two shallowest layers
(Conv1 and ReLU1) are shielded.

\noindent \ding{174} \underline{Shielding Large-Magnitude Weights} partitions
the DNN according to the absolute weight value, and then puts the weights with
large magnitudes and ReLU layers in the TEE.
\F~\ref{fig:model_partition_categories} shields partial convolution layers (to
represent large-magnitude weights) and ReLU layers. 

\noindent \ding{175} \underline{Shielding Intermediate Layers} puts randomly
chosen intermediate layers in the TEE. \F~\ref{fig:model_partition_categories}
shields ReLU1 and Conv2 as the random-selected layers.

\noindent \ding{176} \underline{Shielding Non-Linear Layers and Obfuscation}
partitions the DNN by the layer types and shields non-linear (\eg ReLU) layers
using TEE. The offloaded linear layers (\eg convolution layers) are protected by
lightweight obfuscation algorithms (\eg matrix transformation).
\F~\ref{fig:model_partition_categories} shields the ReLU layers and offloads all
convolution layers.

\subsection{Representative Defenses}
\label{subsec:defense-selection}

\parh{Scheme Selection.}~For each of the five \tsdp\ schemes, we select one
representative solution for evaluation. For shielding deep layers (\ding{172}),
we choose DarkneTZ because according to related
work~\cite{hashemi2021darknight,mo2021ppfl}, it is the state-of-the-art (SOTA)
solution for protecting training data privacy. For shielding shallow layers
(\ding{173}) and shielding intermediate layers (\ding{175}), we select
Serdab~\cite{elgamal2020serdab} and SOTER~\cite{shen2022soter} because they are
the most recent papers published in peer-reviewed conferences. For shielding
large-magnitude weights (\ding{174}), we choose Magnitude~\cite{hou2021model}
because it has practical defense cost. The other solution in this category
(NNSplitter~\cite{zhou2023nnsplitter}) introduces significant training cost and
is not suitable for our evaluation. For shielding non-linear layers
(\ding{176}), we choose ShadowNet~\cite{sun2020shadownet} since they are the
only solutions in their respective categories.

\parh{Configuration Setting.}~All these schemes require configurations, e.g.,
for \ding{172}, we need to configure the exact number of ``deep'' layers in the
TEE. Overall, we configure each defense scheme according to their papers. In
particular, for DarkneTZ (\ding{172}), we put the last classification layer into
TEE. For Serdab (\ding{173}), the TEE shields the first four layers. For
Magnitude (\ding{174}), the TEE shields 1\% weights with the largest magnitudes.
For SOTER (\ding{175}), the TEE shields 20\% randomly-selected layers and
multiplies the other offloaded layers with a scalar to conceal the weight values.
For ShadowNet (\ding{176}), the TEE shields all the ReLU layers and obfuscates
all the offloaded convolution layers with matrix transformation and filter
permutation (\iflongappendix
\else
detailed description in our Website~\cite{TEESliceWebsite}\fi).
Furthermore, we note that selecting proper configurations constitutes a key
factor that undermines their security/utility guarantee. We will discuss other
configurations later in \S~\ref{sec:dilemma}.

\subsection{Evaluated Attacks}
\label{sec:evaluated_attacks}

\parh{Attack Selection.}~We consider \ac{ms} and \ac{mia} attacks that can
extract confidential model weights and private training data as the security
benchmark for existing \ac{tsdp} approaches.
For \ac{ms}, we employ standard query-based stealing techniques where the
attacker trains a model from a set of collected data labeled by the
partially-shielded $M_{vic}$. Query-based \ac{ms} has been widely adopted in
literature~\cite{orekondy2019knockoff, jagielski2020high,
shen2022model,orekondy2020prediction}. We leverage the attack implementation
offered by Knockoff Net~\cite{KnockoffNetCode}, a SOTA baseline accepted by
prestigious
publications~\cite{orekondy2020prediction,jagielski2020high,shen2022model}. 
For \ac{mia}, we chose the transfer attack that is designed against the 
label-only scenario~\cite{li2021membership}. Transfer attack builds $M_{\rm
sur}$ to imitate the behavior of $M_{\rm vic}$ and infer the privacy of $M_{\rm
vic}$ from white-box information of $M_{\rm sur}$ (\eg, confidence scores, loss,
and gradients). The intuition is that membership information can transfer from
$M_{\rm vic}$ to $M_{\rm sur}$. We chose the standard confidence-score-based
algorithm to extract membership from $M_{\rm sur}$. In particular, this process
trains a binary classifier, such that given the confidence score of $M_{\rm
sur}$, the classifier predicts if the corresponding DNN input is in the training
data of $M_{\rm vic}$. Confidence-score-based \ac{mia} has been extensively used
in previous attacks~\cite{papernot2016towards, salem2019mlleak,
shokri2017membership,li2021membership,nasr2018machine}, and we reused the attack
implementation from a recent benchmark suite,
\textsc{ML-Doctor}~\cite{MLDoctorCode,liu2022mldoctor}. 
Recent work has consistently employed \textsc{ML-Doctor} for membership
inference~\cite{shen2022model,chen2021when,carlini2022membership}.

\begin{figure}[!t]
\centering
\includegraphics[width=0.65\linewidth]{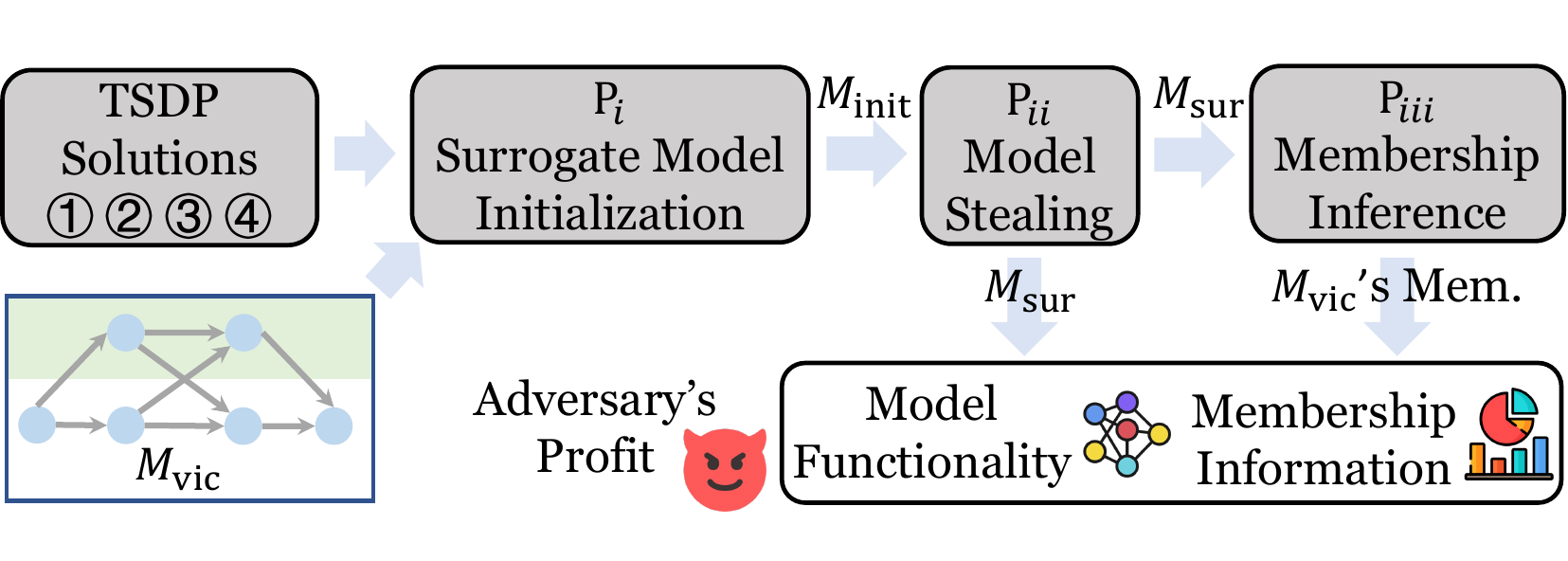}
\caption{A three-phase attack pipeline.}
\label{fig:attack_pipeline}
\vspace{-15pt}
\end{figure}

\parh{Attack Pipeline.}~As in \F~\ref{fig:attack_pipeline}, the attack goal is
to get the victim model $M_{vic}$'s functionality and membership information.
The attack pipeline consists of three phases: surrogate model initialization
(P$_{i}$), \ac{ms} (P$_{ii}$), and \ac{mia} (P$_{iii}$). The attacker first
analyzes the target defense scheme and then conducts P$_{i}$ to get an
initialized surrogate model (denoted as $M_{\rm init}$). P$_{ii}$ trains $M_{\rm
init}$ with queried data and outputs the surrogate model $M_{\rm sur}$ (the
recovered victim model). P$_{iii}$ takes $M_{\rm sur}$ as input, uses \ac{mia}
algorithms and outputs $M_{\rm vic}$'s membership privacy. Specifically, in
P$_{iii}$, the adversary first trains a binary classifier. Then, given an input
$i$ and the $M_{\rm sur}$'s prediction confidence score $p$, the classifier
decides if $i$ belongs to the $M_{\rm vic}$'s training data by taking $p$ as its
input~\cite{shokri2017membership,carlini2022membership,yuan2022membership}.

\parh{Surrogate Model Initialization.}~Steps in our attack pipeline are
automated except for the first step, surrogate model initialization (P$_i$). To
run the attacks, attackers first need to construct $M_{\rm init}$ with the
exposed knowledge in the public part of the \ac{tsdp} protected models. The
high-level process to construct $M_{\rm init}$ has three steps. First, the
attacker infers the architecture of the \tsdp\ protected model based on the
offloaded part and the model output with existing
techniques~\cite{chen2021teacher,chen2022copy}. Then, the attacker chooses a
public model $M_{\rm pub}$ (with the same or an equivalent architecture) as
$M_{\rm init}$. 
Lastly, the attacker transports the model weights in the offloaded part of
$M_{\rm vic}$ to the corresponding parts of $M_{\rm init}$. For DarkneTZ
(\ding{172}), Serdab (\ding{173}), and SOTER (\ding{175}), we use the offloaded
layers to replace the corresponding layers of $M_{\rm init}$. For Magnitude
(\ding{174}), we use the offloaded weights that run on GPUs to replace the
corresponding weights in $M_{\rm init}$. For ShadowNet (\ding{176}), the
attacker uses the public model to decode the obfuscation algorithm (
\iflongappendix
\else
detailed description in our Website~\cite{TEESliceWebsite}
\fi
) and uses the decoded weights to
initialize $M_{\rm init}$.

\parh{Comparison Baselines.}~To ease the comparison, we also provide baseline
evaluation results. Notably, for \ac{ms}, we consider the
\noshield solution (offloading the whole $M_{\rm vic}$ to GPU, referred to as
``No-Shield'') as the easiest baseline because the adversary can directly use
the offloaded $M_{\rm vic}$ as $M_{\rm sur}$ and does not need to train the
model. We consider a black-box (or \shieldwhole) setting (referred to as
``Black-box'') where attackers can only access the model prediction labels and
identify the $M_{\rm vic}$'s corresponding public models. However, the attacker
cannot leverage the $M_{\rm vic}$'s weights offloaded on GPU to construct
$M_{\rm init}$. This is a challenging setting, and the attacker needs to steal
every layer's weights from $M_{\rm vic}$.
As for \ac{mia}, the \noshield setting denotes that the attackers
can directly use the $M_{\rm vic}$'s output confidence score for membership
inference because the whole $M_{\rm vic}$ is offloaded. Recall, as noted in our
threat model (\S~\ref{sec:threat_model}), \ac{tsdp} solutions often refuse to
return confidence scores and only return predicted labels to mitigate
membership inference. In contrast, the black-box setting denotes that the
attacker directly uses $M_{\rm pub}$ as $M_{\rm init}$ and trains $M_{\rm sur}$
from queried data. Apparently, this is comparable to ``random guess'' (success
rate around 50\%\footnote{In the evaluation setting (\S~\ref{subsec:evaluation-setup}), we follow prior \ac{mia} work to set the size of target training dataset equal the size of target testing dataset.}), given that a $M_{\rm init}$ (\ie, $M_{\rm pub}$) contains no
information on the $M_{\rm vic}$'s private training datasets.

\subsection{Evaluation Setting}
\label{subsec:evaluation-setup}

\parh{Datasets.}~We use four different datasets and five DNN models to evaluate
different defenses. The dataset and model selection refers to prior \ac{ms} and
\ac{mia} literatures~\cite{liu2022mldoctor,jagielski2020high}. In particular,
the datasets include CIFAR10, CIFAR100~\cite{krizhevsky2009learning},
STL10~\cite{coates2011an}, and UTKFace~\cite{zhang2017age}. CIFAR10 and CIFAR100
are default choices of prior \ac{ms}/\ac{mia}
literatures~\cite{orekondy2020prediction,jagielski2020high,liu2022mldoctor,carlini2022membership,yuan2022membership}.
STL10 and UTKFace are used by \textsc{ML-Doctor} to quantify model
vulnerability~\cite{liu2022mldoctor}. 
When evaluating \ac{ms}, we use each dataset's
default train/test split to avoid possible bias. To evaluate \ac{mia}, we follow
the setting of \textsc{ML-Doctor} to partition each dataset into four splits:
the target training dataset, the target testing dataset, the shadow training
dataset, and the shadow testing dataset. The model owner uses the target
training and the target testing datasets to train and evaluate the victim model
$M_{\rm vic}$. The adversary uses the shadow training dataset and the shadow
testing dataset to train the binary classifier. This is a common setting for
evaluating
\ac{mia} in prior work~\cite{carlini2022membership,shokri2017membership,shen2022model,li2021membership}.

\parh{Models.}~The benchmark models include ResNet18~\cite{he2016deep},
VGG16\_BN~\cite{simonyan2015very}, AlexNet~\cite{alex2012imagenet},
ResNet34 and VGG19\_BN. These models are widely used in prior security
studies~\cite{liu2022mldoctor,chen2021teacher,yuan2022membership,rakin2022deepsteal}.
\ifentireresult
{}
\fi
We set the hyper-parameters following the paper and released code of prior
works~\cite{MLDoctorCode,KnockoffNetCode}. As introduced in
\F~\ref{fig:attack_pipeline}, for all cases, we use the public models as
initialization to get a better model
performance~\cite{orekondy2019knockoff,orekondy2020prediction}. For the training
in the \ac{ms} part, we follow the hyper-parameter settings of Knockoff
Nets~\cite{KnockoffNetCode}. Accuracies of the trained models $M_{\rm vic}$ are
reported in \T~\ref{tbl:nettailor_accuracy}. The accuracies are generally
consistent with public results~\cite{STL10ACC,CIFAR10ACC,CIFAR100ACC}. For the
training in the \ac{mia} part, we follow the settings of
\textsc{ML-Doctor}~\cite{MLDoctorCode}. 
\iflongappendix
\else
Model accuracies are reported in the Website~\cite{TEESliceWebsite}. 
\fi
All models achieve high accuracy on the
target training dataset, and the accuracies are consistent with prior
works~\cite{liu2022mldoctor}. 
\iflongappendix
\else
We leave detailed settings of hyper-parameters in our Website~\cite{TEESliceWebsite}.
\fi

\parh{Metrics.}~Overall, we systematically evaluate the effectiveness of de
facto \tsdp\ solutions using three \ac{ms} metrics and four
\ac{mia}~\cite{jagielski2020high, rakin2022deepsteal, nasr2019comprehensive,
song2020information, yuan2022membership}. Specifically, we record \ac{ms}
accuracy, fidelity, and \ac{asr} according to prior
work~\cite{jagielski2020high,orekondy2019knockoff,zhu2021hermes,rakin2022deepsteal}.
For \ac{mia}, we use confidence-based \ac{mia} accuracy, gradient-based \ac{mia}
accuracy, generalization gap, and confidence gap following prior literature as
well~\cite{nasr2019comprehensive,song2020information,yuan2022membership}. We
report ``\ac{ms} accuracy'' (denoted as ``Model Stealing'') and
``confidence-based \ac{mia} attack accuracy'' (denoted as ``Membership
Inference'') as the major metrics. \ac{ms} accuracy denotes the prediction
accuracy of the $M_{\rm
sur}$~\cite{zhu2021hermes,orekondy2019knockoff,jagielski2020high,shen2022model}.
A higher accuracy indicates that $M_{\rm sur}$ successfully steals more
functionality from $M_{\rm vic}$. As for the \ac{mia} accuracy, a higher
accuracy denotes that attackers can more accurately decide if a given sample is
in $M_{\rm vic}$'s training dataset.

\subsection{Attack Results}
\label{sec:empirical_evaluation_results}

We aim to answer the following research question primarily:

\vspace{-1pt}
\begin{tcolorbox}[size=small]
    \textbf{RQ1:} How secure are the selected \tsdp\ solutions against
    model and data privacy stealing?
\end{tcolorbox}
\vspace{-3pt}

\ifentireresult
We report the evaluation results over five models (AlexNet, ResNet18,
VGG16\_BN, \diff{ResNet34 and VGG19\_BN}) over the two major metrics (\ac{ms} accuracy and MI accuracy) in \T~\ref{tbl:evaluate_solution_MS} and \T~\ref{tbl:evaluate_solution_MI} respectively (total 20 cases). As
aforementioned, we also report the baseline settings (``No-Shield'' and
``Black-box'') for comparison. 
\fi
To compare the performance between different defenses, for each case, we compute
a relative accuracy as the times of the accuracy over the accuracy of the
black-box baseline. We report the average relative accuracy in the last row of
\T~\ref{tbl:evaluate_solution_MS} and \T~\ref{tbl:evaluate_solution_MI}. A defense scheme is considered more effective if
its corresponding attack accuracy is closer to the black-box baseline (the
relative accuracy is closer to 1.00$\times$). For the \noshield baseline (the whole $M_{\rm vic}$ is offloaded to the
GPU), the relative accuracies are {$4.57\times$} for model stealing in \T~\ref{tbl:evaluate_solution_MS} and
{$1.37\times$} for membership inference in \T~\ref{tbl:evaluate_solution_MI}.

\ifentireresult
\colorlet{best}{green!30}
\colorlet{high}{red!30}
\colorlet{low}{yellow}

\begin{table*}[!h]
\caption{\diff{Model Stealing Attack accuracies regarding representative defense schemes. ``C10'',
``C100'', ``S10'', and ``UTK'' represent CIFAR10, CIFAR100, STL10, and UTKFace,
respectively. The last row reports the average accuracy toward each
defense relative to the baseline black-box solutions. For each setting, we mark
the highest attack accuracy in \colorbox{high}{red} and the lowest accuracy in
\colorbox{low}{yellow}. Attack accuracy toward our approach
(\S~\ref{sec:approach}) is marked with \colorbox{best}{green}.}}
\label{tbl:evaluate_solution_MS}
\setlength{\tabcolsep}{1.5pt}
\begin{adjustbox}{max width=0.65\linewidth}

    \begin{tabular}{cccccccccc}
    \hline
                               &      & No-Shield & \ding{172}Dark. & \ding{173}Serdab  & \ding{174}Mag. & \ding{175}SOTER & \ding{176}Shadow.  & Ours     & Black-box \\ \hline
    \multirow{4}{*}{\rotatebox{90}{AlexNet}}   & C10  & 83.72\%   & 77.15\%  & \cellcolor{low}63.58\% & 65.97\%   & 76.90\% & \cellcolor{high}83.57\% & \cellcolor{best}19.04\% & 24.38\%     \\
                               & C100 & 56.60\%   & \cellcolor{low}41.57\%  & 46.48\% & 47.86\%   & 50.83\% & \cellcolor{high}56.43\% & \cellcolor{best}8.27\% & 10.68\%      \\
                               & S10  & 76.55\%   & \cellcolor{high}75.17\%  & 69.06\% & 73.67\%   & 37.60\% & \cellcolor{low}35.98\% & \cellcolor{best}24.15\% & 15.26\%      \\
                               & UTK  & 90.01\%   & \cellcolor{high}88.74\%  & 82.92\% & 86.65\%   & \cellcolor{low}58.86\%  & 73.93\% & \cellcolor{best}52.27\% & 48.62\%     \\\hline
    \multirow{4}{*}{\rotatebox{90}{ResNet18}}  & C10  & 95.91\%   & 87.55\%  & \cellcolor{high}93.94\% & \cellcolor{low}89.92\%   & 92.61\% & 91.58\% & \cellcolor{best}31.40\% & 19.88\%     \\
                               & C100 & 81.63\%   & \cellcolor{low}70.11\%  & 78.01\% & 74.84\%   & \cellcolor{high}79.28\% & 78.51\% & \cellcolor{best}10.90\% & 15.41\%    \\
                               & S10  & 87.45\%   & \cellcolor{high}86.03\%  & 85.05\% & \cellcolor{low}77.08\%   & 80.83\% & 84.38\% & \cellcolor{best}29.19\% & 21.66\%    \\
                               & UTK  & 90.78\%   & 85.65\%  & 84.65\% & \cellcolor{low}64.99\%   & 76.43\% & \cellcolor{high}89.42\% & \cellcolor{best}51.95\% & 45.41\%     \\\hline
    \multirow{4}{*}{\rotatebox{90}{VGG16\_BN}} & C10  & 92.95\%   & 87.76\%  & \cellcolor{high}91.34\% & 87.35\%   & \cellcolor{low}81.52\% & 90.67\% & \cellcolor{best}30.87\% & 14.62\%     \\
                               & C100 & 72.78\%   & \cellcolor{low}63.68\%  & 72.19\% & 68.82\%   & 66.06\% & \cellcolor{high}72.85\% & \cellcolor{best}9.78\% & 10.93\%     \\
                               & S10  & 90.03\%   & 89.17\%  & 89.33\% & \cellcolor{low}84.33\%   & \cellcolor{high}89.46\% & 89.43\% & \cellcolor{best}32.92\% & 18.97\%     \\
                               & UTK  & 91.51\%   & 87.60\%  & 89.60\% & 90.28\%   & \cellcolor{low}87.30\% & \cellcolor{high}91.14\% & \cellcolor{best}48.37\% & 45.46\%     \\ \hline

        \multirow{4}{*}{\rotatebox{90}{ResNet34}}  & C10  & 91.05\%   & 86.28\%  & 30.88\%   & \cellcolor{low}14.31\%   & \cellcolor{high}93.10\% & 91.03\% & \cellcolor{best}26.43\% & 12.07\%     \\
                                   & C100 & 80.85\%   & \cellcolor{low}73.64\%  & 76.22\% & 74.68\%   & \cellcolor{high}81.47\% & 81.34\% & \cellcolor{best}10.67\%   & 17.22\%    \\
                                   & S10  & 88.29\%   & 86.74\%  & 84.09\% & \cellcolor{low}70.94\%   & 81.81\% & \cellcolor{high}88.79\%  & \cellcolor{best}34.19\% & 20.77\%   \\
                                   & UTK  & 87.65\%   & \cellcolor{high}87.83\%  & 78.07\% & \cellcolor{low}45.87\%   & \cellcolor{high}87.83\%  & 86.97\% & \cellcolor{best}48.50\% & 46.16\%    \\ \midrule
        \multirow{4}{*}{\rotatebox{90}{VGG19\_BN}} & C10  & 92.59\%   & 90.34\%  & 86.87\% & \cellcolor{low}82.66\%   & 87.34\% & \cellcolor{high}92.89\%  & \cellcolor{best}24.08\% & 10.90\%    \\
                                   & C100 & 71.17\%   & \cellcolor{low}64.30\%  & 69.23\% & 66.55\%   & 66.32\%  & \cellcolor{high}72.48\%  & \cellcolor{best}11.47\% & 10.54\%   \\
                                   & S10  & 89.62\%   & 89.31\%  & 88.99\% & \cellcolor{low}85.26\%   & \cellcolor{high}88.41\%  & 87.36\%  & \cellcolor{best}36.11\% & 19.93\%    \\
                                   & UTK  & 90.55\%   & 90.51\%  & 89.33\% & 89.83\%   & \cellcolor{high}90.74\% & 88.26\%  & \cellcolor{best}47.09\% & 46.44\%    \\ \midrule
        \multicolumn{2}{c}{Average}       & 4.57$\times$      & 4.27$\times$     & 4.10$\times$    & 3.85$\times$     & 4.20$\times$ & 4.56$\times$ &  1.32$\times$    & 1.00$\times$       \\ \hline
    \end{tabular}

\end{adjustbox}
\vspace{-10pt}

\end{table*}

\colorlet{best}{green!30}
\colorlet{high}{red!30}
\colorlet{low}{yellow}

\begin{table*}[!h]
\caption{\diff{Membership Inference Attack accuracies regarding representative defense schemes. ``C10'',
``C100'', ``S10'', and ``UTK'' represent CIFAR10, CIFAR100, STL10, and UTKFace,
respectively. The last row reports the average accuracy toward each
defense relative to the baseline black-box solutions. For each setting, we mark
the highest attack accuracy in \colorbox{high}{red} and the lowest accuracy in
\colorbox{low}{yellow}. Attack accuracy toward our approach
(\S~\ref{sec:approach}) is marked with \colorbox{best}{green}.}}
\label{tbl:evaluate_solution_MI}
\setlength{\tabcolsep}{1.5pt}
\begin{adjustbox}{max width=0.65\linewidth}

    \begin{tabular}{cccccccccc}
    \hline
                              
                               &      & No-Shield & \ding{172}Dark. & \ding{173}Serdab  & \ding{174}Mag. & \ding{175}SOTER & \ding{176}Shadow.  & Ours     & Black-box \\ \hline
    \multirow{4}{*}{\rotatebox{90}{AlexNet}}   & C10  & 67.25\%   & 57.67\%  & 62.96\% & \cellcolor{low}52.67\%   & 62.18\% & \cellcolor{high}69.43\% & \cellcolor{best}50.00\% & 50.00\%   \\
                               & C100 &  78.32\%   & \cellcolor{low}63.27\%  & 72.20\% & 71.31\%   & 63.39\% & \cellcolor{high}81.23\% & \cellcolor{best}50.00\% & 50.00\%   \\
                               & S10  & 65.12\%   & \cellcolor{low}58.49\%  & 61.51\% & \cellcolor{high}66.26\%   & 59.72\% & 65.57\% & \cellcolor{best}50.00\% & 50.00\%   \\
                               & UTK  &  62.97\%   & 55.84\%  & \cellcolor{low}55.43\% & 56.28\%   & 55.52\% & \cellcolor{high}63.53\% & \cellcolor{best}50.00\% & 50.00\%   \\\hline
    \multirow{4}{*}{\rotatebox{90}{ResNet18}}  & C10  &  70.37\%   & 65.01\%  & 66.59\% & 59.12\%   & \cellcolor{low}52.67\% & \cellcolor{high}69.53\% & \cellcolor{best}50.00\% & 50.00\%   \\
                               & C100 &  82.75\%   & 81.10\%  & 82.92\% & \cellcolor{low}67.55\%   & 76.31\% & \cellcolor{high}83.73\% & \cellcolor{best}50.00\% & 50.00\%   \\
                               & S10  &  76.09\%   & 65.98\%  & \cellcolor{high}74.22\% & 64.29\%   & \cellcolor{low}59.83\% & 74.07\% & \cellcolor{best}50.00\% & 50.00\%   \\
                               & UTK  &  62.87\%   & 56.33\%  & 59.25\% & 54.53\%   & \cellcolor{low}51.69\% & \cellcolor{high}63.62\% & \cellcolor{best}50.00\% & 50.00\%   \\\hline
    \multirow{4}{*}{\rotatebox{90}{VGG16\_BN}} & C10  &  63.17\%   & \cellcolor{high}64.03\%  & 62.44\% & 58.63\%   & \cellcolor{low}55.20\% & 62.14\% & \cellcolor{best}50.00\% & 50.00\%   \\
                               & C100 &  81.22\%   & 78.63\%  & \cellcolor{high}81.34\% & 71.25\%   & \cellcolor{low}50.10\% & 81.13\% & \cellcolor{best}50.00\% & 50.00\%   \\
                               & S10  &  66.08\%   & \cellcolor{high}68.20\%  & 66.20\% & 66.97\%   & \cellcolor{low}58.22\%  & 65.85\% & \cellcolor{best}50.00\% & 50.00\%   \\
                               & UTK  &  58.73\%   & 52.79\%  & 58.48\% & \cellcolor{high}58.93\%   & \cellcolor{low}51.34\%  & 57.17\% & \cellcolor{best}50.00\% & 50.00\%   \\ \hline

        \multirow{4}{*}{\rotatebox{90}{ResNet34}}  & C10  &  67.72\%   & 61.19\%  & 56.41\% & 55.68\%   & \cellcolor{low}50.16\% & \cellcolor{high}66.74\% & \cellcolor{best}50.00\% & 50.00\%   \\
                                   & C100 &  65.56\%   & \cellcolor{high}67.78\%  & 66.06\% & 67.29\%   & 66.84\%  & 65.63\%  & \cellcolor{best}50.00\% & 50.08\%   \\
                                   & S10  &  72.03\%   & 63.85\%  & \cellcolor{high}69.40\% & 64.95\%   & \cellcolor{low}63.42\% & 67.52\& & \cellcolor{best}50.00\% & 50.00\%   \\
                                   & UTK  &  61.33\%   & 56.80\%  & 56.91\% & 55.93\%   & \cellcolor{low}52.92\%  & \cellcolor{high}62.85\% & \cellcolor{best}50.00\% & 50.00\%   \\ \midrule
        \multirow{4}{*}{\rotatebox{90}{VGG19\_BN}} & C10  &  63.57\%   & 62.28\%  & 62.52\% & \cellcolor{low}60.74\%   & \cellcolor{high}63.15\% & 62.94\%  & \cellcolor{best}50.00\% & 50.00\%   \\
                                   & C100 &  77.94\%   & 77.58\%  & \cellcolor{high}78.22\% & 72.91\%   & \cellcolor{low}52.53\%  & 78.16\% & \cellcolor{best}50.00\% & 50.01\%   \\
                                   & S10  &  66.35\%   & 63.14\%  & 66.62\% & \cellcolor{high}67.25\%   & \cellcolor{low}58.06\%  & 66.17\% & \cellcolor{best}50.00\% & 49.91\%   \\
                                   & UTK  &  61.50\%   & 59.27\%  & 60.28\% & 54.55\%   & \cellcolor{low}54.13\%  & \cellcolor{high}61.74\% & \cellcolor{best}50.00\% & 50.00\%   \\ \midrule
        \multicolumn{2}{c}{Average}       &  1.37$\times$      & 1.28$\times$     & 1.32$\times$    & 1.25$\times$      & 1.16$\times$  & 1.36$\times$   & 1.00$\times$    & 1.00$\times$      \\ \hline
    \end{tabular}

\end{adjustbox}
\vspace{-10pt}

\end{table*}

\fi

\diff{From \T~\ref{tbl:evaluate_solution_MS} and \T~\ref{tbl:evaluate_solution_MI},} we can observe that the defense
effectiveness of all solutions is limited. It is evident that even the lowest
attack accuracy (marked in \colorbox{low}{yellow}), indicating the highest
defense effectiveness, among each setting, is still \textit{much higher} than
the black-box baseline: the lowest attack accuracies are averagely {3.85$\times$}
higher than black-box for \ac{ms} and {1.16$\times$} higher for \ac{mia}. Even
worse, the highest attack accuracies (marked in \colorbox{high}{red}) are
similar to that of the \noshield baseline, indicating that the defense schemes
are ineffective.

The relative accuracy (w.r.t.~black-box baselines) of DarkneTZ (\ding{172}) for
\ac{ms} is \diff{4.27$\times$} and \diff{1.28$\times$} for \ac{mia}. For Serdab (\ding{173}),
the relative attack accuracies are \diff{4.10$\times$} and \diff{1.32$\times$}. Since the
attack performance toward both Serdab and DarkneTZ is high, we interpret that
shielding a limited number of deep layers (\ding{172}) or shallow layers
(\ding{173}) facilitates limited protection. Magnitude (\ding{174}) achieves a
similar defense effect with DarkneTZ, with \diff{3.85$\times$} higher accuracy for
\ac{ms} and \diff{1.25$\times$} higher accuracy for \ac{mia}. 
Though the DarkneTZ and Magnitude papers empirically demonstrate the defense
effectiveness against naive adversaries (without the surrogate model initialized
by a pre-trained model or public data), we depict that well-designed and
practical attacks can crack such empirical settings. 

SOTER (\ding{175}) offers best protection across all solutions for \ac{mia}:
in \diff{13 (out of 20)} cases, SOTER achieves the lowest \ac{mia} accuracy among
the five solutions. However, its higher security strength does not come for
free. SOTER shields the largest number of layers using TEEs (20\% layers)
compared with other solutions. That is, SOTER has a much higher inference
latency and utility cost. 

Among the five schemes, ShadowNet (\ding{176}) shows the weakest
protection and most ``\colorbox{high}{red cases}'' in
both \T~\ref{tbl:evaluate_solution_MS} and \T~\ref{tbl:evaluate_solution_MI}: \diff{eight (out of 20)} for \ac{ms} and \diff{nine} for
\ac{mia}. The average attack accuracy against ShadowNet is similar to that of
the No-Shield baseline. According to our evaluation~\cite{TEESliceWebsite}, the
adversary can recover 95\% of the obfuscated weights. This indicates that its
lightweight obfuscation (matrix obfuscation and filter permutation) is
insufficient to protect the target model in front of well-designed attacks.

\diff{For ResNet34, the relative attack accuracy of DarkneTZ (\ding{172}) for
\ac{ms} is 3.47$\times$ and 1.25$\times$ for \ac{mia}. Of Serdab (\ding{173}) and Magnitude (\ding{174}), the relative accuracies are 2.79$\times$, 2.14$\times$ for \ac{ms} and 1.24$\times$, 1.22$\times$ for \ac{mia}. SOTER (\ding{175}) still offers best protection across all solutions for \ac{mia} and ShadowNet (\ding{176}) shows the weakest
protection for ResNet34. These results are consistent with the results of other models with only few differences in the specific number. For VGG19\_BN, although SOTER (\ding{175}) shows best protection for \ac{mia}, it behaves as poorly as ShadowNet (\ding{176}) when defensing \ac{ms}.}

\diff{From the results of ResNet34 and VGG19\_BN, it can be seen that the specific performance of different solutions may vary for models of the same architecture but different sizes. However, all the results show that when being exploited by \ac{ms}
and \ac{mia}, existing \tsdp\ solutions
do not provide a black-box level security guarantee. The results of our experiments are consistent among models of different architectures and sizes.}

\begin{tcolorbox}[size=small]
\textbf{Answer to RQ1}: Contrary to our expectation, existing \tsdp\ solutions
do not provide a black-box level security guarantee when being exploited by \ac{ms}
and \ac{mia}. That is, their shielded model weights and private training data
are vulnerable to attackers with well-prepared $M_{\rm init}$ on hand.
\end{tcolorbox}

\section{Challenges of Straightforward Mitigations}
\label{sec:dilemma}

Our study and observation in answering \textbf{RQ1} show that straightforward
mitigation to the attacks for existing \ac{tsdp} approaches is to put a larger proportion of a DNN model into TEEs to improve the protection effectiveness.
However, this straightforward solution needs to address the \textit{Security vs.
Utility Trade-off}: putting more portions of a DNN model into TEEs boosts
security but presumably diminishes utility (e.g., increases prediction latency).
Thus, the objective is to find a ``sweet spot'' configuration (i.e., how large
the TEE protected part should be) that satisfies the security requirement while
minimizing the utility overhead. This section will address the
following research question:

\begin{tcolorbox}[size=small]
    \textbf{RQ2:} 
 For each of the five \tsdp\ solutions evaluated in
\S~\ref{sec:evaluate_existing_solutions}, is there a systematic approach to
identify its ``sweet spot'' configuration that simultaneously achieves high
utility and security?

\end{tcolorbox}
\vspace{-5pt}

\subsection{Problem Formalization}
To systematically find the ``sweet spots'' of the optimal size of the TEE
shielded part for the \tsdp\ solutions in
Section~\ref{sec:evaluate_existing_solutions}, we first formalize the problem as
an optimization problem. Formally, let $P$ be a \tsdp\ solution that splits a
DNN model into TEE-shielded and GPU-offloaded portions. Let $C$ denote a
configuration instance of $P$ that specifies to what degree the model is
shielded. We define two evaluation functions, $Security(C)$ and $Utility(C)$,
which quantify the security risk and the utility cost of $C$, respectively. We
also define $Security_{black}$ as the security risk baseline of a black-box setting, which puts the whole DNN model in TEE. As noted in
\S~\ref{sec:evaluated_attacks}, $Security_{black}$ denotes the lower bound of
the security risk (the strongest protection \tsdp\ can offer). Then, given the
security requirement $\Delta$, we formulate the ``sweet spot'' configuration
$C^*$ that satisfies $|\ Security(C) - Security_{black} \ | < \Delta$ with the
minimal $Utility(C)$.  Alternatively, the  ``sweet spot'' is the solution to
Equation~\ref{equ:optimal_config}.
\begin{equation}
    C^* = \mathop{\arg\min}_{ \substack{ | \ Security(C) - Security_{black} \ | < \Delta  }} \ Utility(C)
    \label{equ:optimal_config}
\end{equation}

\subsection{Experimental Settings}
\label{sec:optimal_config_setting}

To answer \textbf{RQ2}, we empirically identify the ``sweet spots'' for the five
\tsdp\ defenses evaluated in \S~\ref{sec:evaluate_existing_solutions}
w.r.t.~different security risk metrics, datasets, and shielded models. We
discuss the details of the experiment setup below.

\parh{Security Risk Metric.}~Consistent with \S~\ref{subsec:evaluation-setup},
we implement $Security(C)$ using seven security metrics. For \ac{ms}, we use
model stealing accuracy, fidelity, and \ac{asr}. For \ac{mia}, we use
confidence-based \ac{mia} accuracy, gradient-based \ac{mia} accuracy,
generalization gap, and confidence gap. 
\ifentireresult
{}
\fi

\parh{Utility Cost Metric.}~As a common setup, we use FLOPs to measure the
utility cost of DNN
models~\cite{hou2021model}. FLOPs is
a platform-irrelevant metric to assess the utility cost $Utility(C)$ by counting
the total number of multiplication and addition operations conducted inside
TEEs. We define $\% FLOPs$ as the ratio of FLOPs in the TEE over the total FLOPs
of the DNN model. According to prior work~\cite{tramer2019slalom}, the
computation speed inside TEE is about 30$\times$ slower than GPUs. Thus, a
larger $\% FLOPs(C)$ indicates fewer computations are offloaded on GPUs, leading
to higher utility costs.

We compute the FLOPs of different layers as follows. For a DNN layer, suppose
the input channel size is $c_{in}$, the output channel size is $c_{out}$,  and
the width and height of the output are $w$ and $h$. The FLOPs of a linear layer
is computed as $2 \times c_{in} \times c_{out}$. The FLOPs of a batch
normalization layer are computed as $2 \times c_{in} \times h \times w$. For a
convolution layer, suppose the kernel size is $k$, the FLOPs are computed as $2
\times c_{in} \times k^2 \times h \times w \times c_{out}$.

\begin{figure}[!t]
\centering
\includegraphics[width=0.7\linewidth]{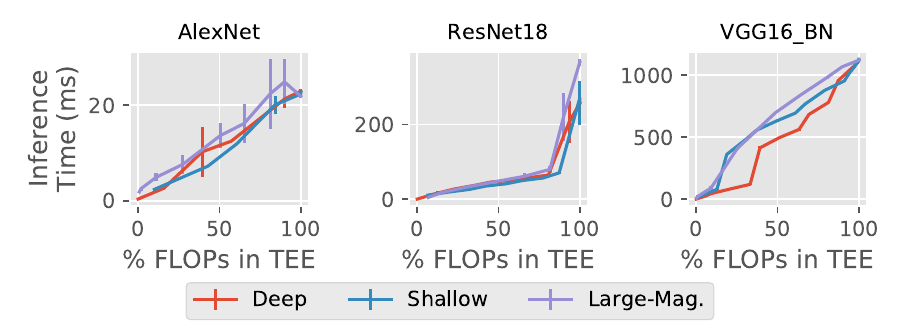}
\caption{Correlation between $\% FLOPs$ and inference latency on
Occlum~\cite{shen2020occlum}. Inference time is averaged over 10 runs.}
\label{fig:occlum_time_flops}
\vspace{-10pt}
\end{figure}

To validate the correlation between $\% FLOPs$ and utility cost, \diff{we did an initial
experiment on the real-world commercial TEE platform,
Occlum~\cite{shen2020occlum}, based on Intel SGX. We evaluate the model
inference time on three defense mechanisms: shielding deep layers (\ding{172}),
shielding shallow layers (\ding{173}), and shielding large-magnitude layers
(\ding{174}). We repeat each experiment ten times and plot the average inference
time and standard error on three models in \F~\ref{fig:occlum_time_flops}. As
the figure shows, for all TSDP solutions, inference time monotonically increases
as $\% FLOPs$ increases. Thus using $\% FLOPs$ to represent inference time and
utility cost is reasonable.}

\parh{Datasets and Models.}~\diff{The dataset and model selection are the same as
\S~\ref{subsec:evaluation-setup}. }
\ifentireresult
{}
\fi

\parh{Configurations.}~For each \tsdp\ defense benchmarked in
\S~\ref{sec:evaluate_existing_solutions}, we iterate possible configurations to
identify $C^*$. In particular, for the defense that shields deep layers
(\ding{172}), we shield different numbers of consecutive ``deep'' layers
starting from the output layer with TEEs. Similarly, for \ding{173}, which
shields shallow layers, we put different amounts of consecutive layers starting
from the DNN input layer. For ResNet models, we use the residual layers as the
dividing boundaries. For VGG models and AlexNet models, we use convolution
layers as boundaries.

For shielding large-magnitude weights (Magnitude; \ding{174}), the number of protected weights is controlled by a configuration parameter
\texttt{mag\_ratio}. We set the range of \texttt{mag\_ratio} as $\{0, 0.01, 0.1, 
0.3, $\\$ 0.5, 0.7, 0.9, 1\}$, whereas $0.01$ is the recommended setting of
Magnitude. For shielding intermediate layers (SOTER; \ding{175}), the number of
shielded layers is also defined by a configuration parameter,
\texttt{soter\_ratio}. We set the range of \texttt{soter\_ratio} as $\{0, 0.1,
0.2, 0.3, 0.5, 0.7, 0.9, 1\}$ and $0.2$ is the recommended setting of the
original paper. For both \ding{174} and \ding{175}, setting \texttt{mag\_ratio}
(and \texttt{soter\_ratio}) to $0$ represents the \noshield baseline while
setting the parameters to $1$ is the black-box baseline. For shielding
non-linear layers (ShadowNet; \ding{176}), we clarify that it does not
require any configuration.

\parh{Attack Implementation.}~We re-run the same \ac{ms} and \ac{mia} as in
\S~\ref{sec:evaluated_attacks}. That is, we re-launch the three-phase attack
pipeline, which comprises surrogate model initialization (P$_{i}$), model
stealing (P$_{ii}$), and membership inference (P$_{iii}$).

\ifentireresult

\begin{figure*}[htp]
\centering
\includegraphics[width=\linewidth]{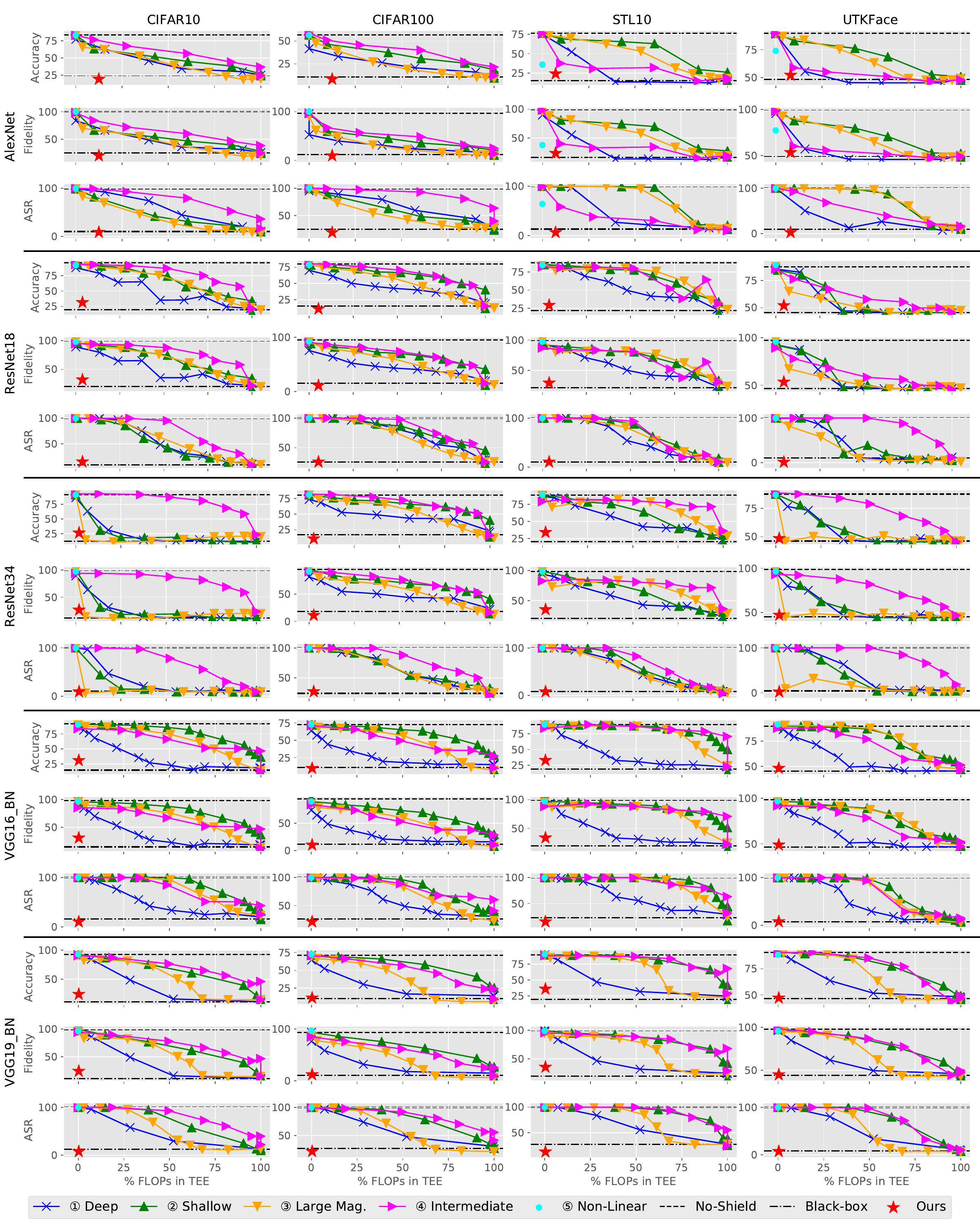}
\caption{\diff{Model stealing results in terms of accuracy, fidelity, and ASR.}}
\label{fig:acc_flops_append_one_fig}
\end{figure*}

\begin{figure*}[htp]
\centering
\includegraphics[width=\linewidth]{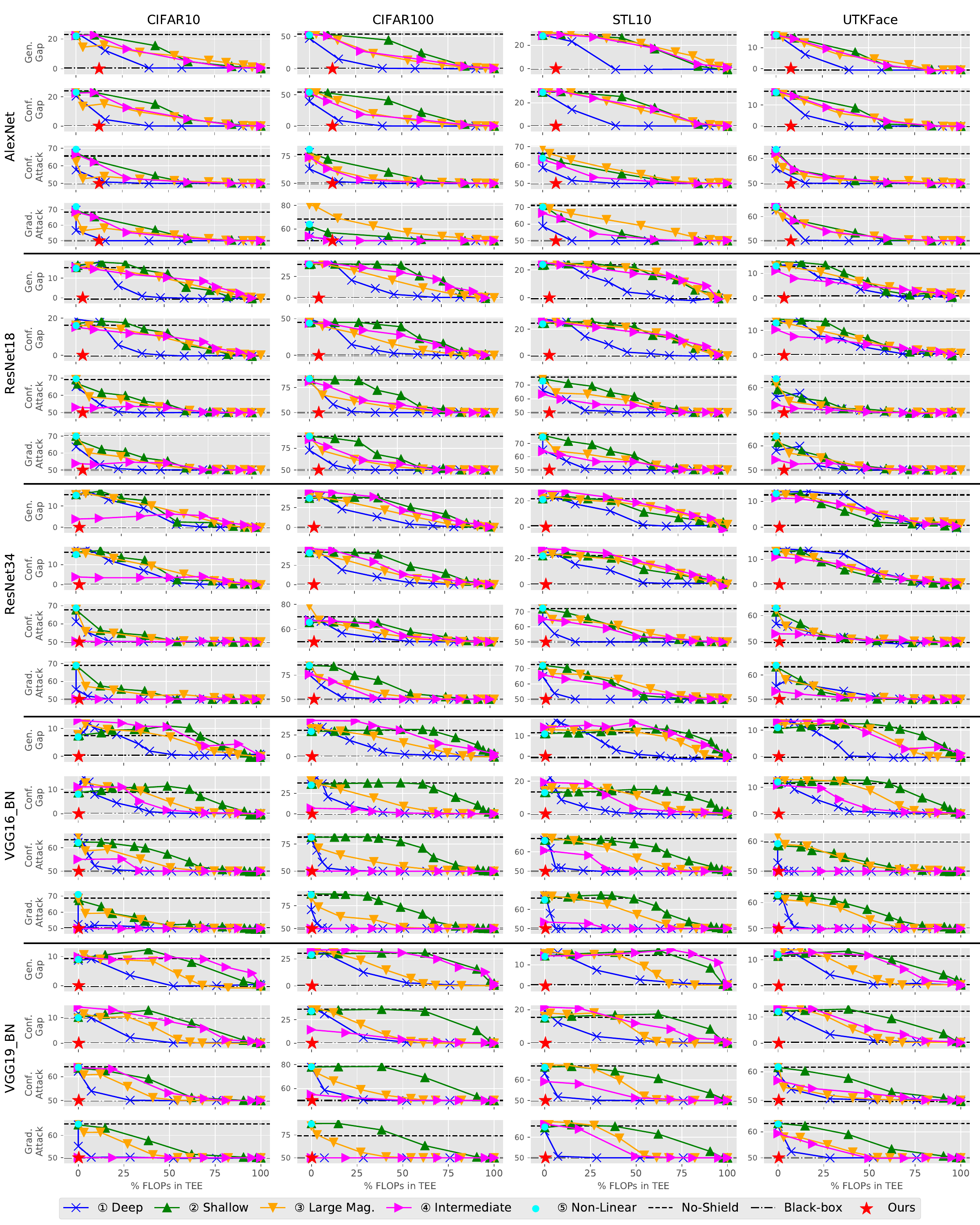}
\caption{\diff{Membership inference results of generalization gap, confidence gap, confidence-based membership inference attack accuracy, and gradient-based membership inference attack accuracy.}}
\label{fig:mia_flops_append_one_fig}
\end{figure*} 
\fi

\subsection{Qualitative and Quantitative Results}
\label{sec:evaluation_dilemma_results}

We compute both qualitative and quantitative results to explore if the ``sweet
spot'' configuration exists for each scheme and the characteristics of the sweet
spots. For qualitative results, {\F~\ref{fig:acc_flops_append_one_fig} and \F~\ref{fig:mia_flops_append_one_fig} presents the
relationship between $Security$ and $Utility$ of different configurations. 
\ifentireresult
{}
\fi
In \F~\ref{fig:acc_flops_append_one_fig} and \F~\ref{fig:mia_flops_append_one_fig}, the x-axis shows $\%FLOPs$ ($Utility$) of
each partition configuration, and the y-axis shows the metrics of \ac{ms} and
\ac{mia} ($Security$). For each model and dataset, every metric is displayed in a
sub-figure. In each sub-figure, shielding deep layers
(\ding{172}), shielding shallow layers (\ding{173}), shielding large-magnitude
weights (\ding{174}), shielding intermediate layers (\ding{175}), and shielding
non-linear layers (\ding{176}) are represented by a blue line with crosses, a
green line with up triangles, an orange line with down triangles, a pink line
with right triangles, and an aquamarine circle, respectively. 
We also plot the performance of
the \noshield baseline and the black-box baseline using horizontal black lines. 
\ifentireresult
{}
\else
\fi
}

In a holistic sense, \diff{\F~\ref{fig:acc_flops_append_one_fig} and \F~\ref{fig:mia_flops_append_one_fig} suggest that there is no
systematic and automated approach to identifying the ``sweet spot'' for the five
representative defenses. The shapes of the lines are substantially different
across datasets and models. Given a requirement $\Delta$ for security risk,  it
is tough to set a uniform threshold for $Utility(C^*)$ without a comprehensive
empirical measurement of both $Security$ and $Utility$. For example, for
shielding deep layers (\ding{172}) of AlexNet for model stealing (the first three row
in \F~\ref{fig:acc_flops_append_one_fig}), the shape of the curves for CIFAR10 and
CIFAR100 are very different from the curve for STL10. Given a requirement
$\Delta$ and a model to protect, the locations of ``sweet spots'' are random for
different datasets. }

\diff{Comparing the curves of ResNet34 with those of ResNet18, and VGG16\_BN with VGG19\_BN, the conclusion comes that even for models of the same architecture but different sizes, the shape of curves for the same dataset are very different. This further verifies the randomness of the locations of ``sweet spots''.}

For quantitative results, we measure the values of the $Utility(C^*)$ as defined
in Equation~\ref{equ:optimal_config} (the smallest value of $Utility$ to achieve
$Security_{black}$) for different defenses. \T~\ref{tbl:optimal_config} reports
the ratio of TEE-shielded FLOPs ($\%FLOPs$) to achieve $Security_{black}$ for
each setting for \ac{ms} and \ac{mia}. We omit the approach of shielding non-linear layers
(ShadowNet, \ding{176}) because it does not require configurations. 
Overall, \T~\ref{tbl:optimal_config}
implies that \textit{the $Utility$ values to achieve $Security_{black}$ are
distinct across protected models and datasets.} For example, to protect AlexNet
from \ac{ms} with shielding deep layers (\ding{172}), we need to put 100\% of
the protected model in TEE to achieve $Security_{black}$ for CIFAR10 (C1) and
CIFAR100 (C100). However, for STL10 (S10) and UTKFace (UTK), we only need to put
39.44\% of FLOPs in TEE to achieve $Security_{black}$. Further, it is tough, if
not impossible, to predict the actual value of $Utility(C^*)$ before we run the
models empirically because the numbers are irregular. We also observe similar
irregularity for other defense methods when protecting different models and
datasets.

\ifentireresult
\colorlet{best}{green!30}
\colorlet{high}{red!30}
\colorlet{low}{yellow}

\begin{table*}[h]
\caption{Different $Utility(C^*)$ ($\% FLOPs(C^*)$) values of ``sweet spot'' in front of \ac{ms} and \ac{mia}. 
A lower value represents a lower utility cost. The $\% FLOPs(C^*)$ for \noshield and black-box baselines are 0\% and 100\%, respectively.
For each \tsdp\ solution (column), we mark the lowest $Utility(C^*)$ with \colorbox{low}{yellow} and the 
highest value with \colorbox{high}{red}.
For each case (model and dataset, column), we mark the lowest $Utility(C^*)$ across all solutions with \colorbox{best}{green}.
The last column is the average utility cost for each solution.
We omit shielding non-linear layers
(ShadowNet, \ding{176}) because it does not require configurations.
}
\label{tbl:optimal_config}
\vspace{-5pt}
\setlength{\tabcolsep}{1.5pt}
\centering
\begin{adjustbox}{max width=0.85\linewidth}

\begin{tabular}{cccccclcccclc}
\hline
                           &      & \multicolumn{4}{c}{\begin{tabular}[c]{@{}c@{}}Model Stealing $\downarrow$\end{tabular}} &  & \multicolumn{4}{c}{\begin{tabular}[c]{@{}c@{}}Membership Inference $\downarrow$\end{tabular}} &  & \multirow{2}{*}{Ours} \\ \cline{3-6} \cline{8-11}
                           &      & \ding{172}Deep             & \ding{173}Shallow          & \ding{174}Large Mag.        & \ding{175}Intermediate        &  & \ding{173}Deep              & \ding{173}Shallow           & \ding{174}Large Mag.          & \ding{175}Intermediate          &  &                       \\ \hline
\multirow{4}{*}{\rotatebox{90}{AlexNet}}   & C10  & \cellcolor{high}100.0\%         & \cellcolor{high}100.0\%         & 81.18\%           & \cellcolor{high}100.0\%            &  & 15.78\%           & 60.56\%           & 34.51\%             & 60.69\%               &  & \cellcolor{best}12.48\%               \\
                           & C100 & \cellcolor{high}100.0\%         & \cellcolor{high}100.0\%         & 90.58\%           & \cellcolor{high}100.0\%            &  & 15.78\%           & 84.22\%           & 53.17\%             & 60.69\%               &  & \cellcolor{best}12.48\%               \\
                           & S10  & 39.44\%          & \cellcolor{high}100.0\%         & \cellcolor{high}100.0\%          & 84.31\%             &  & 15.78\%           & 60.56\%           & 71.82\%             & 60.69\%               &  & \cellcolor{best}7.12\%                \\
                           & UTK  & 39.44\%          & \cellcolor{high}100.0\%         & 71.82\%           & \cellcolor{low}60.69\%             &  & 15.78\%           & 42.81\%           & 34.51\%             & 27.95\%               &  & \cellcolor{best}8.01\%                \\ \hline
\multirow{4}{*}{\rotatebox{90}{ResNet18}}  & C10  & \cellcolor{high}100.0\%         & \cellcolor{high}100.0\%         & \cellcolor{high}100.0\%          & \cellcolor{high}100.0\%            &  & 23.97\%           & 62.54\%           & 61.48\%             & 72.80\%               &  & \cellcolor{best}3.80\%                \\
                           & C100 & \cellcolor{high}100.0\%         & \cellcolor{high}100.0\%         & 94.71\%           & 100.00              &  & 23.97\%           & 86.52\%           & 61.48\%             & 72.80\%               &  & \cellcolor{best}5.33\%                \\
                           & S10  & \cellcolor{high}100.0\%         & \cellcolor{high}100.0\%         & \cellcolor{high}100.0\%          & \cellcolor{high}100.0\%            &  & 23.97\%           & 76.03\%           & \cellcolor{high}76.61\%             & 72.80\%               &  & \cellcolor{best}3.80\%                \\
                           & UTK  & \cellcolor{low}37.46\%          & \cellcolor{low}38.55\%          &61.48\%           & \cellcolor{high}100.0\%            &  & 23.97\%           & \cellcolor{low}38.55\%           & 44.93\%             & 11.02\%               &  & \cellcolor{best}4.58\%                \\ \hline
\multirow{4}{*}{\rotatebox{90}{VGG16\_BN}} & C10  & \cellcolor{high}100.0\%         & \cellcolor{high}100.0\%         & \cellcolor{high}100.0\%          & \cellcolor{high}100.0\%            &  & 9.03\%            & 66.90\%           & 50.56\%             & 33.84\%               &  & \cellcolor{best}0.34\%                \\
                           & C100 & \cellcolor{high}100.0\%         & \cellcolor{high}100.0\%         & 87.43\%           & \cellcolor{high}100.0\%            &  & 21.07\%           & 90.97\%           & 66.47\%             & \diagfil{2.1cm}{low}{best}{0.10\%}                &  & 0.47\%                \\
                           & S10  & \cellcolor{high}100.0\%         & \cellcolor{high}100.0\%         & \cellcolor{high}100.0\%          & \cellcolor{high}100.0\%            &  & 6.02\%            & 90.97\%           & 66.47\%             & 33.84\%               &  & \cellcolor{best}0.40\%                \\
                           & UTK  & 69.23\%          & \cellcolor{high}100.0\%         & \cellcolor{high}100.0\%          & \cellcolor{high}100.0\%            &  & \cellcolor{low}3.01\%            & 60.88\%           & 50.56\%             & \diagfil{2.1cm}{low}{best}{0.10\%}              &  & 0.60\%                \\ \hline
\multirow{4}{*}{\rotatebox{90}{ResNet34}}  & C10  & 80.01\%          & 75.50\%          & 20.35\%           & \cellcolor{high}100.0\%            &  & 18.01\%           & 56.05\%           & 40.84\%             & \cellcolor{best}0.60\%              &  & 1.83\%                \\
                           & C100 & \cellcolor{high}100.0\%         & \cellcolor{high}100.0\%         & 94.96\%           & \cellcolor{high}100.0\%            &  & \cellcolor{high}55.49\%           & 87.03\%           & 58.31\%             & \cellcolor{high}83.89\%               &  & \cellcolor{best}2.56\%                \\
                           & S10  & \cellcolor{high}100.0\%         & \cellcolor{high}100.0\%         & \cellcolor{high}100.0\%          & \cellcolor{high}100.0\%            &  & 18.01\%           & 75.50\%           & 74.82\%             & \cellcolor{high}83.89\%               &  & \cellcolor{best}1.73\%                \\
                           & UTK  & 37.47\%          & 56.05\%          & \cellcolor{low}5.22\%            & \cellcolor{high}100.0\%            &  & 37.47\%           & 38.02\%           & 40.84\%             & 36.36\%               &  & \cellcolor{best}1.92\%                \\ \hline
\multirow{4}{*}{\rotatebox{90}{VGG19\_BN}} & C10  & 75.79\%          & \cellcolor{high}100.0\%         & \cellcolor{high}100.0\%          & \cellcolor{high}100.0\%            &  & 28.42\%           & 62.11\%           & 40.90\%             & 69.26\%               &  & \cellcolor{best}0.27\%                \\
                           & C100 & \cellcolor{high}100.0\%         & \cellcolor{high}100.0\%         & 67.68\%           & \cellcolor{high}100.0\%            &  & 28.42\%           & \cellcolor{high}97.63\%           & 54.49\%             & 19.05\%               &  & \cellcolor{best}0.37\%                \\
                           & S10  & \cellcolor{high}100.0\%         & \cellcolor{high}100.0\%         & \cellcolor{high}100.0\%          & \cellcolor{high}100.0\%            &  & 7.11\%            & \cellcolor{high}97.63\%           & 54.49\%             & 49.82\%               &  & \cellcolor{best}0.35\%                \\
                           & UTK  & 75.79\%          & \cellcolor{high}100.0\%         & 67.98\%           & 95.30\%             &  & 28.42\%           & 90.53\%           & \cellcolor{low}27.02\%             & 69.26\%               &  & \cellcolor{best}0.37\%                \\ \hline
\multicolumn{2}{c}{Average}       & 82.73\%          & 93.50\%          & 82.18\%           & 97.02\%             &  & 20.97\%           & 71.30\%           & 53.21\%             & 45.98\%               &  & 3.44\%                \\ \hline
\end{tabular}

\end{adjustbox}
\vspace*{-15pt}

\end{table*}

\fi

\begin{tcolorbox}[size=small]
    \textbf{Answer to RQ2}: It is difficult to systematically
    identify the ``sweet spots'' configuration $C^{*}$ for prior TSDP solutions. 
    \end{tcolorbox}
    \vspace{-5pt}

\section{Design of \tool}
\label{sec:approach}

Besides systematically benchmarking de facto \tsdp\ solutions (\textbf{RQ1}) and
summarizing their common drawbacks (\textbf{RQ2}), we conclude the root cause
of the \tsdp solutions' vulnerabilities and propose a novel partition scheme,
which alleviates the security vs. utility trade-off and can automatically find
the ``sweet spot'' configuration.

The root cause of \tsdp solutions' weakness is that all the \tsdp approaches
follow a \textit{\trainbeforepartition} strategy, which first trains $M_{vic}$
using private data and then partitions the private model. After training, all
the model weights (including the offloaded part) are updated by the private data
and thus contain private information. During deployment, the private information
in the offloaded weights is exposed to the untrusted environment. As
demonstrated in \textbf{RQ1} and \textbf{RQ2}, attackers could effectively
recover private information of $M_{vic}$ from offloaded privacy-related weights
with the help of public knowledge, \textit{i.e.}, a well-prepared $M_{\rm init}$ on hand.
With this regard, we champion that an ideal \tsdp\ solution should ensure
\textit{the offloaded DNN weights on GPUs are never trained using private data}
and thus, no information is leaked to the untrusted environment.

\begin{figure}[!h]
  \centering
  \includegraphics[width=0.65\linewidth]{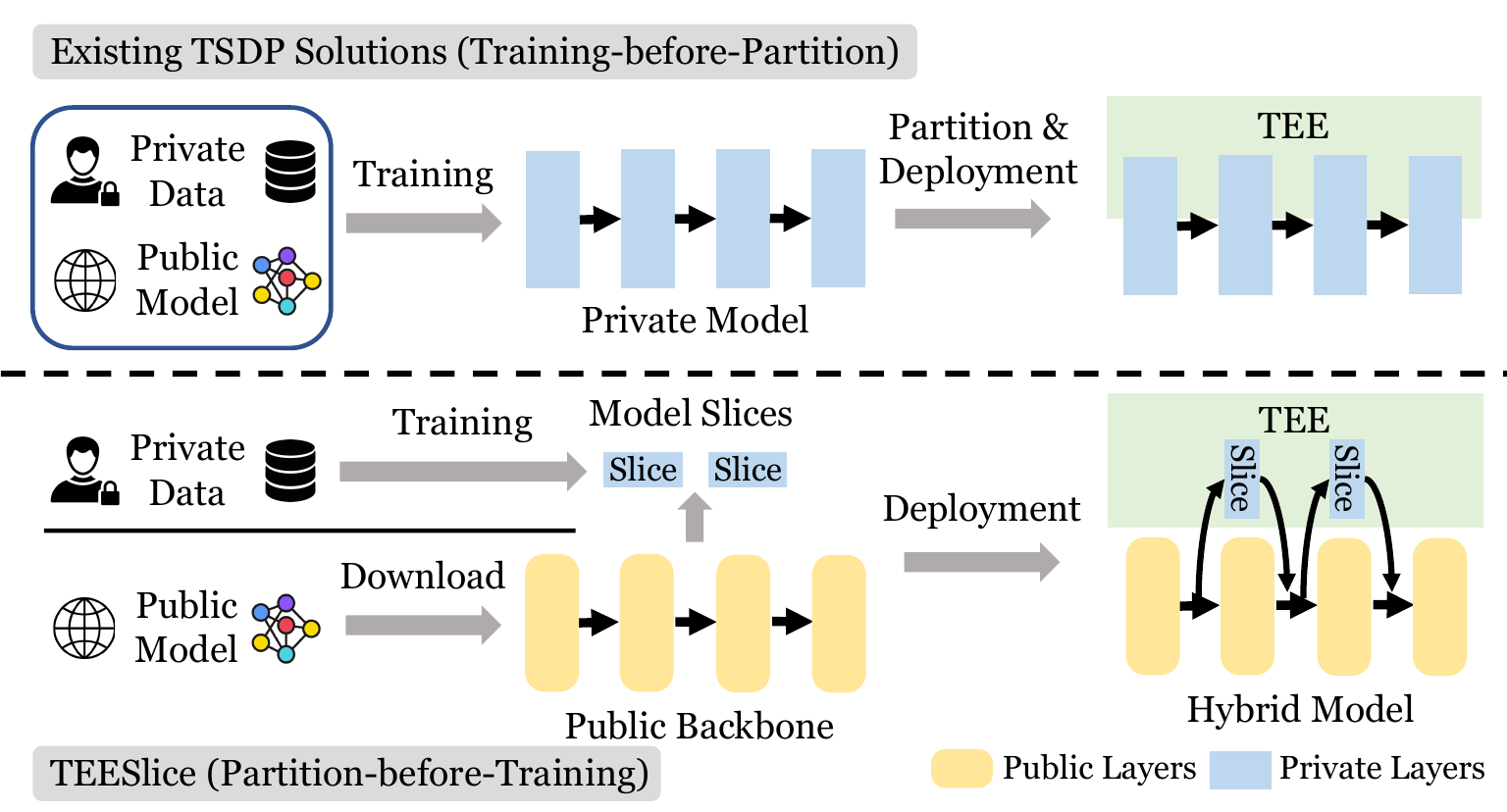}
  \vspace{-10pt}
  \caption{The comparison between \sys and prior \tsdp\ solutions. For \tool, all information generated by private data will be handled in the TEE.}
  \label{fig:teeslice_compare}
  \vspace{-10pt}
\end{figure}

\subsection{Approach Overview}

We propose \sys, a novel partitioning strategy that offloads DNN layers with no
private information to GPUs at the inference stage. Prior solutions use public
pre-trained model and private data to train $M_{\rm vic}$, use heuristic designs
to partition $M_{\rm vic}$, and shield a subset of model parts. On the contrary,
\tool uses a \textit{\partitionbeforetrain} paradigm, which partitions private data from
the pre-trained model and then separately trains privacy-related layers.
\F~\ref{fig:teeslice_compare} compares existing \tsdp\
solutions (\trainbeforepartition) and \tool (\partitionbeforetrain). 

We term the public pre-trained model as \textit{backbone}, and the
privacy-related layers as \textit{model slices}~\cite{zhang2020dynamic,zhang2022remos,zhang2022teeslice,zhang2023fedslice}. The model slices are
lightweight and contain the private knowledge of $M_{\rm vic}$. The backbone and
model slices are combined to form a \textit{hybrid model} (denoted as $M_{\rm
hyb}$) that imitates the behavior of $M_{\rm vic}$. Each slice takes the output
of the prior layer (of the backbone) as its input and produces the input for the
next layer (of the backbone). A detailed illustration of \tsdp is shown in
\F~\ref{fig:model_partition_categories}, where the yellow rounded blocks (Layer1
to Layer4) represent layers of the backbone and the blue squares (Slice1 and
Slice2) are privacy-related model slices. The arrows between backbone layers and
model slices indicate the data flow of internal DNN features. For example, the
output of Layer1 is fed to Slice1, and Layer4 takes the outputs of Slice2 and
Layer3 as input.

The key challenge for \sys is to generate small slices that can run in TEEs with
enough little or no accuracy lost. To this end, \sys leverages a two-staged
approach. First, it builds an instance of $M_{\rm hyb}$, a densely sliced model
(denoted as $M_{\rm dense}$), with substantial private slices that can achieve
high accuracy. However, $M_{\rm dense}$ cannot fit into TEEs due to its large
size. Then, \sys prunes $M_{\rm dense}$ with a self-adaptive, iterative pruning
strategy that produces fewer slices without losing the $M_{\rm hyb}$'s
performance. The pruned model is another instance of $M_{\rm hyb}$, which we
call a sparsely sliced model (denoted as $M_{\rm sparse}$). As the pruning is
conducted simultaneously with the training phase, \tool\ can prune slices with
little accuracy drop. Note that the iterative pruning strategy is specifically
designed for the resource-constrained TEE environment.

\sys is partially motivated by \nettailor~\cite{morgado2019nettailor}, which
implements a training framework of $M_{\rm dense}$. However, \nettailor does not
aim to protect model privacy with TEEs and generates a large number of slices. To
meet the constraint of TEEs, \tool generates $M_{\rm sparse}$ with an adaptive
pruning strategy to reduce the computation cost of slices. Besides,
\textsc{NetTailor} lacks cryptographic primitives to securely transmit DNN
intermediate data between GPUs and TEEs. \tool\ employs a one-time pad
(OTP)~\cite{OneTimePad} and Freivalds'
algorithm~\cite{freivalds1977probabilistic} to secure TEE-GPU transmission.

\subsection{Detailed Design}
\label{sec:teeslice_detail}

\sys consists of two stages: model slice extraction (training phase) and hybrid
model deployment (inference phase). The slice extraction stage automatically
finds the ``sweet spot'' by minimizing the utility cost while maintaining
accuracy and security. \sys trains $M_{\rm dense}$ from the public backbone and
then prunes $M_{\rm dense}$ to get $M_{\rm sparse}$. 
In the hybrid model deployment stage, $M_{\rm sparse}$ is deployed across the
TEE and GPU. Model slices and non-linear layers (of the backbone) are deployed
inside TEEs, whereas the other part of the backbone is offloaded on GPUs.

\subsubsection{Model Slice Extraction}

\parh{Densely Sliced Model Generation.}
\label{subsec:dense_model_generation}
Let the $i$-th layer of the public backbone be $L_i$. The private slice is
represented as $ A^i_p $, which connects layer $L_p$ with layer $L_i$. $ A^i_p $
is designed to be lightweight and is $18\times$ smaller than $L_i$. The slices
in $M_{\rm dense}$ connect a layer pair from backbone whenever the distance of
the layers in this pair is less than three. 
During the training stage, \sys assigns one importance scalar $\alpha_p^i$ to
each slice $ A^i_p $ following the same strategy in related
work~\cite{morgado2019nettailor}. The output of $ A^i_p $ is multiplied by
$\alpha_p^i$ and sent to the next layer (of the backbone). A smaller
$\alpha_p^i$ diminishes the influence of $ A^i_p $. The model slices and the
scalars are optimized simultaneously with a loss function that penalizes both
model performance and complexity. The output of this step is $M_{\rm dense}$
with close performance as $M_{\rm vic}$.

\parh{Iterative Slice Pruning.}~The pruning algorithm is guided by the
importance scalar $\alpha_p^i$, which controls the impact of $A^i_p$ on the
model prediction. We iteratively prune the slices with the smallest $\alpha_p^i$
and re-train the hybrid model to maintain the desired accuracy. 
A pre-defined threshold $\delta$ (defined as 1\%) governs the tolerable accuracy
loss during pruning. Let $ACC_{\rm vic}$ represent the accuracy of $M_{\rm
vic}$. The tolerable accuracy is $ACC_{\rm tol} = (1-\delta) \cdot ACC_{\rm
vic}$, and the accuracy of the final $M_{\rm hyb}$ should be greater than
$ACC_{\rm tol}$.

\A~\ref{alg:pruning_pipeline} depicts the pruning pipeline. $\alpha_{\rm setup}$ determines how to prune unimportant slices during the setup phase. The slice layers with $\alpha^l_p < \alpha_{\rm setup}$ are pruned. As a heuristic, we set $\alpha_{\rm init}$ to be $0.05$ according to \textsc{NetTailor}. Iterative pruning requires two hyper-parameters: the number of the pruned slices in each round $n$ and the total number of training rounds $rounds$. Each round begins with an evaluation of the model's accuracy $ACC_r$. If the current model satisfies the performance requirement ($ACC_r > ACC_{\rm tol}$), \sys prunes $n$ model slices with the smallest $\alpha^l_p$ and trains the pruned model. Otherwise, \sys skips the pruning operation and continues training the model. 

\begin{algorithm}
  \footnotesize
    \SetAlgoLined
    \SetKwProg{Fn}{Function}{:}{} \SetKwFunction{FIterativePrune}{Iterative
    Slice Pruning} \KwIn{The densely sliced model $M_{\rm dense}$, pre-defined
    parameters $\alpha_{\rm setup}$, $n$, and $rounds$} \KwOut{The hybrid model
    $M_{\rm hyb}$}
    
    Prune $M_{\rm dense}$ by $\alpha_{\rm setup}$ to get $M_{1}$ \;
    
    \For{$r\leftarrow 1$ \KwTo $rounds$}{
      Compute the accuracy $ACC_r$ of $M_r$ \;
      \If{$ACC_r > ACC_{\rm tol}$}{
          Store the model $M_{\rm hyb} = M_r$ \;
          Select $n$ slices with smallest $\alpha^l_p$ \;
          Prune the selected slices $A^l_p$ \;
      }
      Re-train $M_r$ to get $M_{r+1}$  \;
    }
    \KwRet The hybrid model $M_{\rm hyb}$
    
    \caption{Iterative Slice Pruning.}\label{alg:pruning_pipeline}
  \end{algorithm}

\parh{Automatically Find the Sweet Spot.}~The iterative slice pruning is indeed
an optimization process that automatically finds the ``sweet spot.'' Given the
constraints of model security and accuracy, the iterative slice pruning
optimizes the size of the private model slices to reduce the utility cost.
\A~\ref{alg:pruning_pipeline} only explicitly considers the threshold
$ACC_{tol}$ for accuracy lost because all the private information is in the
slices, which will run in TEEs. Therefore, the confidentiality of $M_{\rm vic}$
remains intact after offloading the backbone. Unlike prior \tsdp solutions,
where it is difficult to find the sweet spot configuration without a
comprehensive evaluation of both security and utility (\S~\ref{sec:dilemma}),
\sys does not have this shortcoming.

\subsubsection{Hybrid Model Deployment}
\label{sec:model_deployment}
When deploying $M_{\rm hyb}$, the private slices and non-linear layers (of the backbone) are shielded by the TEE, while the GPU hosts the backbone's linear layers. Shielding non-linear layers of the backbone is a common practice for prior \tsdp solutions~\cite{tramer2019slalom,hashemi2021darknight,lucien2021goten,hou2021model} because non-linear layers are hard to securely offload to GPUs and only occupy a small fraction (about 1.5\%) of the DNN's computation cost~\cite{tramer2019slalom}. In the illustration figures (\F~\ref{fig:model_partition_categories} and \F~\ref{fig:teeslice_compare}) we omit the non-linears in TEE for simplicity. There are two security challenges to deploy $M_{\rm hyb}$: 1) how to encrypt features transmitted between GPU and TEE, and 2) how to verify the correctness of computations offloaded on GPUs. These two challenges can be solved separately using one-time pad (OTP) and Freivalds' algorithm~\cite{freivalds1977probabilistic}.

\parh{Feature Encryption.}~For a backbone linear layer $g(\cdot)$, let
$\textbf{h}$ be the plaintext input shielded by TEE. We first quantize
$\textbf{h}$ into a 8-bit representation following prior
literature~\cite{tramer2019slalom, lucien2021goten} and get $\hat{\textbf{h}}$.
Then, we select a large prime value $p$, generate a random mask $\mathbf{r}$ (as
the OTP), and encrypt the feature by
\begin{equation}
    \mathbf{h}_e = (\hat{\mathbf{h}} + \mathbf{r}) \ \% \ p. 
\end{equation}

GPU receives $\mathbf{h}_e$, computes $g(\mathbf{h}_e)$, and returns the result
back to TEE. \sys decrypts the result by computing $g(\hat{\mathbf{h}}) =
g(\mathbf{h}_e) - g(\mathbf{r})$ because
\begin{equation}
  \begin{aligned}
  & \ g( \mathbf{h}_e) - g( \mathbf{r}) 
      =  g( (\hat{\mathbf{h}} + \mathbf{r})\ \% \ p) - g( \mathbf{r}\ \% \ p) \\
      = & \ g( (\hat{\mathbf{h}} + \mathbf{r})\ \% \ p - \mathbf{r}\ \% \ p ) 
      =  \ g( (\hat{\mathbf{h}} + \mathbf{r} - \mathbf{r})\ \% \ p ) \\
      = & \ g( \hat{\mathbf{h}}\ \% \ p ) \ = \ g( \hat{\mathbf{h}} ).
  \end{aligned}
  \label{equ:one_time_pad}
  \end{equation}
The last equation holds as long as $p > 2^8$. Note that following prior
work~\cite{tramer2019slalom}, computing $g(\mathbf{r})$ can be conducted by the
model provider or inside TEE in an offline phase. Both strategies do not
increase the overhead of online inference and do not impede its utility.

\parh{Result Verification.}~Freivalds' algorithm can periodically verify the
computation results on GPUs on all linear layers. Let the weight of the linear
layer $g(\cdot)$ be $\mathbf{W}$ and $g(\mathbf{h}) = \mathbf{h}^\mathsf{T}
\mathbf{W}$, \sys samples a random vector $\mathbf{s}$ that has the same shape
as $g(\mathbf{h})$. \sys then pre-computes $\tilde{\mathbf{s}} = \mathbf{W}
\mathbf{s}$. The verification can be conducted by checking
$g(\mathbf{h})^\mathsf{T} \mathbf{s} = \mathbf{h}^\mathsf{T}
\tilde{\mathbf{s}}$.

\subsection{Design for Large Language Models (LLMS)}
\label{sec:design_for_llm}
\diff{
The \partitionbeforetrain strategy is also applicable to LLMs to protect the
sensitive model privacy with TEEs. It can be integrated with recent LLMs'
parameter-efficient training techniques~\cite{peft} (\eg LoRA~\cite{hu2022lora})
to efficiently shield LLMs' critical privacy-related slices in the TEEs.
However, the original design of \tool in Section~\ref{sec:teeslice_detail} is
not directly applicable to LLMs because the training process of LLMs is much
more complex and the importance scalar $\alpha_p^i$ is difficult to train.
Therefore, we propose a new design for the slice extraction of LLM models, which
approximates the importance scalar $\alpha_p^i$ with the weight magnitude of
the slice. In this section, we will take LoRA~\cite{hu2022lora} as an example to
illustrate the design of LLMs. The phase of deployment of the hybrid model remains the
same as the original design.}

\parh{LLM Architecture.}
Existing large language models follow the transformer-based architecture. A LLM is composed of multiple transformer blocks. Each transformer block consists of two modules: an attention module and a feed-forward (FFN) module. We illustrate the composition of these two modules as follows.
For the attention module, let $x_a$ be the module input. The module computes three features: a key $x_k$, a query $x_q$, and a value $x_v$ by
\begin{equation}
    x_k = W_k \cdot x_a, x_q = W_q \cdot x_a, x_v = W_v \cdot x_a.
\end{equation}
The attention score is computed by 
\begin{equation}
\label{equ:attention_score}
    Attention(x_q, x_k, x_v) = softmax( \frac{x_q \cdot x_k}{\sqrt{d}}) x_v,
\end{equation}
where $d$ is the dimension of the hidden state. 
The existing attention module also uses a multi-head attention mechanism. This mechanism first projects the query, key, and value vectors into multiple lower dimensions, performs the attention operation within each dimension, concat the attention, and projects to the final values. This multi-head attention can be formalized as
\begin{equation}
    head^i = Attention(x_q \cdot W_q^i, x_k \cdot W_k^i, x_v \cdot W_v^i), \\
\end{equation}
\begin{equation}
        MultiHead(x_q, x_k, x_v) = Concat(head^1, \dots, head^n) \cdot W_o.
\end{equation}
In the above equations, $i$ represents the head index. $W_q^i$, $W_k^i$, and $W_v^i$ are the projection matrix for the head $i$, and $W_o$ is the output projection matrix. The FFN module is not as complex as the attention module. This module consists of three layers: a linear layer, an activation layer, and another linear layer. Let $x_f$ represent the input of the FFN module. The module can be formulated as:
\begin{equation}
    FFN(x_f) = act( x_f \cdot W_1 + b_1 ) \cdot W_2 + b_2,
\end{equation}
where $act$ represents the activation layer. 

From the introduction of LLM architecture, we can find that there are several linear operations in each transformer block that we can add a slice to. In the attention module, the three projection matrixes ($W_q$, $W_k$, and $W_v$) and the output projection matrix ($W_o$) are linear layers. In the FFN module, there are two linear operations ($W_1$ and $W_2$).

\diff{\parh{LoRA.}~LoRA (Low Rank Adaption) aims to reduce the computational cost of
training LLMs on downstream tasks. Instead of updating the large number of LLM model weights, LoRA adds low-rank and lightweight layers in addition to the original
linear layers of LLMs. During the training process, LoRA only updates the
weights of the low-rank layers and keeps the original weights of the LLMs unchanged.
Let $L_i$ represent the $i$-th linear layer of LLM, $d_{M}$ represent the
input/output dimension of $L_i$ and $r$ represent the reduced dimension of the
low-rank layers. The low-rank layers for $L_i$ are designed to be two sequential
linear layers with dimensions $(d_{M}, r)$ and $(r, d_{M})$, respectively. We
denote the combination of these two sequential layers by $A_i^{i+1}$ because
they connect $L_i$ and $L_{i+1}$. The output of $A_i^{i+1}$ is added to the
output of $L_i$. }

\diff{\parh{Densely Sliced Model Generation.}~During the private training phase, the
LoRA layers are one type of private model slices. However, it is difficult
to directly apply the importance scalar $\alpha$ (of
\S~\ref{subsec:dense_model_generation}) to LoRA layers because the LLM training process is much more complex than CNNs. Therefore, we use the weight
information in the unmodified LoRA layers to approximate the importance scalar
$\alpha$. Specifically, for $A_i^{i+1}$, we use the weight magnitude of the LoRA
layers as the importance scalar $\alpha_i^{i+1}$. The weight magnitude of
$A_i^{i+1}$ is defined as the sum of the absolute value of all the weights in
$A_i^{i+1}$ and is denoted as $Mag(A_i^{i+1})$. The output of this step is also
$M_{\rm dense}$ with a close performance to $M_{\rm vic}$.}

\diff{\parh{Iterative Slice Pruning.}~The pruning algorithm is guided by
$Mag(A^{i+1}_i)$, which controls the impact of $A^{i+1}_i$ on the prediction of the model. We iteratively remove $n$ slices with the smallest $M(A^i_j)$ and
retrain the hybrid model to maintain the desired accuracy. Tolerable accuracy $ACC_{\rm tol}$ is also governed by the predefined threshold $\delta$,
which is the same as the description in \S~\ref{subsec:dense_model_generation}.
\A~\ref{alg:pruning_pipeline_llm} depicts the pruning pipeline for LLMs. Similar
to \A~\ref{alg:pruning_pipeline}, the algorithm requires two hyperparameters:
the number of the pruned slices in each pruning step $n$ and the number of
training rounds $rounds$. The difference compared to
\A~\ref{alg:pruning_pipeline} is that the pruning step is guided by the
magnitude $Mag(A^i_j)$.}

\begin{algorithm}
  \footnotesize
    \SetAlgoLined
    \SetKwProg{Fn}{Function}{:}{} \SetKwFunction{FIterativePrune}{Iterative
    Slice Pruning for LLM} \KwIn{The densely sliced LLM model $M_{\rm dense}$, pre-defined
    parameters $n$, and $rounds$} \KwOut{The hybrid model
    $M_{\rm hyb}$}

    \For{$r\leftarrow 1$ \KwTo $rounds$}{
      Compute the accuracy $ACC_r$ of $M_r$ \;
      \If{$ACC_r > ACC_{\rm tol}$}{
          Store the model $M_{\rm hyb} = M_r$ \;
          Select $n$ slices with smallest $Mag(A_{i}^{i+1})$ \;
          Prune the selected slices $A_i^{i+1}$ \;
      }
      Re-train $M_r$ to get $M_{r+1}$  \;
    }
    \KwRet The hybrid model $M_{\rm hyb}$
    
    \caption{\diff{Iterative Slice Pruning for LLM.}}\label{alg:pruning_pipeline_llm}
  \end{algorithm}

\subsection{Outsource Computation for Attention}

Although TEESlice can apply to LLM to generate privacy-related LoRA slices and reduce the amount of computation of slices. Applying TSDP to LLM faces another problem: \textbf{How to outsource the attention computation securely?}. In this section, we propose a dynamic attention substitution solution to securely the attention computation process based on the One-Time-Pad. Our solution dynamically replaces the hard-to-protect attention modules with easy-to-protect linear attention modules. We can use One-Time-Pad to efficiently protect the linear attention modules and outsource their computation to GPUs.

\parh{Limitation of One-Time-Pad.}
In \E~\ref{equ:one_time_pad}, we analyze the feature encryption technique based on One-Time-Pad. One key reason why \E~\ref{equ:one_time_pad} holds is that the linear layer $g(\cdot)$ has the distributive property. The computation process of the one-time-pad vector $\mathbf{r}$ does not rely on the input feature $\mathbf{h}$, thus it can be precomputed in the offline phase. Assume that $g(\mathbf{h}) = W \cdot \mathbf{h}$, the distributive property means: 
\begin{equation}
    g(\mathbf{h} + \mathbf{r}) = W \cdot (\mathbf{h} + \mathbf{r}) = W \cdot \mathbf{h} + W \cdot \mathbf{r} = g(\mathbf{h}) + g(\mathbf{r}).
\end{equation}
Note that $g(\mathbf{r}) = W \cdot \mathbf{r}$. Both $W$ and $\mathbf{r}$ are not related to the input feature $\mathbf{h}$, thus {the computation process is independent from the input feature.}

\parh{Difficulty to Protect Attention Modules.}
However, the matrix multiplication in the attention module does not have this property. The goal of TSDP is to outsource the computation-intensive matrix multiplication to low-end GPUs. The attention module contains the inter-feature matrix multiplication, which should also be outsourced to GPU and differs from the typical weight-feature matrix multiplication in CNN. However, the inter-feature multiplication can not be protected by a One-Time-Pad, unlike the weight-feature multiplication. It is because, after splitting the attention equation via the distributive property, \textbf{all the operands rely on the input feature}, thus we can not precompute any results in the offline phase because we don't know the value of the input feature in the offline phase. Take the attention score of \E~\ref{equ:attention_score} as the example. For the $x_q \cdot x_k$ operation, after we use One-Time-Pad to protect, the outsourced operation becomes
\begin{equation}
    ( x_q + r_q ) \cdot ( x_k + r_k).
\end{equation}
Following the distributive property, this equation can be decomposed as
\begin{equation}
    ( x_q + r_q ) \cdot ( x_k + r_k) = x_q \cdot x_k + x_q \cdot r_k + r_q \cdot x_k + r_q \cdot r_k.
\end{equation}
In this equation, only $r_q \cdot r_k$ can be precomputed. To decrypt $x_q \cdot x_k$, we also need to compute $x_q \cdot r_k$ and $x_k \cdot r_q$. The computation overhead of these two factors is the same as the overhead of computing the original operation $x_q + r_q$. So, the One-Time-Pad encryption technique can not protect the inter-feature matrix multiplication in the attention module, and the TSDP existing solution can not protect the attention module.

\parh{Our insight: Linear Attention.}
To address the challenge of protecting attention modules, we propose to utilize linear attention to substitute the inter-feature matrix multiplication~\cite{shen2021efficient,katharopoulos2020transformers}. Given the key $x_k$, value $x_v$, and query $x_q$, the linear attention can be represented as
\begin{equation}
\begin{aligned}
    Attention\_Weight = softmax( W \cdot Concat(x_q, x_k) ), \\
    Linear\_Attention(x_q, x_k, x_v) = W \cdot (Attention\_Weight, x_v).
\end{aligned}
\end{equation}
The advantage of linear attention is two-fold. First, the linear attention does not have inter-feature multiplication. Instead, it uses weight-feature multiplication, which can be efficiently protected by One-Time-Pad. Second, the matrix multiplication can build the long-distance relationship between tokens and capture the important tokens in the input sentence.

\parh{Dynamic Adjustment}
One disadvantage of linear attention is that it performs worse than inter-feature attention in \E~\ref{equ:attention_score}. Directly replacing all the attention with linear attention will harm the model performance by a large margin. To mitigate the performance degradation, we introduce a trainable scalar factor $\beta$ to dynamically control how many inter-feature attention modules should be replaced by the linear attention modules. During the training phase, the attention output is interpolated by $\beta$:
\begin{equation}
    Dynamic\_Attention(x_q, x_k, x_v) = \beta \cdot Attention(x_q, x_k, x_v) + (1-\beta) \cdot Linear\_Attention(x_q, x_k, x_v)
\end{equation}
The scalar factor is added to the loss function as a regulation term to encourage the model to use more linear attention modules. Let $\beta_l$ represent the $\beta$ for the $l$-th layer. The training loss can be formulated as:
\begin{equation}
    loss = CE\_loss + \sum_l \beta_l,
\end{equation}
where $CE\_loss$ is the cross-entropy loss to train the whole model.
After training, the attention modules with small $\beta$ are replaced with linear attention modules. These linear attention modules can be efficiently outsourced to GPU using One-Time-Pad.

\section{Experiments}
\label{sec:teeslice_experiment}

We implement \sys with PyTorch 1.7 and we select ResNet18 as the public backbone
following \textsc{NetTailor}. We have the flexibility to choose the backbone as
commonly-used models without affecting the security
guarantee~\cite{morgado2019nettailor}. We select ResNet18 because it is the
default setting of \textsc{NetTailor}. As a fair setting, we set the
training time for $M_{\rm dense}$ and $M_{\rm sparse}$ to half the time required
to train $M_{\rm vic}$, respectively. Hence, the overall training time for
$M_{\rm sparse}$ and $M_{\rm vic}$ are equivalent. We apply \sys to all the
datasets and victim models in \S~\ref{subsec:evaluation-setup}. The experiments
aim to answer the following RQs:

\begin{tcolorbox}[size=small]
    \textbf{RQ3:} How does \sys compare with representative defenses w.r.t. security and utility? 

    \textbf{RQ4:} Does \sys sacrifice the accuracy of the original model?

    \textbf{RQ5:} What is performance of \tool on real-world devices? How much can
    \tool speed up compared to the \shieldwhole baseline?

    \textbf{RQ6:} How is \tool's performance on LLMs?

    \textbf{RQ7:} How is \tool's scalability to NLP tasks.

\end{tcolorbox}

\subsection{Security Guarantee and Utility Cost}
\label{sec:experiment:security_utility}

\parh{Security Guarantee.}~We follow the same experiment protocol in
Section~\ref{sec:evaluate_existing_solutions} to compare \sys with five representative \tsdp\ schemes. Specifically, for \sys, we assume the attacker
knows the architecture of $M_{\rm hyb}$ (default assumption), including which
public backbone it uses and the structure of privacy-related model slices. The
attack pipeline is the same as in Section~\ref{sec:evaluate_existing_solutions}.

\diff{\T~\ref{tbl:evaluate_solution_MS} and \T~\ref{tbl:evaluate_solution_MI}} reports the results, with attack accuracy against
our approach marked in \colorbox{best}{green}. The results are highly promising:
in all cases, the attack accuracies are comparable with black-box protection and
are better than the best of existing defenses (marked with
\colorbox{low}{yellow}). For \ac{ms}, the relative accuracy of \sys compared to
the black-box baseline is 1.32$\times$, while the relative value of the best
xdefense, Magnitude (\ding{174}), is 3.85$\times$. 
For \ac{mia}, the attack accuracy of \sys is similar to the black-box baseline
(random guess). It is because all the feature communications are encrypted, and
the TEE shields all the privacy-related slices. 

\parh{Security Under Other Assumptions of $M_{\rm sur}$.}
\label{sec:experiment:other_assumption}
This section evaluates the security guarantee of \sys with two different
assumptions. The first assumption, \textit{backbone-only}, is a weaker one that
assumes the attacker only knows the public backbone of \sys and does not know the
slice information (slice positions and structures). The second assumption,
\textit{victim-knowing}, is a stronger assumption that assumes the attacker
knows the structure of the original $M_{vic}$. Note that the victim-knowing is
\textit{not} a realistic assumption, and we only evaluate it to show the
performance of \sys under different settings. We also follow the evaluate
protocol and attack pipeline in \S~\ref{sec:evaluate_existing_solutions}
for a fair experiment.

We show the \ac{ms} accuracies of the two additional assumptions with the
default assumption (knowing the structure of $M_{\rm hyb}$) in
\T~\ref{tbl:other_assumption}. 
\ifentireresult
Note we omit \ac{mia} because the additional assumptions do not introduce new
information of $M_{\rm vic}$'s training data. Thus the results of \ac{mia} are
the same as the default assumption. For each model and dataset
\T~\ref{tbl:other_assumption}, we mark the highest accuracy with
\colorbox{high}{red} and the lowest accuracy with \colorbox{best}{green}. \diff{From
\T~\ref{tbl:other_assumption}, we can see that for AlexNet, ResNet18, VGG16\_BN
and VGG19\_BN,} victim-knowing has lower accuracies than the other two
assumptions (\diff{eight green cells and two red cells}). The default assumption
x(Hybrid $M_{\rm hyb}$) and backbone-only perform similarly (has six and eight
red cells, respectively). We suspect that the reason for victim-knowing's low
accuracy is that the $M_{\rm vic}$ of the four model has a smaller model
capacity than the backbone. In \T~\ref{tbl:other_assumption}, $M_{\rm vic}$ with
larger capacity (ResNet34) has the highest \ac{ms} accuracy for all datasets.

\begin{table}[htp]
    \caption{\diff{Comparison of model stealing accuracy between different attack assumptions of $M_{\rm sur}$.}}
    \vspace{-5pt}
    \label{tbl:other_assumption}
    \centering
    \begin{adjustbox}{max width=0.5\linewidth}
    \begin{tabular}{ccccc}
    \hline
    \multicolumn{1}{l}{}       & \multicolumn{1}{l}{} & Hybrid $M_{\rm hyb}$  & Backbone & Victim $M_{\rm vic}$ \\ \hline
    \multirow{4}{*}{AlexNet}   & CIFAR10                  & \cellcolor{best}19.04\% & 19.56\%  & \cellcolor{high}23.71\% \\
                               & CIFAR100                 & \cellcolor{best}8.27\%  & \cellcolor{high}14.48\%  & 11.9\%  \\
                               & STL10                  & 24.15\% & \cellcolor{high}32.75\%  & \cellcolor{best}17.14\% \\
                               & UTKFace                  & \cellcolor{high}52.27\% & 51.32\%  & \cellcolor{best}47.0\%  \\ \hline
    \multirow{4}{*}{ResNet18}  & CIFAR10                  & \cellcolor{high}31.4\%  & 25.63\%  & \cellcolor{best}17.33\% \\
                               & CIFAR100                 & 10.9\%  & \cellcolor{high}18.33\%  & \cellcolor{best}7.78\%  \\
                               & STL10                  & \cellcolor{best}29.19\% & 32.77\%  & \cellcolor{high}32.86\% \\
                               & UTKFace                  & \cellcolor{high}51.95\% & \cellcolor{best}50.86\%  & 51.63\% \\ \hline
    \multirow{4}{*}{VGG16\_BN} & CIFAR10                  & \cellcolor{high}30.87\% & 25.65\%  & \cellcolor{best}20.69\% \\
                               & CIFAR100                 & 9.78\%  & \cellcolor{high}18.44\%  & \cellcolor{best}6.38\%  \\
                               & STL10                  & \cellcolor{high}32.92\% & 32.51\%  & 3\cellcolor{best}1.75\% \\
                               & UTKFace                  & \cellcolor{best}48.37\% & \cellcolor{high}52.54\%  & 51.04\% \\ \hline
    \multirow{4}{*}{ResNet34}  & CIFAR10                  & 26.43\% & \cellcolor{best}24.01\%  & \cellcolor{high}26.45\% \\
                               & CIFAR100                 & \cellcolor{best}10.67\% & 16.46\%  & \cellcolor{high}20.54\% \\
                               & STL10                  & 34.19\% & \cellcolor{best}32.71\%  & \cellcolor{high}42.02\% \\
                               & UTKFace                  & \cellcolor{best}48.5\%  & 51.63\%  & \cellcolor{high}52.18\% \\ \hline
    \multirow{4}{*}{VGG19\_BN} & CIFAR10                  & 24.08\% & \cellcolor{high}25.73\%  & \cellcolor{best}22.09\% \\
                               & CIFAR100                 & \cellcolor{best}11.47\% & \cellcolor{high}18.19\%  & 14.69\% \\
                               & STL10                  & \cellcolor{high}36.11\% & \cellcolor{best}32.74\%  & 37.67\% \\
                               & UTKFace                  & \cellcolor{best}47.09\% & \cellcolor{high}52.72\%  & 51.54\% \\ \hline
    \end{tabular}

\end{adjustbox}
\vspace{-5pt}

\end{table}

\fi

\parh{Security Under Other Assumptions of Data.}
In \S~\ref{sec:evaluate_existing_solutions} and \S~\ref{sec:dilemma}, we study a
realistic adversary that has a small amount of data. Although our assumption on
the adversary is
realistic~\cite{rakin2022deepsteal,hua2018reverse,yan2020cache}, we still study
the security of \sys with an ideal adversary who has a large amount of data to
verify if \sys ensures the security of DNN models under extreme conditions.
We compare \ac{ms} accuracies on various $M_{\rm sur}$ and a large range
of queried data size between our approach and black-box protection. The goal is
to study if our approach increases \ac{ms} accuracy under the new assumption.
The $M_{\rm sur}$ includes $M_{\rm hyb}$, the backbone (\ie, ResNet18;
backbone-only), and all the $M_{\rm vic}$ in \S~\ref{subsec:evaluation-setup}
(victim-knowing). Following prior work~\cite{orekondy2019knockoff}, we set the
queried data sizes as \{50, 100, 300, 500, 1K, 3K, 5K, 10K, 15K, 20K, 25K,
30K\}. 
\ifentireresult

\begin{figure*}[ht]
    \centering
    \includegraphics[width=\linewidth]{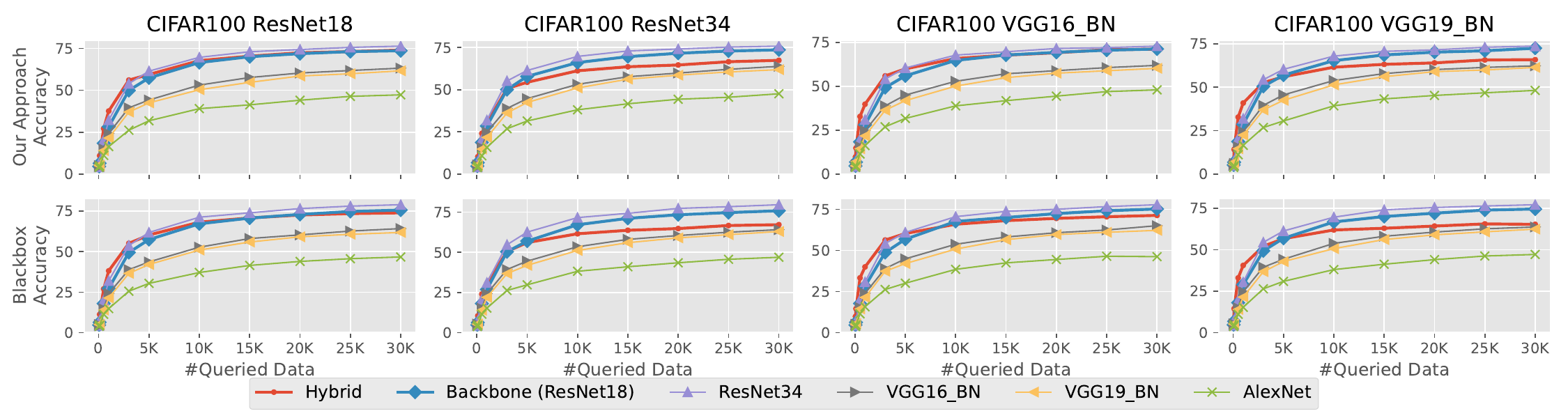}
    \caption{\diff{Comparison of \sys and the black-box protection against \ac{ms}
    attacks with different sizes of queried data. We report the accuracy of
    $M_{\rm sur}$, where the first row represents \sys and the second row is for the black-box baseline. The \ac{ms} attack performance is indistinguishable between TEESlice (the upper row) and black-box protection (the lower row).}}
    \label{fig:multi_arch_cifar100_accuracy}
    \end{figure*} 

\fi

\ifentireresult
We display \ac{ms} accuracy on CIFAR100 and \diff{four models (ResNet18, ResNet34, VGG16\_BN and VGG19\_BN)} in \F~\ref{fig:multi_arch_cifar100_accuracy}. \diff{Each column of \F~\ref{fig:multi_arch_cifar100_accuracy} means $M_{\rm vic}$ is respectively a ResNet18, ResNet34, VGG16\_BN and VGG19\_BN model.} The first row of \F~\ref{fig:multi_arch_cifar100_accuracy} displays the results of \tool, and the second row shows the results of the black-box defense. 

We can observe {from \F~\ref{fig:multi_arch_cifar100_accuracy} }that,
for all cases, the \ac{ms} accuracy against \tool is similar to that of
black-box baseline. According to the Wilcoxon signed-rank
test~\cite{wilcoxon1992individual}, the null hypothesis is that there is no
difference in accuracy. The p-value is $0.81$, which cannot reject the null
hypothesis. This result indicates that the differences between the accuracy of
\sys and the black-box baseline have no statistical significance. To summarize,
under a different assumption of more queried data, the \ac{ms} accuracy has no
difference between \tool and the black-box baseline. 
    
\fi

\parh{Utility Cost.}~We qualitatively compare \tool\ with the security-utility
curves of other defenses in {\F~\ref{fig:acc_flops_append_one_fig} and \F~\ref{fig:mia_flops_append_one_fig}}. For both \ac{ms}
and \ac{mia}, \sys achieves a similar level of black-box defense
with a distinguishably smaller $\% FLOPs$ than other defenses.
\textcolor{red}{\ding{72}}, denoting \sys, locates at the bottom left corners for
all cases in {\F~\ref{fig:acc_flops_append_one_fig} and \F~\ref{fig:mia_flops_append_one_fig}}. We also quantitatively compare
the $Utility(C^*)$ of \sys with other defenses in \T~\ref{tbl:optimal_config}.
For each case (column), we mark the lowest value of $Utility(C^*)$ with
\colorbox{best}{green}. \sys achieves the lowest utility cost in seventeen out of 20
cases. The average utility cost of \sys is 3.44\%. On the contrary, the average
utility cost of other defenses ranges {from 45.98\% to 97.02\%}. That is, \sys
takes $10\times$ less utility cost to achieve the highest (black-box) defense
level.

\diff{The average utility cost of \sys is 2.01\% for ResNet34 and 0.34\% for VGG19\_BN. Compared to 4.38\% and 0.45\% for ResNet18 and VGG16\_BN, the average utilities of deeper models of the same architecture are lower. This can be due to the Slice Pruning phase of \sys. With more layers, deeper models need smaller percentages of slices to maintain the accuracies on private datasets. }

\begin{tcolorbox}[size=small]
\textbf{Answer to RQ3}: \sys\ features promising, black-box-level security
guarantees under different attack assumptions. The utility cost ($Utility(C^*)$)
of \sys is $10\times$ less than other \tsdp\ solutions. 
\end{tcolorbox}
\vspace{-5pt}

\subsection{Accuracy Loss}
\label{sec:experiment:trade-off}
To answer this research question, we compare the accuracy between $M_{\rm vic}$
and their derived hybrid models $M_{\rm hyb}$ trained by \sys. The result is in
\T~\ref{tbl:nettailor_accuracy}. In general, \sys does not lead to a
considerable loss of accuracy. For AlexNet, \sys achieves a higher accuracy
because the model capacity of the backbone (i.e., ResNet18) is larger than
AlexNet. This phenomenon is consistent with the finding in
\S~\ref{sec:experiment:security_utility} that AlexNet (the lowest model
complexity) always has low accuracy. \diff{For ResNet34, the general accuracy is similar to ResNet18. For VGG19\_BN, it is similar to VGG16\_BN. This is because of the same architectures of the victim models. Despite the varying architecture and sizes, the accuracy loss is negligible.}
To statistically understand the accuracy loss, we compute the statistical
significance using the Wilcoxon signed-rank test~\cite{wilcoxon1992individual}
across all models except AlexNet. The null hypothesis is that the accuracies
between $M_{\rm vic}$ and $M_{\rm hyb}$ have no difference. The p-value is 0.75
and provides little statistical significance to reject the null hypothesis.
Thus, the pruning processes have little effect on the accuracy of $M_{\rm hyb}$.

We also measure the accuracy loss of eNNclave~\cite{schlogl2020ennclave}, a
recent work sharing similar concepts, by putting $M_{\rm vic}$'s last layer into
TEE while replacing the GPU-offloaded shallow layers with a public backbone. As
clarified in \S~\ref{sec:literature_category}, eNNcalve suffers from low
accuracy. We evaluate the accuracies of eNNclave over all models and datasets
and find that eNNclave has an average downgrade of 34\%, higher than
(about $10\times$) our approach.

\begin{table}[]

\caption{\diff{The accuracy comparison between the victim model and the hybrid model
trained by \tool in the form of $M_{\rm vic}$/$M_{\rm hyb}$.
Except for AlexNet where \sys has a higher accuracy due to a larger
model capacity, by average, \sys's relative accuracy loss (the ratio between
the accuracy of $M_{\rm hyb}$ and the accuracy of $M_{\rm vic}$) is 0.34\%.}}
\label{tbl:nettailor_accuracy}
\vspace{-5pt}
\setlength{\tabcolsep}{3.0pt}
\begin{adjustbox}{max width=0.6\linewidth}
    \begin{tabular}{@{}ccccc@{}}
    \toprule
              & CIFAR10         & CIFAR100        & STL10           & UTKFace         \\ \midrule
    AlexNet   & 83.71\%/86.37\% & 56.46\%/61.96\% & 76.54\%/80.17\% & 89.42\%/88.92\% \\
    ResNet18  & 95.47\%/93.65\% & 79.94\%/76.79\% & 87.51\%/86.22\% & 86.97\%/88.24\% \\
    ResNet34  & 91.11\%/91.75\% & 81.00\%/76.53\% & 88.22\%/86.15\% & 87.69\%/89.55\% \\
    VGG16\_BN & 91.62\%/93.06\% & 73.03\%/73.11\% & 89.67\%/89.42\% & 89.19\%/89.46\% \\
    VGG19\_BN & 92.48\%/92.70\% & 71.38\%/73.15\% & 89.62\%/90.70\% & 89.96\%/89.46\% \\ 
    \bottomrule
    \end{tabular}
\end{adjustbox}

\end{table}

\begin{tcolorbox}[size=small]
\textbf{Answer to RQ4}: Besides achieving a principled security guarantee, \sys
doesn't undermine model accuracy.

\end{tcolorbox}
\vspace{-5pt}

\subsection{Performance on Real-World Devices}

We created a prototype framework on a Desktop PC with Intel Core i7-8700 3.20GHz
CPU and NVIDIA GeForce GTX 1080 GPU to evaluate \tool's speed-up on real
devices. The framework has two parts: SGX's shielded section and the GPU's
offloaded part. The SGX component is developed in C++ and is compiled using
Intel SGX SDK 2.6 and GCC 7.5. The GPU component is built in PyTorch
1.7 and is supported by CUDA 11.7. We reused code from
Goten~\cite{lucien2021goten} and Slalom~\cite{tramer2019slalom} and implemented
other \tool's operations, such as convolution, the OTP-based feature encryption,
and hybrid model architecture. We emulate production conditions by switching SGX
to hardware mode with all its protection. We ran all experiments ten times and
got the average inference time. We verify that the running time deviates less
than 10\% from the average. We mainly report the throughput (images per second)
as it is more straightforward to evaluate the speed of ML
systems~\cite{tramer2019slalom}. The lowest required throughput for a real-time
on-device ML service is 30 (a latency of 33ms)~\cite{bateni2020neuos}. The
throughput is computed by $1000/average\_latency$.

\begin{table}[]
\caption{The throughput comparison between \shieldwhole (in the TEE), no-shield, and \tool on
a real desktop with SGX and GPU. For the no-shield baseline, we report both the performance on the CPU and GPU to give a fair comparison. We switch SGX to the hardware mode to enable
all protections. In parentheses, we present the speedup w.r.t. the \shieldwhole
baseline.}
\label{tbl:realdevice}
\centering
\small
\begin{adjustbox}{max width=\linewidth}

\begin{tabular}{@{}ccccccc@{}}
\toprule
                              &             & AlexNet         & ResNet18        & ResNet34        & VGG16\_BN        & VGG19\_BN       \\ \midrule
\multicolumn{2}{c}{Black-box (TEE)}         & 6.56            & 7.67            & 3.74            & 1.55            & 1.27           \\
\multicolumn{2}{c}{No-Shield (CPU)}         & 60.97 (9.30$\times$)   & 79.54 (10.37$\times$)  & 45.52 (12.16$\times$)  & 11.22 (7.22$\times$)   & 9.42 (7.40$\times$)   \\
\multicolumn{2}{c}{No-Shield (GPU)}         & 495.27 (75.53$\times$) & 288.56 (36.56$\times$) & 101.51 (29.25$\times$) & 103.10 (66.42$\times$) & 83.59 (65.66$\times$) \\ \midrule
\multirow{4}{*}{TEESlice-CPU} & CIFAR10     & 36.40 (5.55$\times$)   & 48.70 (6.35$\times$)   & 48.67 (13.00$\times$)  & 52.90 (34.08$\times$)  & 53.22 (41.80$\times$) \\
                              & CIFAR100    & 38.05 (5.80$\times$)   & 37.58 (4.90$\times$)   & 37.83 (10.10$\times$)  & 44.33 (28.56$\times$)  & 41.41 (32.53$\times$) \\
                              & STL10       & 58.12 (8.86$\times$)   & 48.50 (6.32$\times$)   & 49.28 (13.16$\times$)  & 53.42 (34.41$\times$)  & 45.93 (36.08$\times$) \\
                              & UTKFaceRace & 35.25 (5.38$\times$)   & 39.14 (5.10$\times$)   & 46.76 (12.49$\times$)  & 35.39 (22.80$\times$)  & 44.65 (35.07$\times$) \\ \midrule
\multirow{4}{*}{TEESlice-GPU} & CIFAR10     & 44.67 (6.78$\times$)   & 63.81 (8.32$\times$)   & 58.22 (15.55$\times$)   & 72.80 (46.90$\times$)  & 71.19 (55.92$\times$) \\
                              & CIFAR100    & 47.36 (7.22$\times$)   & 46.63 (6.08$\times$)   & 40.42 (10.79$\times$)  & 58.69 (37.81$\times$)  & 51.99 (40.83$\times$) \\
                              & STL10       & 85.79 (13.08$\times$)  & 65.24 (8.50$\times$)   & 58.10 (15.52$\times$)  & 71.35 (45.97$\times$)  & 59.20 (46.50$\times$) \\
                              & UTKFaceRace & 41.29 (6.30$\times$)   & 58.03 (6.26$\times$)   & 61.21 (16.35$\times$)  & 42.34 (27.28$\times$)  & 56.69 (44.53$\times$) \\ \bottomrule
\end{tabular}

\end{adjustbox}

\end{table}

\T~\ref{tbl:realdevice} presents the throughput of \tool on the five models, as well as two baselines \shieldwhole (in the SGX) and no-shield. For a fair comparison, we run the no-shield twice. One runs the whole model on the CPU (outside of SGX), and the other is on the GPU. We also run two variants of \tool to show the source of improvement. One variant of \tool is \tool-CPU, in which we run the \tool-d model only on the CPU. The backbone is run on the CPU, and the private slices are shielded by the SGX. The other variant is \tool-GPU, where we offload the backbone to the GPU. Shielding-whole-model is the throughput lower bound, and no-shield is the upper bound. 
For \shieldwhole, the
throughputs on the three models range from $1.55$ to $7.67$, far from the
required throughput of real-time service ($30$). The throughputs are improved by a large margin for the two no-shield variants. The throughput of the no-shield (GPU) baseline ranges from $101.51$ to $495.27$, much faster than the real-time requirement. For no-shield (CPU), the throughput is up to $79.54\times$ with an average of $40.83\times$. Although the computation units between TEE and CPU are the same, the CPU is in a much richer environment with no constrain on the available memory and CPU cores. Thus, the performance of no-shield (CPU) is better than that of black-box (TEE).
For \tool-CPU, we can observe that the partition mechanism speeds up the inference because the backbone can be run in a rich environment. The speed is accelerated by an average of $18.11\times$. For \tool-GPU, the benefits are even larger because of the strong computation ability of GPUs. The inference speed is improved by $23.32\times$ average. Someone may observe that the improvement from \tool-GPU to \tool-CPU is less than the comparison between no-shield (GPU) and no-shield (CPU). The throughput of no-shield (GPU) is averagely $6.62\times$ than that of no-shield (CPU). We think it is because offloading the computation to the GPU introduces more communication than offloading to the CPU. To transfer the computation matrix to GPU, \tool-GPU needs to use the PCIe bus to communicate with external devices. On the contrary \tool-CPU only needs to operate on the memory and do not need to communicate with external devices.

\diff{For ResNet34, \sys achieves 12.19$\times$ and 14.55$\times$ average speedup in TEESlice-CPU and TEESlice-GPU variants. They are larger than those for ResNet18 because of the larger model sizes. We also observe the same phenomenon when comparing VGG19\_BN with VGG16\_BN.}

To further analyze the performance of \tool, we also logged the latency of
different parts during the inference phase. We break down the inference latency
of \tool into four parts: \texttt{Data Transfer}, \texttt{Slice in TEE},
\texttt{Backbone on GPU}, and \texttt{Non-Linear in TEE}. \texttt{Data Transfer}
is the time to transfer internal results between SGX and GPU. \texttt{Slice in
TEE} is the time to compute the private slices inside SGX. \texttt{Backbone on
GPU} is the time to compute the convolution layers of the backbone on the GPU.
\texttt{Non-Linear in TEE} is the time to compute the non-linear layers (\eg
ReLU) inside SGX (recall \S~\ref{sec:teeslice_detail} that the ReLU layers of
the backbone are computed inside TEE). \T~\ref{tbl:realdevice_breakdown} displays
the percentage of each part over the total inference latency. From the table, we
can see that \texttt{Slice in TEE} occupies 40.49\% of the inference time due to
the constrained computation resources inside SGX. \texttt{Data Transfer} and
\texttt{Non-Linear in TEE} occupy 35.61\% and 20.96\% of the inference time
because all the non-linear layers of the backbone are computed inside SGX.
\texttt{Backbone on GPU} only occupies 2.84\% of the time due to the strong
computation ability of the GPU. Note that although \tool introduces the additional
overhead of \texttt{Data Transfer}, \tool still accelerates the overall
inference time by a large margin.

\begin{table}[]
    \caption{\tool inference time breakdown.}
    \label{tbl:realdevice_breakdown}
    \vspace{-5pt}
    \centering
    \small
    \begin{adjustbox}{max width=0.6\linewidth}

    \begin{tabular}{@{}cccc@{}}
    \toprule
    \texttt{Data Transfer} & \texttt{Slice in TEE} & \texttt{Backbone on GPU} & \texttt{Non-Linear in TEE} \\ \midrule
    35.61\%       & 40.49\%      & 2.84\%          & 20.96\%           \\ \bottomrule
    \end{tabular}

\end{adjustbox}

\end{table}

\minor{We also record the memory consumption of \tool in the SGX, and the results are displayed in \T~\ref{tbl:memory}. Because the generated slices are much smaller than the original private model and the public model backbone, the SGX memory consumption is reduced by a large margin. The most enormous memory consumption is 63.79 MB, which is much smaller than the limitation of SGX memory (about 93MB). The deeper analysis of the SGX memory consumption demonstrates that \tool can effectively reduce the utility cost of \tsdp by reducing the memory usage and computation cost in TEE.}

\begin{table}[]
\caption{Memory consumption in the SGX of on different models and datasets.}
\label{tbl:memory}
\begin{tabular}{@{}cccccc@{}}
\toprule
            & AlexNet & ResNet18 & ResNet34 & VGG16\_BN & VGG19\_BN \\ \midrule
CIFAR10     & 63.79MB & 41.64MB  & 41.64MB  & 40.31MB   & 40.31MB   \\
CIFAR100    & 56.92MB & 56.92MB  & 56.92MB  & 46.01MB   & 45.02MB   \\
STL10       & 43.31MB & 41.64MB  & 44.30MB  & 55.21MB   & 44.98MB   \\
UTKFaceRace & 39.91MB & 31.92MB  & 46.72MB  & 35.26MB   & 34.17MB   \\ \bottomrule
\end{tabular}
\end{table}

\begin{tcolorbox}[size=small]
\textbf{Answer to RQ5}: \tool accelerates the throughput by an average of
$18.37\times$ compared with the \shieldwhole baseline and satisfies the
real-time requirement. 
\end{tcolorbox}
\vspace{-5pt}

\subsection{Scalability to Large Language Models (LLMs)}
\label{subsec:scalability_to_llm}
\diff{
In this part, we evaluate the scalability of \tool to large language models. We will first introduce the evaluation protocol and then present the results.}

\subsubsection{Evaluation Protocol}
\diff{In \S~\ref{sec:design_for_llm} we describe the design of \tool for LLMs. By iteratively pruning the \lora layers and retraining the $M_{hyb}$, we can obtain a hybrid model with the desired accuracy. To demonstrate the effectiveness of the design, we evaluate the algorithm in Vision Transformer (ViT)~\cite{vit} with different image classification data sets. We compare the accuracy of the hybrid model with the original model. Furthermore, we compare the security cost between \tool and \shieldwhole. 
The security cost of the private slices is represented as the \%FLOPs of the remaining \lora layers. The complexity of the entire model is represented as the \%FLOPs of the entire model.}

\parh{Datasets.} \diff{We use three datasets that are widely used to evaluate vision transformer models: CIFAR10, CIFAR100~\cite{krizhevsky2009learning} and STL10~\cite{coates2011an}. }

\parh{Models.} \diff{Following the common practice in the AI community, we use the pretrained ViT B\_16, ViT B\_32, ViT L\_16 and ViT L\_32 as the backbones to show the scalability to LLMs. All the models are pretrained on ImageNet-21k and finetuned on ImageNet-1k\cite{pretrained-vit}. For ViT-Base, \ie ViT B\_16 and ViT B\_32, the dimension of $W_q$ and $W_v$ in the self-attention module is $(768 \times 768)$ (\ie $d_{\rm model} = 768$). The number of blocks is 12, so the number of initial \lora layers (\ie private slices) is 24. For ViT-Large, \ie ViT L\_16 and ViT L\_32, $d_{\rm model} = 1024$ , the number of blocks is 24 and the number of initial \lora layers is 48. The reduced dimension of \lora layer $r$ we chose is 4.}

\parh{Metrics} \diff{We report two metrics to compare the $M_{\rm vic}$ and $M_{\rm hyb}$: the accuracy and security cost. Following \S~\ref{sec:evaluation_dilemma_results}, we use \%FLOPs that are shielded by TEE to represent the security cost. The accuracy comparison shows how much performance \tool sacrifices. The security cost shows how much \tool can reduce the computation required in the slow TEE. For $M_{\rm vic}$, \%FLOPs equals 100\% because, as indicated \S~\ref{sec:empirical_evaluation_results}, without \tool, only placing the entire model in TEE can achieve the security of the upper bound. For $M_{\rm hyb}$, \%FLOPs is the ratio of private slices because only the slices contain information related to private. }

\subsubsection{Results}

\begin{table}[htp]
    \caption{\diff{The comparison of accuracy and security cost between $M_{\rm vic}$ and $M_{\rm hyb}$. In each cell, we present the results in the form of $M_{\rm vic}$/$M_{\rm hyb}$ to give a better comparison.}}
    
    \label{tbl:llm_accuracy_complexity}
    \vspace{-5pt}
    \centering
    \small
    \begin{adjustbox}{max width=0.85\linewidth}

            \begin{tabular}{ccccc}
\hline
\multicolumn{2}{c}{}                                           & CIFAR10          & CIFAR100         & STL10            \\ \hline
\multicolumn{1}{c}{\multirow{4}{*}{Accuracy}}      & ViT B\_16 & 97.69\%/97.85\%  & 89.03\%/89.40\%   & 99.35\%/99.35\%  \\ \cline{2-5} 
\multicolumn{1}{c}{}                               & ViT B\_32 & 98.30\%/98.39\%  & 90.31\%/90.72\%  & 99.26\%/99.15\%  \\ \cline{2-5} 
\multicolumn{1}{c}{}                               & ViT L\_16 & 98.64\%/98.92\%  & 92.16\%/92.83\%  & 99.67\%/99.70\%  \\ \cline{2-5} 
\multicolumn{1}{c}{}                               & ViT L\_32 & 98.73\%/98.85\%  & 91.50\%/92.28\%  & 99.57\%/99.58\%  \\ \hline
\multicolumn{1}{c}{\multirow{4}{*}{Security Cost}} & ViT B\_16 & 0.084\%/100.00\% & 0.096\%/100.00\% & 0.077\%/100.00\% \\ \cline{2-5} 
\multicolumn{1}{c}{}                               & ViT B\_32 & 0.105\%/100.00\% & 0.112\%/100.00\% & 0.105\%/100.00\% \\ \cline{2-5} 
\multicolumn{1}{c}{}                               & ViT L\_16 & 0.072\%/100.00\% & 0.102\%100.00\%  & 0.072\%/100.00\% \\ \cline{2-5} 
\multicolumn{1}{c}{}                               & ViT L\_32 & 0.082\%/100.00\% & 0.090\%/100.00\% & 0.077\%/100.00\% \\ \hline
\end{tabular}

\end{adjustbox}
\vspace{-5pt}

\end{table}

\diff{We display the results in \T~\ref{tbl:llm_accuracy_complexity}, and accuracy and security cost are displayed in two rows. In general, \tool does not introduce a significant loss of accuracy. For STL10, \tool achieves a comparable accuracy with the original model. In fact, we observe that $M_{hyb}$ achieves a slightly higher accuracy for ViT B\_32 and during the initial pruning rounds for other models. This can be due to that the pruning procedure reduces the model complexity and alleviates the overfitting effect. For CIFAR10 and CIFAR100, the accuracy of $M_{hyb}$ is slightly lower than the victim model. The relative accuracy loss (the ratio between the accuracy of $M_{\rm hyb}$ and the accuracy of $M_{\rm vic}$) is only 0. 235\% among all models and datasets. Therefore, for ViT architecture, \tool does not introduce a noticeable accuracy loss.
For security cost, we can observe that the \%FLOPs of private slices are very small compared to shielding the entire model ($M_{vic}$). The slight higher cost for CIFAR10 dataset than other two datasets can be because of its relative complexity. The average security cost of \tool is only 0.089\% among all models and datasets. It shows that \tool can reduce the amount of computation for large language models by a large margin. The performance gain is even larger than traditional DNN.}

\begin{tcolorbox}[size=small]
\textbf{Answer to RQ6}: \tool can protect large language models such as vision transformers.
\end{tcolorbox}

\subsection{Scalability to NLP Tasks}

The design of \tool is applicable to protect various DNN models. Thus, findings
on protecting computer vision models in \S~\ref{sec:teeslice_experiment} are also
applicable to NLP models. To demonstrate the generalization of \tool, we
evaluate \tool on a representative NLP model, BART~\cite{lewis2019bart}, and
three NLP datasets (SST-2, MRPC, and RTE) from the popular GLUE
dataset~\cite{wang2018glue}. We mainly report \ac{ms} accuracy in
\T~\ref{tbl:nlp}, and omit \ac{mia} accuracy as the partition strategy of \tool
does not leak additional membership information. The comparison baselines are
``No-Shield'' and ``Black-box'', aligned with
\S~\ref{sec:evaluate_existing_solutions}. The results are generally consistent
with \S~\ref{sec:teeslice_experiment}. The attack accuracies of \tool are
comparable with ``Black-box'' and are lower than ``No-Shield''. Besides, the
FLOPs of shielded layers by \tool are significantly lower than ``Black-box''
(over 10$\times$). Thus, we interpret that NLP models can also be effectively
shielded by \tool, and our findings in \textbf{RQ4} are generalizable to NLP
models.

\begin{table}[h]
    \caption{\ac{ms} accuracy on NLP tasks against \tool.}
    \label{tbl:nlp}
    \vspace{-5pt}
    \centering
    \small
    \begin{adjustbox}{max width=0.4\linewidth}

    \begin{tabular}{@{}ccccc@{}}
    \toprule
              & SST-2   & MRPC    & RTE     & Average \\ \midrule
    Black-box & 50.92\% & 68.87\% & 48.38\% & 56.05\% \\
    No-Shield & 92.55\% & 85.05\% & 66.79\% & 81.46\% \\
    TEESlice  & 50.92\% & 68.64\% & 46.93\% & 55.49\% \\ \bottomrule
    \end{tabular}

\end{adjustbox}
\vspace{-15pt}

\end{table}

\begin{tcolorbox}[size=small]
\textbf{Answer to RQ7}: \tool applies to NLP tasks and manifests
consistently high effectiveness.
\end{tcolorbox}

\section{Threats to Validity}
\label{sec:threats_to_validity}

\parh{Internal Validity.}
\diff{The setting of hyper-parameters for different model partition solutions is one threat to internal validity. We mitigate this threat by adopting the recommended setting of the original paper and iterate all possible model partition settings (\F~\ref{fig:acc_flops_append_one_fig} and \F~\ref{fig:mia_flops_append_one_fig}). For the training hyper-parameters of the two attacks, we follow the settings of prior literature~\cite{liu2022mldoctor,orekondy2019knockoff}.}

\parh{External Validity.}
\diff{The choices of datasets, models, and attacks are threats to external validity. 
The dataset and model selection refers existing benchmark~\cite{liu2022mldoctor} and prior work~\cite{orekondy2019knockoff}. We tried our best to cover more datasets and models. There are total $4$ (datasets) $\times$ $5$ (models) $= 20$ settings in our experiment, which is comparable to the previous benchmark~\cite{liu2022mldoctor}.
For the attacks, we mitigate this threat by using implementations of previous work. The implementation of model stealing is widely used in prior security work~\cite{jagielski2020high,chen2022copy}. The privacy inference attack is referred to previous work~\cite{liu2022mldoctor,yuan2022membership,carlini2022membership}. We choose the membership inference attack to measure the privacy leakage of prior model partition solutions because it is widely studied in the security community.}

\parh{Construct Validity.}
\diff{The choice of evaluation metrics is one threat to construct validity. We
mitigate this threat by selecting a wide range of diverse metrics. There are
three metrics for model stealing and four metrics for membership inference. We
use FLOPs as a platform-irrelevant metric to evaluate the utility cost inside
TEEs. We mitigate the threat of FLOPs by running different models on the
industrial-level SGX platform and record the relationship between FLOPs and
model run-time. As \F~\ref{fig:occlum_time_flops} shows, the value of FLOPs
within the TEE is positively correlated with the inference time. We omit the
inference time on the GPU because the time is
negligible~\cite{tramer2019slalom}. }

\section{Other Related Work}
\label{sec:other_related_work}

In addition to the \tsdp approaches in
Section~\ref{sec:evaluate_existing_solutions}, we notice the following areas related
to the security of DNN models with TEEs.

\parh{GPU TEE.}~\diff{Integrating TEE in GPUs and extending CPU TEEs to use GPU resources has become a hot
topic in both academia~\cite{volos2018graviton,
hua2020guardnn,wang2023building,shwetashinde2023acai} and industry~\cite{NvidiaH100}. Such solutions usually require customizing hardware and are designed for server centers. For example, Nvidia H100 GPU has integrated TEE to
support confidential computation~\cite{NvidiaH100}. Researchers also explored building TEE for
ARM GPU~\cite{wang2023building} and integrating GPU to ARM
CCA~\cite{shwetashinde2023acai}. Moreover, Intel TDX can connect the TEE of Nvidia
H100~\cite{NvidiaH100}. }

\diff{These advances may make it possible to shield the
entire DNN model in TEE. However, we believe that our \tool solution is still important even when GPU TEE is available
because of the following reasons. 
First, GPU TEE is still not widely available on conmercial GPUs. Currently, only high-level GPUs incorporate TEEs, and many academic solutions are not implemented on commercial GPUs. In contrast, \tool serves as a software-based alternative, adaptable to existing low-end GPUs, offering a more flexible approach to confidential computation.
Second, GPU TEE is not applicable to multi-GPU scenarios, such as model training
and inference on large language models. In this context, our solution becomes an important solution
for multi-GPU scenarios. }

Moreover, our solution can also act as a complementary solution to H100-based solutions. Thus,
this paper employs TEE in GPUs as an ``out-of-the-box'' manner.

\parh{Side Channels.}~\diff{One of the most important threats to TEEs is side-channel attacks~\cite{nilsson2020a,bulck2018nemesis,kocher2019spectre,chen2020sgxpectre,schaik2019ridl,murdock2020plundervolt}.}
While side channels may threaten DNN privacy, various defensive methods have
been proposed to mitigate side channel
breaches~\cite{nilsson2020a,lee2017inferring,chen2018racing,gruss2017strong}. \diff{Gruss \etal use hardware transactional memory to prevent cache-based side-channel attacks~\cite{gruss2017strong}. Lee \etal propose a software-based countermeasure to mitigate one of the critical side-channel attack, branch shadowing attack~\cite{lee2017inferring}. Chen \etal address hyper-threading side-channel threats by creating a special shadow thread for each enclave thread~\cite{chen2018racing}. Other possible solutions include incorporating some cryptographic
protections in the TEE to protect data confidentiality in the worst case. For
example, recent work combines low-level TEE and cryptographic solutions to
provide a strong security guarantee with less side-channel
leakage~\cite{wu2022hybrid,huang2022stamp,dong2023poster}.}
\sys can be integrated with these various defenses to reduce side channel
leakages.

\parh{Model Stealing Attacks.}~\diff{De facto model stealing techniques can be divided
into two camps: hardware-based attacks and query-based attacks. For the first
camp, the adversary uses side channels of the system to steal model weights and
architectures~\cite{zhu2021hermes,rakin2022deepsteal,hua2018reverse,yan2020cache,batina2019csi,hu2020deepsniffer}. For example, Batina \etal show that side-channel attack can reveal much important information of a neural network~\cite{hua2018reverse}, including number of layers, activation functions and weights \etal These information can be used to effectively reverse engineer networks.  
For the second camp, people aim to train a surrogate model that mimic the victim
model~\cite{tramer2016stealing,papernot2017practical,jagielski2020high,orekondy2019knockoff,chandrasekaran2020exploring}. For example, Jagielski \etal propose a practical functionally-equivalent extraction attack for directly extracting a model's weights~\cite{jagielski2020high}.
Model stealing attacks can also be boosted with the help of reinforcement
learning~\cite{orekondy2019knockoff} and active
learning~\cite{chandrasekaran2020exploring}. Zhu~\etal use unencrypted PCIe
traffic to steal the precise value of model weights~\cite{zhu2021hermes}.
Rakin~\etal propose a framework to effectively steal DNN weights with the aid
of memory side-channel attack~\cite{rakin2022deepsteal}. 
Hua~\etal infer
partial model weights when an accelerator performs dynamic zero pruning for
on-chip memory accesses~\cite{hua2018reverse}. 
Existing countermeasures include
perturbing the model output~\cite{orekondy2020prediction,
tramer2016stealing,lee2019defending}, returning the
label~\cite{tramer2016stealing, chandrasekaran2020exploring}, and analyzing
suspicious queries~\cite{juuti2019prada}.}

\parh{Membership Inference Attack.}~\diff{Membership inference attacks can be
conducted in both black-box settings and white-box
settings~\cite{shokri2017membership,salem2019mlleak,nasr2019comprehensive,song2021systematic}. Specifically, Nasr \etal propose a novel white-box membership to trace model's training data records~\cite{nasr2019comprehensive}, while Shokri \etal train shadow models to inference membership given only a black-box access to a model\cite{shokri2017membership}. Moreover, Song and
Mittal conduct a systematic evaluation on the membership leakage of DNN
models~\cite{song2021systematic}.
In addition to the standalone model paradigms, Melis~\etal found that model
updates conducted during collaborative learning may leak membership
information~\cite{melis2019exploiting}. Prior work has shown that overfitting is
one primary cause of membership
leakage~\cite{salem2019mlleak,shokri2017membership,yuan2022membership}, therefore regularization technology can certainly defend against such attacks~\cite{kaya2020effectiveness}. Other countermeasures include using transfer learning, information perturbation and generative model-based method~\cite{hu23survey}. Huang \etal leverage domain adaptation to defense membership inference attack~\cite{huang22mid}.
Hu \etal defense such attacks by utilizing the data generated by GANs to get a surrogate model~\cite{hu23gan}.}

\parh{Shielding-Whole-Model by TEE.}~We have reviewed existing \tsdp\ solutions
in \S~\ref{sec:literature_category}. In addition to splitting DNNs and
offloading certain parts of the model on GPUs to speedup model inference, we
also notice existing works explore putting the entire DNN models into
TEEs~\cite{lee2019occlumency, hanzlik2021mlcapsule, li2021lasagna,
kim2020vessels, shen2020occlum}. \diff{Occlum is an industrial-level library OS for
Intel SGX and provides various DNN applications~\cite{shen2020occlum}. MLCapsule
implemented defenses against various attacks inside SGX, such as model stealing,
reverse engineering and membership inference~\cite{hanzlik2021mlcapsule}. To
improve the efficiency of TEE-shielded solutions, Lasagna implemented a task
scheduler~\cite{li2021lasagna} and Vessels optimizes memory
usage~\cite{kim2020vessels}.  Nevertheless, even after efficiency optimization,}
these works still notably sacrifice the utility of the protected DNN models due
to the lack of GPU support and rich computation resources.

\parh{TSDP for DNN Training.}~Researchers have proposed various TSDP solutions
for DNN training to protect the privacy of training data on the cloud
server.
Hashemi\etal create input obfuscation within a TEE and offload the data to GPUs for fast computation~\cite{hashemi2021darknight}. \diff{Lucien \etal use additive secret
sharing to protect privacy ~\cite{lucien2021goten}. Tramer \etal outsource computation from a TEE to a co-located GPU~\cite{tramer2019slalom}.} These
solutions are different from \tool because \tool is designed to
protect the model inference stage.

\section{Discussion}
\label{sec:discussion}

\parh{Other Choices of $Security$ and $Utility$.}~This paper aims to
comprehensively evaluate \ac{tsdp} with seven empirical metrics of $Security$
from well-developed attack toolkits~\cite{liu2022mldoctor,orekondy2019knockoff}.
These seven metrics cover the majority of \ac{ms} and \ac{mia} in literature. We
notice that differential privacy (DP) can also theoretically quantify
$Security$~\cite{dwork2014DP}, but we decide not to use it due to its
prohibitively high computational cost for large models. 

An intuitive choice of $Utility$ is the model inference latency. However, as
there are various TEE architectures on the market, \textit{e.g.}, Intel
SGX~\cite{mckeen2013innovative}, AMD SEV~\cite{kaplan2016amd}, Intel
TDX~\cite{IntelTDX}, ARM CCA~\cite{ArmCCA}, and
TrustZone~\cite{alves2004trustzone}, evaluating the latency on all DNN models,
TEE architectures and possible configurations is difficult. We leverage FLOP, a
platform-irrelevant function, to form $Utility$, and therefore, our conclusion
should not be affected by the TEE implementation details, and is generally
applicable to the wide range of TEE architectures.

\parh{Attacks to Black-box Models.}~Attackers can still compromise \sys with
black-box attacks~\cite{juuti2019prada, orekondy2019knockoff,
orekondy2020prediction, papernot2017practical,
tramer2016stealing,li2021membership,mehnaz2022are}. However, the black-box
attacks are \textit{much less effective} due to the lack of information about
model architectures and model weights. That is, we deem black-box attacks as the
upper bound security guarantee that can be offered by TEEs. Several methods are
proposed to mitigate black-box
attacks~\cite{juuti2019prada,orekondy2020prediction};
we view those defenses are orthogonal to TEE-based defenses.

\parh{Hyper-Parameters of \sys.}
We clarify that although \sys has involved hyper-parameters in
the training phase, those parameters are merely used for reducing the computation cost
inside TEE (amount of privacy-related slices) instead of influencing privacy
leakage. The training phase of \sys relies on several hyper-parameters,
including $\delta$, $\alpha_{\rm setup}$, $n$, and $rounds$. They are mundane in
training our model setup. Thus, we clarify that it is unnecessary to tune those
parameters and benchmark if their different values may influence privacy leakage.

\parh{New TEE Architectures.}
Desipte the traditional TEE architectures (\eg Intel
SGX~\cite{mckeen2013innovative} and ARM TrustZone~\cite{alves2004trustzone})
that have been widely used in the industry, new TEE architectures are still
emerging (\eg Intel TDX~\cite{IntelTDX} and ARM CCA~\cite{ArmCCA}). Such new
architectures may have stronger computation abilities. For example, Intel TDX
has a larger encrypted memory of 1 TB~\cite{IntelTDX}. Although such new TEEs
may mitigate the performance overhead of \shieldwhole solutions, they do not
harm the practicality of \tool because the computation speed of such new TEEs is
still not comparable with GPUs, not to say the GPU architectures are also
evolving. We believe \tool can be a promising solution to bridge the gap between
the new TEEs and the evolving GPUs.

\parh{Application Scope of \tool.}
This paper focuses on an important application scope: protecting DNN privacy on the user's
end devices with TEEs. With the development of hardware architectures, many mobile/IoT devices are already equipped with TEEs by default, such as TrustZone in Raspberry Pi~\cite{TrustZoneRP3} and Android 7~\cite{TrustZoneAndroid}. With the increasing awareness on the user privacy and companies' intellectual property embedded in the DNN model, we believe this topic will attract more attention.

\parh{Differential Privacy (DP).}
DP is a promising technique to theoretically quantify the privacy leakage of DNN
training data to defense against \ac{mia}~\cite{dwork2014DP}. Nevertheless, DP
is not designed to defense \ac{ms}. Besides, recent works show that DP may
provide insufficient privacy~\cite{mo2021ppfl}, harm
utility/fairness~\cite{bagdasaryan2019differential}, or degrade
performance~\cite{subramani2021enabling}.

\parh{Black-box Model Output.}
Following prior \ac{tsdp} papers and also real-world
productions~\cite{mo2020darknetz,hou2021model,shen2022soter,sun2020shadownet},
we assume that the deployed models only generate labels 
to users (i.e., ``label-only outputs''). In other words, we assume that the confidence of the model
prediction is an intermediate result, therefore, can be protected by TEEs.
This assumption is supported by a comprehensive survey on the output type of
on-device ML systems~\cite{sun2021mind}. Further, we also surveyed the eight
most important on-device ML tasks. For each task, we collect the three most
downloaded Android applications (24 apps in total) over three different
application markets (Google Play, Tencent My App, and 360 Mobile Assistant). We
manually checked the output type of the applications and found that \textit{all}
of the 24 applications only return the prediction label and keep the confidence of models in the intermediate results. 

\parh{DNN/LLM Quantization.}
Recently, some work aims to quantize DNN and LLM so that these models can be directly run on CPUs for inference without the need for GPU~\cite{shen2023efficient,zhou2024survey,xiao2023smoothquant,lin2024awq}. After careful algorithm design and engineering optimization, the inference speed on the CPU is accelerated by a large margin. 
However, both CPU and GPU can benefit from the quantization, and the gap between CPU and GPU still exists~\cite{cpu_gpu_inference}. For a widely-used language model, Llama-7B~\cite{touvron2023llama}, the inference speed of a low-grade GPU, GeForce RTX 4090, is over one hundred tokens per second. On the contrary, the inference speed of a high-end CPU, Apple M1 Pro, on the same model is only about fifteen tokens per second, which is $6.6\times$ slower than GeForce RTX 4090~\cite{cpu_gpu_inference}. Thus, even after model quantization, a GPU is also desired for fast computation. 
We regard this research line as an orthogonal direction of \tool, and the quantization technique can be integrated with \tool. This is because both GPU, CPU and TEE inference can be boosted by quantization. The backbone and private slices of \tool can be further optimized by quantization techniques to improve the inference speed.

\parh{TSDP for LLM.}
Recently, some concurrent work has also applied TSDP to LLM to protect the model security and training data privacy~\cite{huang2024fast,li2024translinkguard}. They propose to use SGX and TDX to protect the non-linear layers of LLM. The difference between their work and \tool is that they do not optimize the slices in the TEE. The computation amount in the TEE of \cite{huang2024fast} is still larger than \tool and leads to more overhead.

\parh{Future Work.}
In the future, we will focus on expanding the evaluation metrics and exploring new approaches to enhance the security and utility of TSDP. First, while this paper relies on eight major metrics for assessing security and utility, exploring more representative metrics is a promising direction for future exploration. Additionally, the continual emergence of new TEE architectures, such as Intel TDX and ARM CCA, presents opportunities to mitigate performance overheads and enhance computational efficiency. We will further investigate these architectures to optimize TSDP performance. Furthermore, to apply the TSDP to larger models, such as LLMs, our goal is to develop techniques to boost inference speed across CPUs, GPUs, and TEEs. By addressing these aspects, our future work aims to refine and expand the applicability of TSDP, ensuring robust protection for diverse DNN models, especially LLMs, across various hardware configurations.

\section{Conclusion}

\diff{We have conducted a thorough examination of the existing \tsdp\ solutions and have identified their vulnerabilities to privacy breaches. Additionally, we discuss the challenges associated with determining the optimal DNN partition configurations, known as the "sweet spot", which can vary significantly across different models and datasets. Drawing insights from the weaknesses observed in previous \tsdp\ approaches, we introduce \tool, an innovative \tsdp\ technique that utilizes the \partitionbeforetrain approach. This method achieves a high level of accuracy and security protection (similar to the \shieldwhole baseline) with significantly lower computational overhead (approximately $10\times$ less). \minor{We also demonstrate that \tool is scalable to large language models. After extracting the private functionalities into lightweight slices, \tool can provide black-box level protection by only shielding slices in TEE.}
}

\bibliographystyle{ACM-Reference-Format}
\bibliography{reference}

\end{document}